\documentclass[lettersize,journal]{IEEEtran}

\usepackage[justification=centering]{caption}
\usepackage{subcaption}
\usepackage[caption=false,font=normalsize,labelfont=sf,textfont=sf]{subfig}
\usepackage{bbm}
\usepackage{amsmath,amsfonts}
\usepackage{algorithmic}
\usepackage{algorithm}
\usepackage{textcomp}
\usepackage{stfloats}
\usepackage{url}
\usepackage{verbatim}
\usepackage{graphicx}
\usepackage{cite}
\usepackage{array}
\usepackage{booktabs, caption}
\usepackage{multirow, makecell}
\usepackage{lipsum}

\usepackage{xcolor}
\usepackage{amssymb}
\usepackage{hhline}
\usepackage{tikz}

\newcommand\encircle[1]{%
	\tikz[baseline=(X.base)] 
	\node (X) [draw, shape=circle, inner sep=0] {\strut #1};}

\usepackage{enumitem}

\NewExpandableDocumentCommand\XGap{m}{\noalign{\vskip #1}}
\NewExpandableDocumentCommand\Gap{}{\XGap{3pt}}

\newtheorem{lemma}{Lemma}

\newtheorem{remark}{Remark}

\DeclareMathOperator*{\argmin}{argmin}

\begin{document}
	
	\title{Cyber Attacks Prevention Towards Prosumer-based EV Charging Stations: An Edge-assisted Federated Prototype Knowledge Distillation Approach}
	
	\author{Luyao Zou, Quang Hieu Vo, Kitae Kim,  ~\IEEEmembership{Student Member,~IEEE}, Huy Q. Le, Chu Myaet Thwal, Chaoning Zhang and Choong Seon Hong, ~\IEEEmembership{Fellow,~IEEE}
		\thanks{© 2024 IEEE.  Personal use of this material is permitted.  Permission from IEEE must be obtained for all other uses, in any current or future media, including reprinting/republishing this material for advertising or promotional purposes, creating new collective works, for resale or redistribution to servers or lists, or reuse of any copyrighted component of this work in other works.}
		\thanks{This work was supported by Institute of Information \& communications Technology Planning \& Evaluation (IITP) grant funded by the Korea government(MSIT) (No.2019-0-01287, Evolvable Deep Learning Model Generation Platform for Edge Computing),  the National Research Foundation of Korea(NRF) grant funded by MSIT) (No. RS-2023-00207816), IITP grant No.2021-0-02068, Artificial Intelligence Innovation Hub), the Convergence security core talent training business support program (IITP-2023(2023)- RS-2023-00266615), and the MSIT under the ITRC support program (IITP-2024-RS-2023-00258649) supervised by the IITP, NRF grant funded by the Korea government(MSIT) (No. RS-2024-00352423), IITP grant funded by MSIT (No. RS-2024-00509257, Global AI Frontier Lab), and KRIT grant funded by Defense Acquisition Program Administration (DAPA) (KRIT-CT-24-001). Dr. CS Hong is the corresponding author.}
		\thanks{Luyao Zou, Quang Hieu Vo, Kitae Kim, Huy Q. Le, Chu Myaet Thwal and Choong Seon Hong are with the Department of Computer Science and Engineering, Kyung Hee University, Yongin-si, Gyeonggi-do 17104, Republic of Korea,
			emails: \{zouluyao, 2019310178, glideslope, quanghuy69, chumyaet, cshong\}@khu.ac.kr.}
		\thanks{Chaoning Zhang is with the Department of Artificial Intelligence, School of Computing, Kyung Hee University, Yongin-si, Gyeonggi-do 17104, Republic of Korea, email: chaoningzhang1990@gmail.com.}
	}
	
	\markboth{Journal of \LaTeX\ Class Files,~Vol.~14, No.~8, August~2021}%
	{Shell \MakeLowercase{\textit{et al.}}: A Sample Article Using IEEEtran.cls for IEEE Journals}
	
	\maketitle
	
	\begin{abstract}	
		In this paper, cyber-attack prevention for the prosumer-based electric vehicle (EV) charging stations (EVCSs) is investigated, which covers two aspects: 1) cyber-attack detection on prosumers’ network traffic (NT) data, and 2) cyber-attack intervention. To establish an effective prevention mechanism, several challenges need to be tackled, for instance, the NT data per prosumer may be non-independent and identically distributed (non-IID), and  the boundary between benign and malicious traffic becomes blurred. To this end, we propose an edge-assisted federated prototype knowledge distillation (E-FPKD) approach, where each client is deployed on a dedicated local edge server (DLES) and can report its availability for joining the federated learning (FL) process.  Prior to the E-FPKD approach, to enhance accuracy, the Pearson Correlation Coefficient is adopted for feature selection. Regarding the proposed E-FPKD approach, we integrate the knowledge distillation and prototype aggregation technique into FL to deal with the non-IID challenge. To address the boundary issue, instead of directly calculating the distance between benign and malicious traffic, we consider maximizing the overall detection correctness of all prosumers (ODC), which can mitigate the computational cost compared with the former way. After detection, a rule-based method will be triggered at each DLES for cyber-attack intervention. Experimental analysis demonstrates that the proposed E-FPKD can achieve the largest ODC on NSL-KDD, UNSW-NB15, and IoTID20 datasets in both binary and multi-class classification, compared with baselines. For instance, the ODC for IoTID20 obtained via the proposed method is separately $0.3782\%$ and $4.4471\%$ greater than FedProto and FedAU in multi-class classification.
	\end{abstract}
	
	\begin{IEEEkeywords}
		Cyber-attack prevention, prosumer-based EV charging stations, non-IID data, federated prototype knowledge distillation.
	\end{IEEEkeywords}
	
	\section{Introduction}
	\IEEEPARstart{A}{t} present, electric vehicles (EVs) have gradually penetrated into our daily lives, which have the potential to reduce the emissions of greenhouse gas thereby protecting the environment \cite{L_Zou_Intelligent_EV_Charging_Urban_Prosumer_Communities}. Specifically, there may be around 125 million EVs being expected to be on the road by 2030 \cite{S_Islam_Intelligent_Privacy_Preservation}. By 2040, EVs will account for $55\%$ of the new global vehicle sales and $33\%$ of vehicles on the road estimated by Bloomberg \cite{T_Morstyn_Conic_Optimization_EV}. Owing to the penetration of EVs, the demand for charging facilities has also gradually surged. To this end, prosumers which can be regarded as EV charging stations (EVCSs) \cite{L_Zou_Edge_assisted_Attention_Multi_Step} are focused in this article. Particularly, we concentrate on the smart meter-equipped prosumers. Nevertheless, smart meters are easy to suffer from  cyber-attacks \cite{A_Takiddin_Robust_Detection_Electricity_Theft}, despite their tamper-detection and encrypted communication capabilities \cite{Sook_Chin_Yip_Anomaly_Detection_identifying}. Due to this defect, attackers are capable of hacking into the meters \cite{A_Takiddin_Robust_Detection_Electricity_Theft}. When the meters are hacked, the energy consumption readings can be tampered which can result in electricity bill reduction \cite{A_Takiddin_Robust_Detection_Electricity_Theft}, leading to vast economic losses for the power utilities all over the world \cite{K_Zheng_density_clustering}. Hence, it is pivotal important to prevent cyber attacks on prosumer-based EV charging stations that are equipped with smart meters.
	\par
	To prevent the cyber attacks, it is necessary to identify the cyber attacks, which can be achieved by detecting the network traffic (NT) data. However, as cyber-attacks continue to grow in sophistication \cite{L_Zou_multiple_cloud} and the boundary between malicious and benign traffic becomes fuzzy \cite{Y_Yue_Contrastive_Intrusion}, detecting cyber-attacks remains a huge challenge. Thus, a method that can identify cyber-attacks as accurately as possible needs to be designed. Centralized learning method has the potential to detect the network traffic (NT) for the prosumer-based EV charging stations. However, this method requires the centralized collection of NT data from each distributed prosumer, which may arouse data privacy issue \cite{Y_M_Saputra_Energy_Demand_FL_EV_Networks}. Thanks to federated learning (FL) which is a privacy-preserving decentralized method \cite{S_Abdulrahman_Survey_FL_Journey}, handling that issue becomes possible since FL can train a model without taking the data off-premises. Nevertheless, network traffic data of all prosumers may be non-independent and identically distributed (non-IID), which may cause FL performance degradation \cite{Y_Kim_Dynamic_Clustering}. Besides, in practice, it is unrealistic to presume all the clients can always serve the training process \cite{T_Huang_efficiency_boosting_client_Selection_fairness}. Thus, there may exist a phenomenon that some clients may be reluctant to participate in the training process.
	\par
	To this end, to address the aforementioned issues,	an edge-assisted federated prototype knowledge distillation (E-FPKD) approach is proposed for cyber attack prevention towards prosumer-based EVCSs. To be specific, through the proposed method, firstly, cyber-attacks will be detected by using the network traffic (NT) data. Secondly,  the detected cyber attacks will be intervened via rule-based method \cite{M_Habiba_Edge_Intelligence_Network_Intrusion_Prevention} at each edge server. Additionally, prior to the proposed E-FPKD approach, since feature selection process can increase the accuracy \cite{T_Wisanwanichthan_Double_Layered_Hybrid_Approach_NIDS}, feature selection with Pearson Correlation Coefficient (PCC) is adopted. For the proposed E-FPKD approach, it is designed based on the FL framework so that it permits each client can join in the training process without letting its local data leave the local side. Thus, the data privacy concern regarding submitting data to a central place can be solved. In addition, considering the non-IID NT data, we employ knowledge distillation (KD) for each client, owing to the following benefits: 1)  its easiness of implementation in any off-the-shelf deep learning (DL) architecture without the need for extra software or hardware \cite{F_Tung_Similarity_Preserving_KD}, 2) applying KD into FL is an effective method to deal with the data heterogeneity \cite{J_Tang_FedRAD_Heterogeneous}. In addition to KD, we also employ the prototype aggregation mechanism into the considered FL framework, because of its supportability for heterogeneous clients \cite{Y_Tan_FedProto_Federated_Prototype_Learning}. Finally, rule-based method is triggered to intervene the detected attacks. To be friendly, the clients' willingness for joining the FL process (i.e., participating availability per client) is additionally integrated.
	\par
	The cyber-attack prevention (cyber attack detection and intervention), KD, and prototype aggregation have been studied by some previous studies, respectively. For ease of understanding, brief summary of several previous studies and this article are conducted. The detailed contents will be illustrated in Section \ref{Section_2_related_work}. In regard to the cyber-attack detection for EVCS, \cite{L_Zou_multiple_cloud, ElKashlan_Aslan_Intrusion_EV_Charging_System, M_ElKashlan_ML_IDS_IoT_electric, M_Basnet_DL_Intrusion_EV_Charging} were proposed. However, none of them consider integrating KD technique and prototype aggregation into FL framework to prevent cyber attacks on prosumer-based EV charging stations. Cyber-attack intervention was considered by \cite{M_Habiba_Edge_Intelligence_Network_Intrusion_Prevention, M_A_Rahman_Data_Mining_Cyber_Attack, V_V_Vegesna_Utilising_VAPT_Technologies}. Particularly, the preventive event was adopted by \cite{M_Habiba_Edge_Intelligence_Network_Intrusion_Prevention} for Distributed Denial-of-Service (DDoS) attacks prevention, the expert-based intervention was employed by \cite{M_A_Rahman_Data_Mining_Cyber_Attack}, and the vulnerability assessment and penetration testing was proposed by \cite{V_V_Vegesna_Utilising_VAPT_Technologies} for cyber attack prevention, which differ from this work. In \cite{P_Qi_FedBKD_Heterogenous_FL_BKD, H_Jin_Personalized_self_KD, C_Wu_Communication_efficient_FL_KD, H_Q_Le_layer_wise_KD_cross_device_FL}, KD approach has been considered. Specifically, with respect to \cite{P_Qi_FedBKD_Heterogenous_FL_BKD, H_Jin_Personalized_self_KD, H_Q_Le_layer_wise_KD_cross_device_FL}, the type of KD adopted by those three studies are different from this work. In the case of \cite{C_Wu_Communication_efficient_FL_KD}, although a mentor network and a mentee network are considered for building each client, prototype aggregation is not utilized. For \cite{T_Gao_Model_Heterogeneous_FL, R_Zhao_Semi_Supervised_FL_KD_Intrusion, J_Shen_Effective_ID_Heterogeneous_IoT}, various KD-based FL approaches were proposed for intrusion detection, while the study \cite{Zou_EFCKD_contrastive_KD_Energy_Theft} aims to detect the energy theft via the method that incorporates FL with KD and model contrastive learning. However, none of them consider the prototype aggregation approach. In regard to \cite{D_Cheng_ProtoHAR_Prototype_FL, A_Wang_Heterogeneous_Defect_Prediction, T_Gao_FedMBP_multi_branch_prototype_FL_Heterogeneous_Data, B_Li_Prototype_Decentralized_FL_varying}, despite all of them introducing the prototype aggregation technique, none of them simultaneously introduce feature selection, client availability reported by the client itself, KD, and prototype aggregation to detect the cyber attacks towards the prosumer-based EV charging stations. Thus, unlike \cite{L_Zou_multiple_cloud}, \cite{M_Habiba_Edge_Intelligence_Network_Intrusion_Prevention, M_A_Rahman_Data_Mining_Cyber_Attack},  \cite{ElKashlan_Aslan_Intrusion_EV_Charging_System, M_ElKashlan_ML_IDS_IoT_electric, M_Basnet_DL_Intrusion_EV_Charging, P_Qi_FedBKD_Heterogenous_FL_BKD, H_Jin_Personalized_self_KD, C_Wu_Communication_efficient_FL_KD, H_Q_Le_layer_wise_KD_cross_device_FL,T_Gao_Model_Heterogeneous_FL, R_Zhao_Semi_Supervised_FL_KD_Intrusion, J_Shen_Effective_ID_Heterogeneous_IoT, Zou_EFCKD_contrastive_KD_Energy_Theft} and \cite{D_Cheng_ProtoHAR_Prototype_FL, A_Wang_Heterogeneous_Defect_Prediction, T_Gao_FedMBP_multi_branch_prototype_FL_Heterogeneous_Data, B_Li_Prototype_Decentralized_FL_varying}, we comprehensively employ feature selection and meanwhile, integrate client availability and KD along with prototype aggregation in FL to detect the cyber-attacks for the prosumer-based EVCSs, while we adopt rule-based approach to intervene in the cyber-attacks. The main contributions of this work are illustrated as follows:
	\begin{itemize}
		\item We consider cyber-attacks prevention for prosumer-based EVCSs, where prosumers are regarded as the infrastructure for charging EVs. Particularly, cyber-attack detection via NT data, and cyber-attack intervention are focused. 
		\item For cyber-attack detection, to make the boundary between benign and malicious traffic as clear as possible, maximizing the mean distance between samples from benign and malicious traffic is considered. However, this manner has a high computational cost if each prosumer has many NT data and/or has high dimensional feature space (proved in Section \ref{Malicious_Benign_Boundary_Model_Section_3_B}). Thus, instead, maximizing the overall detection correctness of all prosumers (ODC) is considered, which can achieve lower computational cost and is proven in Section \ref{Problem_Formulation}. Afterward, we consider intervention against the detected cyber-attacks.		
		\item We propose a method named edge-assisted federated prototype knowledge distillation (E-FPKD) approach which is capable of overcoming the non-IID NT data. Before utilizing the proposed E-FPKD method, the Pearson Correlation Coefficient (PCC) is adopted to select features over the NT data. Besides, we give each client the freedom to report their own availability for participating in FL process. In addition, we introduce KD for each client. Motivated by \cite{C_Wu_Communication_efficient_FL_KD} which constructs each client with a mentor and a mentee model, we consider each client owns a teacher and a student network. Moreover, we introduce prototype aggregation \cite{Y_Tan_FedProto_Federated_Prototype_Learning} at each round, while we additionally add model aggregation in the final round to generate global model. Then a rule-based method called \textit{``if...then"} \cite{S_Wang_Integrated_IDS_Cluster_Wireless} is employed for cyber-attacks intervention.
		\item Finally, a comparison experiment is conducted on three datasets, i.e., NSL-KDD \cite{Dataset_NSL_KDD}, UNSW-NB15 \cite{Dataset_UNSW_NB15_1, Dataset_UNSW_NB15_2, Dataset_UNSW_NB15_3, Dataset_UNSW_NB15_4, Dataset_UNSW_NB15_5}, and IoTID20 \cite{Dataset_IoTID20}. Specifically, we compare the proposed method with several base methods, e.g., FedProto \cite{Y_Tan_FedProto_Federated_Prototype_Learning}, pFedSD \cite{H_Jin_Personalized_self_KD}, FedAU \cite{S_Wang_Lightweight_Method_Tackling}, and FedExP \cite{D_Jhunjhunwala_FedExP_Speeding_Up}. Experimental results clarify that the proposed method's effectiveness is notable compared with those baselines. For instance, for IoTID20, in binary classification, the ODC achieved by the proposed method is $1.0810\%$ and $16.10\%$ larger than that of pFedSD and FedExP, respectively.	
\end{itemize}
\par
The remainder of this article is given below. In Section \ref{Section_2_related_work}, we discuss several related works. In Section \ref{Section_3_Nomenclature}, we show the notations of this work. In Section \ref{Section_4_System_Model_PD}, the system model with associated problem formulation are described. Some basic knowledge regarding the proposed method is given in Section \ref{Section_5_Preliminaries}. In Section \ref{Section_6_Methodology}, we describe the proposed E-FPKD approach. In Section \ref{Section_7_experimental_result_analysis}, a comparison experiment is performed. The limitations and future potential area are provided in Section \ref{Section_8_Discussion}. Finally, a conclusion is drawn in Section \ref{Section_9_conclusion}.

\section{Related Work}
\label{Section_2_related_work}
As aforementioned, we focus on preventing cyber attacks by detecting and intervening in the attacks on NT data. To devise the solution, we consider employing knowledge distillation in FL, while adopting the prototype aggregation to deal with the non-IID challenge. In this section, we provide the previous literature on the following aspects: 1) cyber-attack detection for EV Charging Stations (EVCSs), 2) cyber-attack intervention/prevention, 3) knowledge distillation technique in FL, and 4) prototype aggregation technique.

\vspace{-0.15cm}
\subsection{Cyber-attack Detection for EV Charging Station}
	Cyber-attack detection for EV charging stations is one of the aspects considered by this article. Consequently, we list some previous studies and illustrate various points considered by this article. Currently, several studies have been proposed for the EVCS in terms of cyber-attack detection such as \cite{L_Zou_multiple_cloud, ElKashlan_Aslan_Intrusion_EV_Charging_System, M_ElKashlan_ML_IDS_IoT_electric, M_Basnet_DL_Intrusion_EV_Charging} (summarized in Table \ref{related_work_1}). Specifically, in \cite{L_Zou_multiple_cloud}, a multiple cloud-aided FL-based security module was proposed for cyber attack intrusion detection towards EVCS. Nevertheless, prosumer-based EVCS, KD and prototype aggregation were not considered by \cite{L_Zou_multiple_cloud}, whereas those points are the focus of this paper. \cite{ElKashlan_Aslan_Intrusion_EV_Charging_System} aimed to detect the Distributed Denial of Service (DDoS) attack towards the EVCS network. Different from \cite{ElKashlan_Aslan_Intrusion_EV_Charging_System}, we consider binary and multi-class classification scenarios. In binary classification, we do not distinguish the types of attacks; all types of attacks will be regarded as anomalies. In multi-class classification, we consider detecting various attacks. Another different point is that the FL-based approach was not taken into account by \cite{ElKashlan_Aslan_Intrusion_EV_Charging_System}, whereas we consider that approach in this paper. In \cite{M_ElKashlan_ML_IDS_IoT_electric}, since the risk of cyber-attacks is increasing, multiple ML-based classifier algorithms (e.g., Naive Bayes classifier, filtered classifier) were evaluated to detect the malicious traffic of the IoT EVCS. However, prosumer-based EVCS and FL-based approach were not considered by \cite{M_ElKashlan_ML_IDS_IoT_electric}. In \cite{M_Basnet_DL_Intrusion_EV_Charging}, to detect EVCS's denial of service (DOS) attacks, intrusion detection systems based on deep learning was proposed. However, prosumer-based EVCS and FL-based approach were not taken into account by \cite{M_Basnet_DL_Intrusion_EV_Charging}.
\begin{table*}[!t]
	\caption{Summary of \cite{L_Zou_multiple_cloud,ElKashlan_Aslan_Intrusion_EV_Charging_System, M_ElKashlan_ML_IDS_IoT_electric, M_Basnet_DL_Intrusion_EV_Charging} and this work regarding cyber-attack detection for EV charging station}
	\vspace{-0.2cm}
	\begin{center}
		\begin{tabular}{|m{1.2cm}|m{2.5cm}|m{2.8cm}|m{9cm}|}
			\hline
			\hfil \textbf{Ref.} & \textbf{Cyber-attack detection for EVCS} & \hfil \textbf{Prosumer-based EVCS} & \hfil \textbf{Remark} \\ \hline
			\hfil \cite{L_Zou_multiple_cloud} & \hfil Yes & \hfil No &  FL-based security module was proposed for cyber-attacks intrusion detection towards EVCS \\ \hline	
			\hfil \cite{ElKashlan_Aslan_Intrusion_EV_Charging_System} & \hfil Yes & \hfil No & Different machine learning techniques (e.g., decision Table classifiers) were proposed to detect the DDOS attacks for EVCS \\ \hline
			\hfil \cite{M_ElKashlan_ML_IDS_IoT_electric} & \hfil Yes & \hfil No & IoT EVCS's malicious traffic detection via multiple machine learning-based classifier algorithms (e.g., Naive Bayes classifier)\\ \hline		
			\hfil \cite{M_Basnet_DL_Intrusion_EV_Charging} & \hfil Yes & \hfil No &  DOS attacks detection for EVCS via deep learning \\ \hline			
			\hfil This work & \hfil Yes & \hfil Yes &  Federated prototype knowledge distillation approach is proposed to detect cyber attacks for prosumer-based EVCS \\ \hline							
		\end{tabular}
		\label{related_work_1}
	\end{center}
	\vspace{-0.4cm}
\end{table*}

\begin{table*}[!t]
	\caption{Summary of \cite{P_Qi_FedBKD_Heterogenous_FL_BKD,  H_Jin_Personalized_self_KD, C_Wu_Communication_efficient_FL_KD, H_Q_Le_layer_wise_KD_cross_device_FL} and this work regarding KD in FL}
	\vspace{-0.2cm}
	\begin{center}
		\begin{tabular}{|m{1.2cm}|m{1.7cm}|m{5cm}|m{6.5cm}|}
			\hline
			\hfil \textbf{Ref.} & \textbf{Use KD in FL} &  \hfil \textbf{KD Type} &  \hfil \textbf{Remark}\\ \hline
			\hfil \cite{P_Qi_FedBKD_Heterogenous_FL_BKD} & \hfil Yes &  Bidirectional KD-based FL was proposed & Public dataset is required. \\ \hline
			\hfil \cite{H_Jin_Personalized_self_KD} & \hfil Yes & Self-KD was proposed for personalized FL & Public dataset is not required. \\ \hline		
			\hfil \cite{H_Q_Le_layer_wise_KD_cross_device_FL} & \hfil Yes & Layer-wised KD in FL was proposed & Proxy dataset is needed.  \\ \hline		
			\hfil \cite{C_Wu_Communication_efficient_FL_KD} & \hfil Yes & Consider a mentor model and a mentee model to form each client. & Public dataset is not required.  \\ \hline					
			\hfil This work & \hfil Yes & Same as \cite{C_Wu_Communication_efficient_FL_KD}, a teacher model and a client model is considered to form each client. & Public dataset is not required. Besides, feature selection stage and prototype aggregation is considered.  \hfil \\ \hline									
		\end{tabular}
		\label{related_work_2}
	\end{center}
	\vspace{-0.65cm}
\end{table*}

\subsection{Cyber-attack Intervention/Prevention}
In this article, we consider cyber-attack prevention, and the intervention process is one of the steps for the purpose of preventing cyber-attacks. Thus, we provide several previous studies regarding cyber-attack intervention or prevention, such as \cite{M_Habiba_Edge_Intelligence_Network_Intrusion_Prevention, M_A_Rahman_Data_Mining_Cyber_Attack, V_V_Vegesna_Utilising_VAPT_Technologies}. Moreover, we also illustrate the differences between those studies and our proposed solution. In \cite{M_Habiba_Edge_Intelligence_Network_Intrusion_Prevention}, a deep learning-based method was proposed to prevent the network from being attacked by DDoS attacks and insecure data flows along with similar network intrusions, where a preventive event is triggered at the edge service, e.g., blocking data transmission. However, \cite{M_Habiba_Edge_Intelligence_Network_Intrusion_Prevention} does not employ FL, KD and prototype aggregation, which are important techniques considered by this paper. In addition, unlike \cite{M_Habiba_Edge_Intelligence_Network_Intrusion_Prevention}, we consider preventing various attacks by multi-class classification in this paper. In \cite{M_A_Rahman_Data_Mining_Cyber_Attack}, J48 decision tree algorithm was adopted to extract the patterns that are related to the cyber-attacks via historical data, and then a prediction model is built to forecast the future cyber-attacks. Afterwards, an expert will take the cyber-attacks intervention to address the cyber-attack. However, identical to \cite{M_Habiba_Edge_Intelligence_Network_Intrusion_Prevention}, FL, KD and prototype aggregation techniques do not utilized by \cite{M_A_Rahman_Data_Mining_Cyber_Attack}, either.  \cite{V_V_Vegesna_Utilising_VAPT_Technologies} provided vulnerability assessment and penetration testing for cyber attack prevention. Different from \cite{V_V_Vegesna_Utilising_VAPT_Technologies}, we adopt an FL-based deep learning approach to detect the cyber attacks and then we employ the rule-based approach to intervene in the detected cyber-attacks. 

\vspace{-0.35cm}
\subsection{Knowledge Distillation (KD) Technique in FL}
\par
In this work, KD is one of the techniques that we integrate into FL to construct the proposed solution. Thus, in this subsection, we will describe several previous literature regarding integrating KD technique in FL including  \cite{P_Qi_FedBKD_Heterogenous_FL_BKD,  H_Jin_Personalized_self_KD, C_Wu_Communication_efficient_FL_KD, H_Q_Le_layer_wise_KD_cross_device_FL} (summarized in Table \ref{related_work_2}). In \cite{P_Qi_FedBKD_Heterogenous_FL_BKD}, a heterogenous FL framework was proposed, where bidirectional KD (i.e., client-to-cloud and cloud-to-client distillation) was utilized to build each client model and public dataset for server is needed. In particular, for client-to-cloud distillation, global model is treated as the student network, while each local network are regarded as a teacher network. Thus, the global model will unify the knowledge from each local teacher network. For cloud-to-client distillation, the global model's knowledge will be distilled for each local network. In \cite{H_Jin_Personalized_self_KD}, personalized FL based on self-KD was proposed, where the knowledge of previous personal model belongs to a client will be distilled to guide its current local model, and no need to use the public dataset. Different from \cite{P_Qi_FedBKD_Heterogenous_FL_BKD} and \cite{H_Jin_Personalized_self_KD}, we consider each client forms by a teacher model and a student model in this article. This is inspired by \cite{C_Wu_Communication_efficient_FL_KD} that adopt a mentor model and a mentee model to build each client. However, prototype aggregation was not leveraged by \cite{C_Wu_Communication_efficient_FL_KD}, whereas we integrate KD with prototype aggregation in each round and model aggregation in the final round. As for \cite{H_Q_Le_layer_wise_KD_cross_device_FL}, layer-wise KD was adopted in FL, where a single teacher (server) and multiple students (multiple devices) were considered. Besides, \cite{H_Q_Le_layer_wise_KD_cross_device_FL} considered distilling the knowledge from various layers of the global model to teach the local student model. Moreover, \cite{H_Q_Le_layer_wise_KD_cross_device_FL} needs to use proxy dataset to produce knowledge. Different from \cite{H_Q_Le_layer_wise_KD_cross_device_FL}, in this paper, proxy dataset is not required, and each client is constructed by a teacher model and a student model. Besides, we consider distilling the knowledge from the teacher model to the student model of the same client.
\par
KD-based FL has been leveraged by several studies for various detection purposes. For instance, \cite{T_Gao_Model_Heterogeneous_FL, R_Zhao_Semi_Supervised_FL_KD_Intrusion, J_Shen_Effective_ID_Heterogeneous_IoT} have conducted research on KD-based FL approach for intrusion detection (ID), while \cite{Zou_EFCKD_contrastive_KD_Energy_Theft} tried to detect energy theft via KD-based FL method to ameliorate energy management. Particularly, in \cite{T_Gao_Model_Heterogeneous_FL}, dual teacher-student paradigm based FL approach was proposed for ID, where the knowledge of two teacher models will be distilled to guide student model. This structure is different from this article, where we build each client with a teacher and a student model. In \cite{R_Zhao_Semi_Supervised_FL_KD_Intrusion}, a KD-based semi-supervised FL approach was presented for ID, where unlabeled data was utilized by distillation approach for enhancing the performance of classifier. Unlike \cite{R_Zhao_Semi_Supervised_FL_KD_Intrusion}, we use labeled data and supervised learning approach. In \cite{J_Shen_Effective_ID_Heterogeneous_IoT}, FL ensemble KD (FLEKD) approach was proposed for ID, where the server-side model was regarded as student model, while the client models were treated as teacher models. Whereas, a client consists of a teacher and a student model is adopted by this article. In \cite{Zou_EFCKD_contrastive_KD_Energy_Theft}, federated contrastive KD was presented to detect energy theft for an urban area. However, although \cite{Zou_EFCKD_contrastive_KD_Energy_Theft} also considered building each client by a teacher and a student model, cyber-attack detection is not considered, whereas that is a main point of this work. Moreover, in \cite{T_Gao_Model_Heterogeneous_FL, R_Zhao_Semi_Supervised_FL_KD_Intrusion, J_Shen_Effective_ID_Heterogeneous_IoT, Zou_EFCKD_contrastive_KD_Energy_Theft}, prototype aggregation are not considered, whereas we adopt it in the proposed solution.

\vspace{-0.45cm}
\subsection{Prototype Aggregation (PA) Technique}
\par
In this part, we provide some previous studies that employ the prototype aggregation (PA), which is one of the techniques used by this work. Besides, we also point out the difference between those studies and this article. Several studies have considered using PA to address various problems, e.g., \cite{D_Cheng_ProtoHAR_Prototype_FL, A_Wang_Heterogeneous_Defect_Prediction, T_Gao_FedMBP_multi_branch_prototype_FL_Heterogeneous_Data, B_Li_Prototype_Decentralized_FL_varying}. Specifically, in \cite{D_Cheng_ProtoHAR_Prototype_FL}, PA was adopted to guide the FL framework for human activity recognition. To be more specific, \cite{D_Cheng_ProtoHAR_Prototype_FL} took into account transmitting both the prototypes and the representations between server and clients. Unlike \cite{D_Cheng_ProtoHAR_Prototype_FL}, in this paper, until the penultimate round, we only consider prototype transferring for the training process. In the final round, not only prototype transferring, but also model transferring are performed. In addition to this, we concentrate on cyber attack detection for prosumer-based charging stations. In \cite{A_Wang_Heterogeneous_Defect_Prediction}, PA was employed in FL for predicting the heterogeneous defect. \cite{T_Gao_FedMBP_multi_branch_prototype_FL_Heterogeneous_Data} proposed a multi-branch prototype FL for processing the image domains in natural and medical, where PA was executed across each layer of the representation layer. Nevertheless, both \cite{A_Wang_Heterogeneous_Defect_Prediction} and \cite{T_Gao_FedMBP_multi_branch_prototype_FL_Heterogeneous_Data} do not consider prosumer-based EVCSs, feature selection and KD.  PA was also adopted by \cite{B_Li_Prototype_Decentralized_FL_varying}, where prototype per device will be transmitted to the neighbors of that device for aggregation. Differ from this, we consider sending local prototypes to distribution system operator (DSO) for aggregation.

\vspace{-0.5cm}
\section{Nomenclature}
	\label{Section_3_Nomenclature}
	In this section, we illustrate the related notations that are used in this work, which are shown in Table \ref{section_3_tab1}.

\begin{table*}[!t]
	\caption{Summary of Notations}
	\begin{center}
		\begin{tabular}{|m{1.7cm}|m{6cm}||m{1.7cm}|m{6.5cm}|}
			\hline
			\hfil \textbf{Notation} & \hfil \textbf{Definition} & \hfil \textbf{Notation} & \hfil \textbf{Definition} \\ \hline \hline
			\multicolumn{4}{|c|}{\textbf{Section \ref{Section_4_System_Model_PD}}} \\ \hline
			\hfil  $\mathcal{P}$ & Set of prosumers & \hfil $\mathcal{S}$ & Set of dedicated local edge servers (DLES) \\ \hline
			\hfil $\mathcal{T}_p$  & Set of time slots for prosumer $p$  & \hfil $\mathcal{Y}_s$  & Labels for Network traffic (NT) data of prosumer $p$  \\ \hline
			\hfil $\mathcal{X}_s$ &  NT data of prosumer $p$ that connects with DLES $s$ & \hfil $\widehat{y}_s^\textrm{t}$ & Detected result for NT data at time slot $t$ \\  \hline
			\hfil $\Omega_s(\cdot)$ & Representation layer of deep learning (DL) model in the DLES $s$ & \hfil $\Lambda_s(\cdot)$ & Classifier of DL model in the DLES $s$ \\ \hline
			\hfil $\varphi_s^\textrm{t} $ & Features extracted from the representation layer & \hfil $\xi_s^\text{boundary}$ & Boundary between samples from various categories for each prosumer $p$ \\ \hline
			\hfil $\widehat{\mathcal{X}}_s$  & Set of malicious traffic for prosumer $p$ & \hfil $\widetilde{\mathcal{X}}_s$ & Set of benign traffic for prosumer $p$ \\ \hline
			\hfil $DIST(\cdot, \cdot)$ & Distance between output of the representation layer for a malicious traffic and for a benign traffic & $|\widehat{\mathcal{X}}_s|$ \&  $|\widetilde{\mathcal{X}}_s|$ & Number of malicious samples and benign samples, respectively, for prosumer $p$  \\ \hline
			\hfil $\Delta_p$ & Total number of operations of eq. (2)  & \hfil $\mathcal{E}_{dist}$  & Number of mathematical operations for calculating $DIST(\varphi_s^{\textrm{t}_1}, \varphi_s^{\textrm{t}_2})$ \\ \hline
			\hfil $\mathbbm{1}_{y_s^\textrm{t}=0}$  & Indicator function (if the detected result belongs to malicious traffic, then this value will be $1$) & \hfil $\mathbbm{1}_{y_s^\textrm{t}=1}$ & Indicator function (if the detected result is normal, then this value will be $1$) \\ \hline
			\multicolumn{4}{|c|}{\textbf{Section \ref{Section_5_Preliminaries}}} \\ \hline
			\hfil $\omega_m$ & Local model of client $m$ & \hfil $\omega$ & Global model  \\ \hline
			\hfil $\mathcal{D}_m$ & Dataset owned by client $m$ & \hfil $\mathcal{D}$ & Overall dataset  \\ \hline
			\hfil $|\mathcal{D}_{m}|$ & Data size of each client $m$ & \hfil $|\mathcal{D}|$  &  Data size of all clients \\ \hline
			\hfil $\mathcal{L}^{KD}$ \& $\zeta$ & KD loss and Temperature, respectively & \hfil $KL(\cdot, \cdot)$ & KL divergence function \\ \hline
			\hfil $\varrho^{k}(x)$ & Soft probability of Softmax\_T (TNet) & \hfil $\widehat{\varrho^{k}}(x))$ & Soft probability of Softmax\_T (SNet)\\ \hline
			\hfil $b(x)$ & A vector of scores generated by the TNet & \hfil $\overline{N}^\textrm{(k)}$ &  Global prototype of each class $k$ \\ \hline
			\hfil $N_m^{(k)}$ & Prototype of the class $k$ owned by client $m$ &\hfil $\mathcal{M}_k$ &  Set of the clients that own class $k$  \\ \hline
			\hfil $\mathcal{D}_{m,k}$ & Subset of local dataset $\mathcal{D}_{m}$ (belong to $k^{th}$ category)  & \hfil $|\mathcal{D}_{m,k}|$ & Overall data size of $\mathcal{D}_{m,k}$ \\ \hline 
			\multicolumn{4}{|c|}{\textbf{Section \ref{Section_6_Methodology}}} \\ \hline
			\hfil $\mathcal{T}$ & All time slots & \hfil $x^\textrm{t}$  & Network traffic data at time slot $t$  \\ \hline
			\hfil $\mathcal{X}$ & Network traffic data of all time slots & \hfil $i$ and $j$ & Two features of $\mathcal{X}$ \\ \hline
			\hfil $x_{i}$ & Vector of feature $i$  & \hfil $x_{j}$ & vector of feature $j$ \\ \hline
			\hfil $\lambda_{ij}$ & PCC between feature $i$ and feature $j$ & \hfil $\overline{x_{i}}$  & Average value of all the elements in the vector $x_{i}$ \\ \hline
			\hfil $\overline{x_{j}}$ & Average value of all the elements in the vector $x_{j}$ & \hfil $\alpha_s$ & Availability per client \\ \hline
			\hfil $b_s(x_s^t)$ & Output of the teacher network per client (DLES $s$) & \hfil $x_s^t$ & Network traffic data at time slot $t$ of DLES $s$ \\ \hline
			$\varrho_{s}^{k}(x_s^t)$/$\widehat{\varrho_{s}^\textrm{k}}(x_s^\textrm{t})$ & Probability for TNet/SNet by using Softmax\_T & \hfil $K$ & Total number of classes \\ \hline	
			\hfil $b_s^u(x_s^t)$ &  $u$-th element of the vector $b_s(x_s^t)$ & \hfil $\mathcal{L}_{s, SN}^{KD}$ & Kullback-Leibler (KL) divergence of client $s$ \\ \hline
			\hfil $\mathbb{N}_s$ & Prototype set of each participating client $s$ & \hfil $N_s^{(k)}$  & Prototype of class $k$ for the participating client $s$  \\ \hline
			\hfil $\Omega_s(\theta_s; x_s^t)$ & Embedding function for the representation layer of client $s$ with the parameter $\theta_s$ & \hfil $\mathcal{D}_{s,k}$ & Training NT data and labels of client $s$ that belong to the $k^{th}$ category \\ \hline
			\hfil  $\mathcal{D}_{s}$ & Local network traffic dataset  & \hfil  $\mathcal{L}_{s, SN}^R$  & Regularization term \\ \hline
			\hfil $distance(\cdot, \cdot)$ & Distance metric  & \hfil $\mathcal{L}_{s, SN}$ & Local loss function \\ \hline
			\hfil $\mathcal{L}_{s, SN}^{SL}$  &  Supervised learning loss per student network & \hfil $\mathcal{L}_{s, SN}^\textrm{entropy}$ & BCE loss or CE loss \\  \hline
			\hfil $\psi$ \& $\gamma$ & Hyperparameters & \hfil $\vartheta_s$  &  Parameter of classifier of client $s$ \\ \hline
			\hfil $N_s^\textrm{(k)}$ & Prototype of the class K generated by SNet per client & \hfil $|D_{s,k}|$ & Data size of $D_{s,k}$ \\ \hline
			\hfil $\hat{H}$ & Total number of the network traffic data owned by the participating clients & \hfil $Q$ \& $\widehat{\eta}$ & Final round and learning rate, respectively  \\ \hline
			\hfil $\omega_s$ &  Local model for SNet of client $s$  & \hfil $|\mathcal{X}_s|$ & Total number of NT data owned by client $s$  \\ \hline	
			\hfil $U$ & Overall number of all participating clients' instances & \hfil $U_k$ & Instances number of class $k$ over all participating clients \\ \hline
			\hfil $G^\textrm{nf}$ & Number of filters in CNN & \hfil $G^\textrm{ks}$ & Kernel size \\ \hline
			\hfil $G^\textrm{ne}$ & Number of elements of the input vector & \hfil $G^\textrm{ts}$ & Number of time steps \\ \hline
			\hfil $e_{t}$ & $(e_{t})^\textrm{th}$ convolutional layer of teacher network &\hfil $\widehat{e_{t}}$ & $\widehat{e_{t}}^\textrm{th}$ fully connected (FC) layer\\ \hline
			\hfil $G_{\widehat{e_{t}}-1}^{\textrm{nu}}$ &  No. of the neural units (NUs) in the FC layer $\widehat{e_{t}}-1$ &\hfil $G_{\widehat{e_{t}}}^{\textrm{nu}}$ & No. of the NUs in the FC layer $\widehat{e_{t}}$ \\ \hline
			\hfil $\widehat{E}$ & Local iteration &\hfil $\widehat{Z}$ & Time complexity of the backbone network for SNet \\ \hline
			\hfil $K_s$ & No. of prototypes generated by each SNet &\hfil $e_{s}$ & $(e_{s})^\textrm{th}$ convolutional layer of SNet\\ \hline
			\hfil $\widehat{e_{s}}$ & $(\widehat{e_{s}})^\textrm{th}$ FC layer of SNet &\hfil $G_{\widehat{e_{s}}-1}^{\textrm{nu}}$ & NUs in the FC layer $\widehat{e_{s}}-1$\\ \hline
			\hfil $G_{\widehat{e_{s}}}^{\textrm{nu}}$ & NUs in the layer $\widehat{e_{s}}$ of FC network &\hfil $M$ & No. of participating clients in each round\\ \hline
			\multicolumn{4}{|c|}{\textbf{Section \ref{Section_7_experimental_result_analysis}}} \\ \hline
			\hfil $\delta$  & Parameter for Dirichlet Distribution & \hfil TP & Correctly detected as attacks \\ \hline	
			\hfil TN  & Correctly detected as normal traffic & \hfil FP &  Normal traffic is incorrectly detected as attack \\ \hline	
			\hfil FN  & Attack traffic is incorrectly detected as the normal traffic & \hfil $\Upsilon_{Acc}$ & Accuracy \\ \hline	
			\hfil $\overline{\Upsilon}_{Acc}$  & Average accuracy (AA) & \hfil $\Upsilon_{Acc}^s$ & Accuracy of client $s$\\ \hline	
			\hfil $\Upsilon_{Prec}$  & Precision & \hfil $\overline{\Upsilon}_{Prec}$  & Average precision (AP) \\ \hline	
			\hfil $\Upsilon_{Prec}^s$  & Precision of client $s$ & \hfil $\Upsilon_{Rcl}$  & Recall\\ \hline	
			\hfil $\overline{\Upsilon}_{Rcl}$  & Average recall (AR) & \hfil $\Upsilon_{Rcl}^s$ & Recall of each client $s$ \\ \hline	
			\hfil $\Upsilon_{F1\_score}$  & F1 score & \hfil $\overline{\Upsilon}_{F1\_score}$ & Average F1 score \\ \hline	
			\hfil $\Upsilon_{F1\_score}^s$  & F1 score of client $s$ & \hfil $\Upsilon_{FAR}$ & False alarm rate (FAR) \\ \hline	
			\hfil $\Xi_{cdr}$  &  No. of the correct detected results  & \hfil $\Xi_{overall}$ & Overall number of network traffic data in a dataset \\ \hline	
		\end{tabular}
		\label{section_3_tab1}
	\end{center}
\end{table*}

\vspace{-0.3cm}
\section{System Model and Problem Formulation}
	\label{Section_4_System_Model_PD}
\begin{figure} [htbp]
	\centering
	    	\vspace{-0.25cm}
	\includegraphics[scale = .51]{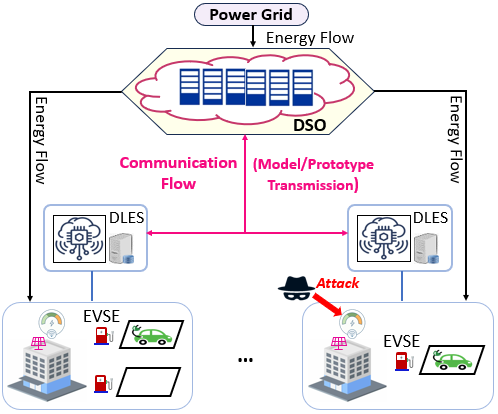}   
	\captionsetup{font=small}    
	\caption{System model of the prosumer-based EV charging stations.}
	\label{fig_system_model} 
		\vspace{-0.7cm}
\end{figure}

Considering the prosumer-based EV charging stations (EVCSs) that are composed of a set $\mathcal{P} = \{1, 2, ..., P\}$ of prosumers, a set $\mathcal{S} = \{1, 2, ..., S\}$ of dedicated local edge servers (DLESs), a set of charging service required EVs, a distribution system operator (DSO) and the power grid, shown as Fig. \ref{fig_system_model}. In particular, we consider each prosumer has a smart meter (SM) that is equipped with tamper-detection and encrypted communication function \cite{Sook_Chin_Yip_Anomaly_Detection_identifying}. Besides, each prosumer is embedded with a solar panel (SP) along with at least an EV supply equipment (EVSE), which is regarded as the physical infrastructure for supplying energy to public EVs. Moreover, each prosumer is considered connecting with the DSO for purchasing power grid energy. Further, we assume each prosumer $p \in \mathcal{P}$ is associated with a DLES $s \in \mathcal{S}$ \cite{L_Zou_Edge_assisted_Attention_Multi_Step}. 

\subsection{Cyber-attack Detection Model via Deep Learning}
	\label{Cyber_Attack_Detection_Model_Subsection}
\vspace{-0.05cm}
For the considered prosumer-based EVCS, in order to prevent cyber attacks, it is necessary to correctly detect cyber attacks. Thus, cyber-attack detection model (CDM) is investigated. Particularly, we focus on training a deep learning (DL)-based CDM by using the network traffic (NT) data of the considered EVCSs, which can distinguish the malicious traffic from the benign traffic. Generally, the DL model is composed of 1) feature extraction layers (i.e., representation layers \cite{Y_Tan_FedProto_Federated_Prototype_Learning}), 2) classification decision-making layers \cite{Y_Tan_FedProto_Federated_Prototype_Learning} (a.k.a. classifier).
\par 
For each prosumer $p \in \mathcal{P}$, we assume its connected DLES owns its NT data. Thus, we deploy the DL model in each connected DLES $s$. In addition, we consider a set $\mathcal{T}_p = \{1, 2, ..., T_p\}$ of $T_p$ time slots. As a result, the NT data of prosumer $p$ at all time slots can be denoted as $\mathcal{X}_s=\{x_s^\textrm{1}, x_s^\textrm{2}, ..., x_s^\textrm{t}\}, x_s^\textrm{t} \in \mathbb{R}^\textrm{d}, t\in \mathcal{T}_p$ and the corresponding labels can be represented as $\mathcal{Y}_s = \{y_s^\textrm{1}, y_s^\textrm{2}, ..., y_s^\textrm{t}\}$. It is worthwhile to mention that $\mathbb{R}^\textrm{d}$ represents the d-dimensional Euclidean space \cite{s_Torquato_random_sequential_d_dimentional_space}. Let $\Omega_s(\cdot)$ and $\Lambda_s(\cdot)$ denote the representation layers and the classifier of DL model in the DLES $s$, respectively. Hence, we can define the extracted features at time slot $t$, $\varphi_s^\textrm{t}$, as $\varphi_s^\textrm{t} = \Omega_s(x_s^\textrm{t})$, which is leveraged as the input of the classifier. For the detected result $\widehat{y}_s^\textrm{t}$, as per \cite{Y_Yue_Contrastive_Intrusion}, it can be obtained by,
\begin{subequations}\label{Opt_1}
		\setlength{\abovedisplayskip}{3.2pt}
		\setlength{\belowdisplayskip}{3.2pt}
		\begin{align}
			\widehat{y}_s^\textrm{t} = (\Lambda_s\circ\Omega_s)(x_s^\textrm{t}) = \Lambda_s\Big(\Omega_s(x_s^\textrm{t})\Big) = \Lambda_s(\varphi_s^\textrm{t}), \tag{1}
		\end{align}	
	\end{subequations}
	where $\circ$ indicates the composition of functions in mathematics.

\subsection{Malicious and Benign Traffic Boundary Model}
	\label{Malicious_Benign_Boundary_Model_Section_3_B}
To detect the cyber-attacks, it is possible to be achieved by distinguishing the malicious traffic (i.e., anomaly) from the benign traffic (i.e., normal). Due to the boundary between those two categories (i.e., inter-class distance \cite{Y_Yue_Contrastive_Intrusion}) becoming ambiguous, many false detection can occur \cite{Y_Yue_Contrastive_Intrusion}. Thus, it is still necessary to be dedicated to distinguish the boundary between those two categories (i.e., malicious category and benign category) as clearly as possible.

\par    
To facilitate the definition of the boundary between benign and malicious traffic per prosumer $p$, we define the malicious traffic set as  $\widehat{\mathcal{X}}_s=\{\widehat{x}_s^\textrm{1}, \widehat{x}_s^\textrm{2}, ..., \widehat{x}_s^{\textrm{t}_1}\}$ and the benign traffic set as $\widetilde{\mathcal{X}}_s=\{\widetilde{x}_s^\textrm{1}, \widetilde{x}_s^\textrm{2}, ..., \widetilde{x}_s^{\textrm{t}_2}\}$, where $\mathcal{X}_s = \widehat{\mathcal{X}}_s \cup \widetilde{\mathcal{X}}_s$. According to \cite{Y_Yue_Contrastive_Intrusion}, the boundary between two categories can be represented as the mean distance between samples from those two categories (i.e., inter-class distance \cite{Y_Yue_Contrastive_Intrusion}). Hence, assuming $\widehat{\mathcal{X}}_s$ and $\widetilde{\mathcal{X}}_s$ are not empty, the mentioned boundary can be given as follows:
\begin{subequations}\label{Opt_2}
	\setlength{\abovedisplayskip}{3.5pt}
	\setlength{\belowdisplayskip}{3pt}
	\begin{align}
		\xi_s^\textrm{boundary} =
		\setlength{\belowdisplayskip}{3.2pt} \frac{1}{|\widehat{\mathcal{X}}_s||\widetilde{\mathcal{X}}_s|}\sum_{t_1=1}^{|\widehat{\mathcal{X}}_s|}\sum_{t_2=1}^{|\widetilde{\mathcal{X}}_s|}DIST(\varphi_s^{\textrm{t}_1},  \varphi_s^{\textrm{t}_2}), \tag{2}
	\end{align}	
\end{subequations}
where $DIST(\varphi_s^{\textrm{t}_1}, \varphi_s^{\textrm{t}_2})$ represents the distance between $\varphi_s^{\textrm{t}_1}$ and $\varphi_s^{\textrm{t}_2}$. Here, $\varphi_s^{\textrm{t}_1}$ and $\varphi_s^{\textrm{t}_2}$ can be obtained by representation layers. As for $|\widehat{\mathcal{X}}_s|$ and $|\widetilde{\mathcal{X}}_s|$, they are used to indicate the number of malicious samples and benign samples, respectively.
\par
For the boundary distinction between malicious category and benign category, as per \cite{Y_Yue_Contrastive_Intrusion}, it can be realized by maximizing the inter-class distance (maximizing $\xi_s^\textrm{boundary}$). However, this method may cost high computational cost, which can be known through the following \textit{Lemma} and its \textit{proof}.
\begin{lemma}
	\textit{The computational cost of eq. \eqref{Opt_2} is high, if there exists many network traffic data per prosumer and/or high-dimensional feature space.}
	\begin{IEEEproof}
		\textit{In eq. \eqref{Opt_2}, for the $DIST(\varphi_s^{\textrm{t}_1},  \varphi_s^{\textrm{t}_2})$, various methods (L1-distance, L2-distance, etc.) can be chosen for calculation. Consequently, the number of mathematical operations relies on the selected method for calculating the distance. For convenience, suppose $\mathcal{E}_{dist}$ denotes the number of mathematical operations for calculating $DIST(\varphi_s^{\textrm{t}_1}, \varphi_s^{\textrm{t}_2})$. Thus, according to eq. \eqref{Opt_2}, $DIST(\varphi_s^{\textrm{t}_1},  \varphi_s^{\textrm{t}_2})$ needs to be calculated $|\widehat{\mathcal{X}}_s|  |\widetilde{\mathcal{X}}_s|$ times. Besides, we also need to compute a multiplication and a division operation for obtaining $\frac{1}{|\widehat{\mathcal{X}}_s|  |\widetilde{\mathcal{X}}_s|}$. Namely, two additional calculations are needed. Accordingly, the total number of operations of eq. \eqref{Opt_2} is $\Delta_p = |\widehat{\mathcal{X}}_s||\widetilde{\mathcal{X}}_s| \mathcal{E}_{dist} + 2$. As a result, if $|\widehat{\mathcal{X}}_s|$, $|\widetilde{\mathcal{X}}_s|$ and/or $\mathcal{E}_{dist}$ are very large, the computational cost for eq. \eqref{Opt_2} will be high.}
	\end{IEEEproof}
\end{lemma}
\subsection{Problem Formulation}
	\label{Problem_Formulation}
In this work, we concentrate on the cyber-attack prevention issue for the considered prosumer-based EVCSs. Specifically, we consider detecting cyber attacks in NT data first, then performing a cyber attack intervention process. The contents regarding intervention is explained in Section \ref{Section_6_Methodology}. Thus, next we will only explain the purpose for cyber-attack detection.

\par
According to \textit{Lemma 1} with its \textit{proof}, when each prosumer owns a large amount of NT data and/or high-dimensional feature space, using eq. \eqref{Opt_2} for malicious and benign traffic data distinguishment will encounter the computational cost issue. Therefore, reducing the computational cost of distinguishing the malicious and benign traffic needs to be tackled.
\par
To this end, in order to mitigate the computational cost, instead of maximizing eq. \eqref{Opt_2}, we consider maximizing the overall detection correctness of all prosumers (ODC). It is worth noting that maximizing the ODC has less computational cost compared with maximizing the eq. \eqref{Opt_2} for all prosumers. To prove this point, we need to define ODC first. To define ODC clearly, two indicator functions, $\mathbbm{1}_{y_s^\textrm{t}=0}$ and $\mathbbm{1}_{y_s^\textrm{t}=1}$, are introduced, which are separately given as follows:
\begin{subequations}\label{Opt_3}
	\begin{align}
		\mathbbm{1}_{y_s^\textrm{t}=0} = \left\{  
		\begin{array}{rcl}
			1, & \!\!\textrm{if $\widehat{y}_s^\textrm{t}=0$}, \\
			0,& \!\!\textrm{otherwise.}  \\
		\end{array} \right. \tag{3}
	\end{align}
\end{subequations} 
\begin{subequations}\label{Opt_4}
	\setlength{\abovedisplayskip}{3pt}
	\setlength{\belowdisplayskip}{3pt}
	\begin{align}
		\mathbbm{1}_{y_s^\textrm{t}=1} = \left\{  
		\begin{array}{rcl}
			1, & \!\!\textrm{if $\widehat{y}_s^\textrm{t}=1$}, \\
			0,& \!\!\textrm{otherwise.}  \\
		\end{array} \right. \tag{4}
	\end{align}
\end{subequations} 
Here, $y_s^\textrm{t}=0$ means the network traffic data at time slot $t$ actually is malicious traffic. If the detected result is also malicious (i.e., $\widehat{y}_s^\textrm{t}=0$), then $\mathbbm{1}_{y_s^\textrm{t}=0}$ is equal to $1$. Otherwise, $\mathbbm{1}_{y_s^\textrm{t}=0} = 0$, which means the detected result is wrong. For $y_s^\textrm{t}=1$, it represents the network traffic data at time slot $t$ is benign traffic in fact. At this situation, if the detected result is correct (the detected result is also benign ($\widehat{y}_s^\textrm{t}=1$)), then $\mathbbm{1}_{y_s^\textrm{t}=1}=1$. Otherwise, $\mathbbm{1}_{y_s^\textrm{t}=1}=0$, which indicates the detected result is malicious traffic. 
\par
Accordingly, the ODC maximization problem can be defined as follows: 
\begin{subequations}\label{Opt_5}
	\setlength{\abovedisplayskip}{3.2pt}
	\setlength{\belowdisplayskip}{3.2pt}
	\begin{align}
		\max \quad \sum_{s=1}^{\left | \mathcal{S} \right |} \sum_{t=1}^{\left | \mathcal{T}_p \right |}(\mathbbm{1}_{y_s^\textrm{t}=0} + \mathbbm{1}_{y_s^\textrm{t}=1}). 
		\tag{5}
	\end{align}	
\end{subequations}
It is noteworthy that since we consider a prosumer $p$ connects with a DLES, we have $\left | \mathcal{S}  \right | = \left | \mathcal{P}  \right |$. Here, $\left | \mathcal{P}  \right |$ means the total number of prosumers in the considered EV charging stations. As for the computational cost of \eqref{Opt_5}, since a traffic data is either normal or abnormal, the calculation cost of all the prosumers by \eqref{Opt_5} is $|\mathcal{S}||\mathcal{T}_p| = |\mathcal{S}|(|\widehat{\mathcal{X}}_s| + |\widetilde{\mathcal{X}}_s|)$. Compared the computational cost of \eqref{Opt_2} and eq. \eqref{Opt_5} for all prosumers, we can get the following \textit{Lemma} with its proof. 
	\begin{lemma}
		\label{lemma_2}
		\textit{The computational cost of \eqref{Opt_5} is less than maximizing eq. \eqref{Opt_2} for all prosumers under some considerations.}
		\begin{IEEEproof}
			\textit{For eq. \eqref{Opt_2}, considering applying it for all prosumers, then the computational cost can be obtained as $|\mathcal{S}|\Delta_p = |\mathcal{S}|(|\widehat{\mathcal{X}}_s||\widetilde{\mathcal{X}}_s| \mathcal{E}_{dist} + 2)$. According to the definition of $|\widehat{\mathcal{X}}_s|$, $|\widetilde{\mathcal{X}}_s|$ and $\mathcal{E}_{dist}$, since we consider $\widehat{\mathcal{X}}_s \neq \varnothing$ and $\widetilde{\mathcal{X}}_s \neq \varnothing$, it can be known that $|\widehat{\mathcal{X}}_s| \geqslant 1$, $|\widetilde{\mathcal{X}}_s| \geqslant 1$ and $\mathcal{E}_{dist} \geqslant 1$. Next, we will illustrate the following cases to prove \textit{Lemma} $2$. 
				\begin{itemize}
					\item \textbf{Case $1$}: If $|\widehat{\mathcal{X}}_s| = 1$, $|\widetilde{\mathcal{X}}_s| = 1$ and $\mathcal{E}_{dist} \geqslant 1$, then the computational cost of applying eq. \eqref{Opt_2} can be calculated as $|\mathcal{S}|(|\widehat{\mathcal{X}}_s||\widetilde{\mathcal{X}}_s| \mathcal{E}_{dist} + 2) = |\mathcal{S}|(\mathcal{E}_{dist} + 2)$. Because $\mathcal{E}_{dist} \geqslant 1$, $|\mathcal{S}|(\mathcal{E}_{dist} + 2) \geqslant 3|\mathcal{S}|$. In the case of the computational cost of \eqref{Opt_5}, it will be  $|\mathcal{S}|(|\widehat{\mathcal{X}}_s| + |\widetilde{\mathcal{X}}_s|) = 2|\mathcal{S}|$. Hence, for this case, \textit{Lemma $2$} can be proved to be true.
					\item \textbf{Case $2$}: If $|\widehat{\mathcal{X}}_s| = 1$, $|\widetilde{\mathcal{X}}_s| > 1$ and $\mathcal{E}_{dist} \geqslant 1$, then we can obtain $|\mathcal{S}|(|\widehat{\mathcal{X}}_s||\widetilde{\mathcal{X}}_s| \mathcal{E}_{dist} + 2) = |\mathcal{S}|(|\widetilde{\mathcal{X}}_s|\mathcal{E}_{dist} + 2) \geqslant |\mathcal{S}|(|\widetilde{\mathcal{X}}_s| + 2)$. The computational cost of \eqref{Opt_5} can be computed as $|\mathcal{S}|(|\widehat{\mathcal{X}}_s| + |\widetilde{\mathcal{X}}_s|) = |\mathcal{S}|(|\widetilde{\mathcal{X}}_s| + 1)$. Thus, in this case, we can know $|\mathcal{S}|(|\widehat{\mathcal{X}}_s||\widetilde{\mathcal{X}}_s| \mathcal{E}_{dist} + 2) > |\mathcal{S}|(|\widehat{\mathcal{X}}_s| + |\widetilde{\mathcal{X}}_s|)$. Accordingly, for this case, \textit{Lemma $2$} is true.
					\item \textbf{Case $3$}: If $|\widehat{\mathcal{X}}_s| > 1$, $|\widetilde{\mathcal{X}}_s| = 1$ and $\mathcal{E}_{dist} \geqslant 1$, then $|\mathcal{S}|(|\widehat{\mathcal{X}}_s||\widetilde{\mathcal{X}}_s| \mathcal{E}_{dist} + 2) = |\mathcal{S}|(|\widehat{\mathcal{X}}_s|\mathcal{E}_{dist} + 2) \geqslant |\mathcal{S}|(|\widehat{\mathcal{X}}_s| + 2)$, and $|\mathcal{S}|(|\widehat{\mathcal{X}}_s| + |\widetilde{\mathcal{X}}_s|) = |\mathcal{S}|(|\widehat{\mathcal{X}}_s| + 1)$. Thus, $|\mathcal{S}|(|\widehat{\mathcal{X}}_s||\widetilde{\mathcal{X}}_s| \mathcal{E}_{dist} + 2) > |\mathcal{S}|(|\widehat{\mathcal{X}}_s| + |\widetilde{\mathcal{X}}_s|)$, thereby, \textit{Lemma $2$} is correct for this case.
					\item \textbf{Case $4$}: If $|\widehat{\mathcal{X}}_s| \geqslant |\widetilde{\mathcal{X}}_s| > 1$, and $\mathcal{E}_{dist} \geqslant 1$, then $|\mathcal{S}|(|\widehat{\mathcal{X}}_s||\widetilde{\mathcal{X}}_s| \mathcal{E}_{dist} + 2) > |\mathcal{S}||\widehat{\mathcal{X}}_s||\widetilde{\mathcal{X}}_s| \mathcal{E}_{dist}$. In addition, in this article,  $|\widetilde{\mathcal{X}}_s|$ is adopted to represent the number of the samples that belong to the benign traffic. Thus, it is integer. Accordingly, since $|\widetilde{\mathcal{X}}_s| > 1$, it can be inferred that  $|\widetilde{\mathcal{X}}_s| \geqslant 2$. Therefore, it can be obtained that $|\mathcal{S}||\widehat{\mathcal{X}}_s||\widetilde{\mathcal{X}}_s| \mathcal{E}_{dist} \geqslant 2|\mathcal{S}||\widehat{\mathcal{X}}_s|$. For the computational cost of \eqref{Opt_5}, because $|\widehat{\mathcal{X}}_s| \geqslant |\widetilde{\mathcal{X}}_s|$, $|\mathcal{S}|(|\widehat{\mathcal{X}}_s| + |\widetilde{\mathcal{X}}_s|) \leqslant |\mathcal{S}|(|\widehat{\mathcal{X}}_s| + \widehat{\mathcal{X}}_s|) = 2|\mathcal{S}||\widehat{\mathcal{X}}_s|$. Accordingly, we can obtain that $|\mathcal{S}|(|\widehat{\mathcal{X}}_s||\widetilde{\mathcal{X}}_s| \mathcal{E}_{dist} + 2) > |\mathcal{S}|(|\widehat{\mathcal{X}}_s| + |\widetilde{\mathcal{X}}_s|)$. Consequently, for this case, \textit{Lemma $2$} is also true.
					\item \textbf{Case $5$}: If $|\widetilde{\mathcal{X}}_s| \geqslant |\widehat{\mathcal{X}}_s| > 1$, and $\mathcal{E}_{dist} \geqslant 1$, since  $|\widehat{\mathcal{X}}_s|$ denotes the number of malicious traffic samples, we can infer $|\widehat{\mathcal{X}}_s| \geqslant 2$. Hence, similar as \textit{Case $4$}, we can prove that $|\mathcal{S}|(|\widehat{\mathcal{X}}_s||\widetilde{\mathcal{X}}_s| \mathcal{E}_{dist} + 2) > |\mathcal{S}||\widehat{\mathcal{X}}_s||\widetilde{\mathcal{X}}_s| \mathcal{E}_{dist} \geqslant 2|\mathcal{S}||\widetilde{\mathcal{X}}_s|$ and $|\mathcal{S}|(|\widehat{\mathcal{X}}_s| + |\widetilde{\mathcal{X}}_s|) \leqslant |\mathcal{S}|(|\widetilde{\mathcal{X}}_s| + \widetilde{\mathcal{X}}_s|) = 2|\mathcal{S}||\widetilde{\mathcal{X}}_s|$. Thus, we can prove \textit{Lemma \ref{lemma_2}} is correct in this case.
				\end{itemize}
				According to the analysis from Case $1$ to Case $5$, we can prove under some considerations, Lemma \ref{lemma_2} is correct.
			}	    		
		\end{IEEEproof}
\end{lemma}

\section{Preliminaries Knowledge}
	\label{Section_5_Preliminaries}
	To prevent the considered prosumer-based EVCSs from suffering cyber-attacks, we propose an edge-assisted federated prototype knowledge distillation (E-FPKD) approach. Before explaining the proposed E-FPKD, as federated learning, knowledge distillation, and prototype aggregation are used to design the proposed solution, for ease of understanding, it is worth to explain those three techniques first before going into the depth of the proposed solution in Section \ref{Section_6_Methodology}.
	
\subsection{Background: Federated Learning (FL)}
\par
The proposed E-FPKD is a FL-based approach. To understand the structure of the proposed solution more easily, we will illustrate the background knowledge of FL in this part.
\par
\begin{figure} [t!]
		\centering
		\includegraphics[scale = .44]{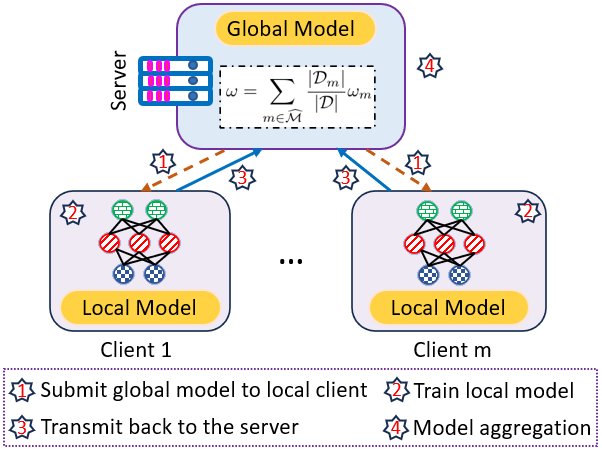}  
		\captionsetup{font=small}
		\caption{Training process of federated learning approach.}
		\label{fig_overview_of_FL_training_process}
		\vspace{-0.5cm}
	\end{figure}

In edge computing, an emerging approach named FL is widely used, which is composed of a server and several clients \cite{G_D_N_Snapshot_Client_Selection_FL}. FL aims to collaboratively train machine learning models among various clients \cite{H_Jin_Personalized_self_KD} without the necessity of letting the data leave their local side. When it comes to FL, it is worth mentioning that a commonly utilized approach is termed \texttt{FederatedAveraging} (i.e., \texttt{FedAvg})  proposed by \cite{H_B_McMahan_Communication_Efficient_Learning_Deep_Networks}. As shown in Fig. \ref{fig_overview_of_FL_training_process}, the procedure of \texttt{FedAvg} consists of $4$ steps, which will be iterated as per \cite{H_B_McMahan_Communication_Efficient_Learning_Deep_Networks}. Those $4$ steps are shown as below:
	\begin{itemize}
		\item \textbf{Step $1$: } \textit{The global model will be distributed to each local client to initialize each local model.}
		\item \textbf{Step $2$: } \textit{Each local client $m$ will perform local training to update local model $\omega_m$.} 		
		\item \textbf{Step $3$: } \textit{The updated local model parameters will be transmitted to the server \cite{J_Geiping_Inverting_Gradients_How_easy}.}
		\item \textbf{Step $4$: } \textit{The model aggregation will be performed on the server to generate new global model.} Particularly, the global model $\omega$ can be obtained as follows \cite{L_Zou_When_HFL_Urban_Prosumers}:
		\begin{subequations}\label{Opt_6}
			\setlength{\abovedisplayskip}{3pt}
			\setlength{\belowdisplayskip}{3pt}
			\begin{align}
				&\label{Opt_6:const1}
				\omega = \sum_{m}\frac{|\mathcal{D}_m|}{|\mathcal{D}|}\omega_m, \tag {6}
			\end{align}	
		\end{subequations}
		where $\mathcal{D} = \sum_{m}\mathcal{D}_m$ indicates the overall dataset and $\mathcal{D}_m$ denotes the dataset of client $m$. $|\mathcal{D}_{m}|$ and $|\mathcal{D}|$ indicate the data size of each client and all the clients, respectively. 
	\end{itemize}
	
\par

\par
	\textit{\textbf{Example 1: }} Consider a FL-based network intrusion detection system (IDS) across various organizations, for this IDS, we consider the model aggregation is executed in the server to generated the global model, and each organization is considered as each client of the FL process and owns a local model. In the beginning, the server will send the global model to each organization to initialize each local model. Then each organization will train the local model via its local network traffic (NT) data such that to update the local model. Next, the updated local model will be transmitted to the server for model aggregation to generate a new global model.		
	
\subsection{Background: Knowledge Distillation (KD)}
	\label{Section_5_B_Backgound_Knowledge_Distillation}
	Considering the NT data of each prosumer may be heterogeneous, since integrating KD into FL is an efficient way for handling the data heterogeneity issue \cite{J_Tang_FedRAD_Heterogeneous}, KD technique is introduced to be a composition of the proposed approach. To make it easier to understand the proposed method, we provide some background knowledge of KD in this part.
	\par
	KD is the approach that can benefit the training process of the smaller student network (SNet) from the supervision of a larger teacher network (TNet) \cite{L_Wang_KD_Outlooks}. Specifically, the knowledge of TNet will be distilled to SNet. To do this, \textit{Kullback-Leibler (KL) divergence} can be employed to enable the SNet to imitate the behavior of the TNet \cite{B_Peng_Correlation_Congruence_ICCV}. Thus, the KD loss, $\mathcal{L}^{KD}$, can be given as below \cite{B_Peng_Correlation_Congruence_ICCV}:
		\begin{subequations}\label{Opt_7}
			\setlength{\abovedisplayskip}{3pt}
			\setlength{\belowdisplayskip}{3pt}
			\begin{align}
				&\label{Opt_7:const1}
				\mathcal{L}^\textrm{KD} = \frac{1}{|\mathcal{D}_m|}\sum_{t=1}^{|\mathcal{D}_m|}\zeta^2 KL(\varrho^\textrm{k}(x), \widehat{\varrho^\textrm{k}}(x)), \tag {7}
			\end{align}	
		\end{subequations}
	where $KL(\cdot, \cdot)$ is the KL divergence function, while $\zeta$ denotes the relaxation hyperparameter (i.e., temperature) \cite{B_Peng_Correlation_Congruence_ICCV}. In \eqref{Opt_7}, $x$ is the input data sample. $\varrho^\textrm{k}(x)$ and $\widehat{\varrho^\textrm{k}}(x)$ are the soft probability of TNet and SNet by using \textit{softmax with temperature scaling} \cite{G_Hinton_Distilling_Knowledge_NN}, respectively.
	\par
	\textit{\textbf{Example 2: }} Consider designing a network intrusion detection system for a mobile device. Suppose there is a pre-trained model that can detect malicious traffic well, however, that model is too large to be deployed on a resource-limited mobile device. Thus, KD is used to create a smaller model. To be specific, assume there is a lightweight model that is deployable to the mobile device. When training that lightweight model, we distill the knowledge from the pre-trained model to the lightweight model by minimizing the KD loss (i.e., eq. \eqref{Opt_7}).
	
	\begin{figure*} [t!]
		\centering
		\includegraphics[scale = .52]{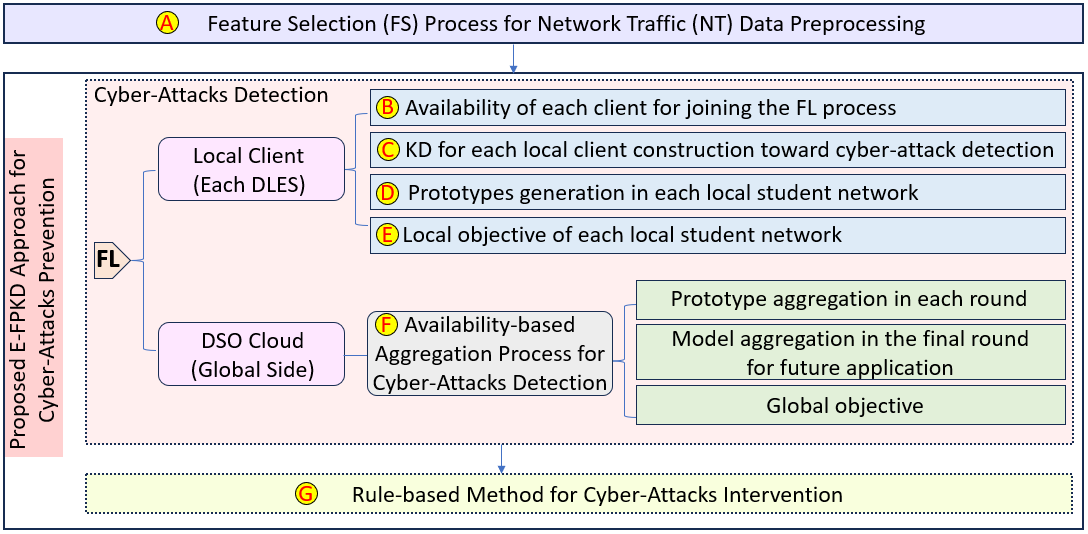}  
		\captionsetup{font=small}  
		\caption{Structure of the proposed solution for cyber-attack prevention towards the considered prosumer-based EVCSs.}
		\label{fig_Overall_Architecture_Proposed_Solution}
			\vspace{-0.4cm}
	\end{figure*}
	
\subsection{Background: Prototype Aggregation (PA)
\label{Section_5_Backgound_Prototype_Aggregation}}
Prototype aggregation (PA) mechanism is another technique that we will integrate into FL to construct the proposed method (Section \ref{Section_6_Methodology}). Thus, for ease of understandability towards the proposed solution, it is valuable to illustrate PA beforehand.
\par
\textit{Prototype} is defined as the average value of the embedding vectors for the instances in class $k$ \cite{Y_Tan_FedProto_Federated_Prototype_Learning}, where the embedding vectors are the extracted features from the representation layers. Notably, according to \cite{Y_Tan_FedProto_Federated_Prototype_Learning}, one prototype is used to indicate one class. Similar to the original \texttt{FedAvg} that adopts model aggregation, for \textbf{prototype aggregation}, the prototypes of each category received from all participated clients will be aggregated to generate new global prototype of each class $k$,  $\overline{N}^\textrm{(k)}$, which can be given as follows \cite{Y_Tan_FedProto_Federated_Prototype_Learning}: 
    \begin{subequations}\label{Opt_8}
			\setlength{\abovedisplayskip}{3pt}
			\setlength{\belowdisplayskip}{3pt}
			\begin{align}
					&\label{Opt_8:const1}
					\overline{N}^\textrm{(k)} = \frac{1}{|\mathcal{M}_k|}\sum_{m\in\mathcal{M}_k} \frac{|D_{m,k}|}{\sum_{m \in \mathcal{M}_k}|D_{m,k}|} N_m^\textrm{(k)},\tag {8}
				\end{align}	
		\end{subequations}
where $N_m^\textrm{(k)}$ denotes the prototype of class $k$ owned by client $m$, $\mathcal{M}_k$ represents a set of the clients who participated in the FL process that own class $k$, and $|D_{m,k}|$ indicates the number of the data of client $m$ that belongs to class $k$. More details will be provided in the next Section.

\section{Overall Architectural Design for the Proposed Methodology}
	\label{Section_6_Methodology}
Based on the background knowledge illustrated in the previous section, we design the method for cyber-attack prevention towards the considered prosumer-based EVCSs.
\begin{figure*} [t!]
	\centering
	\includegraphics[scale = .65]{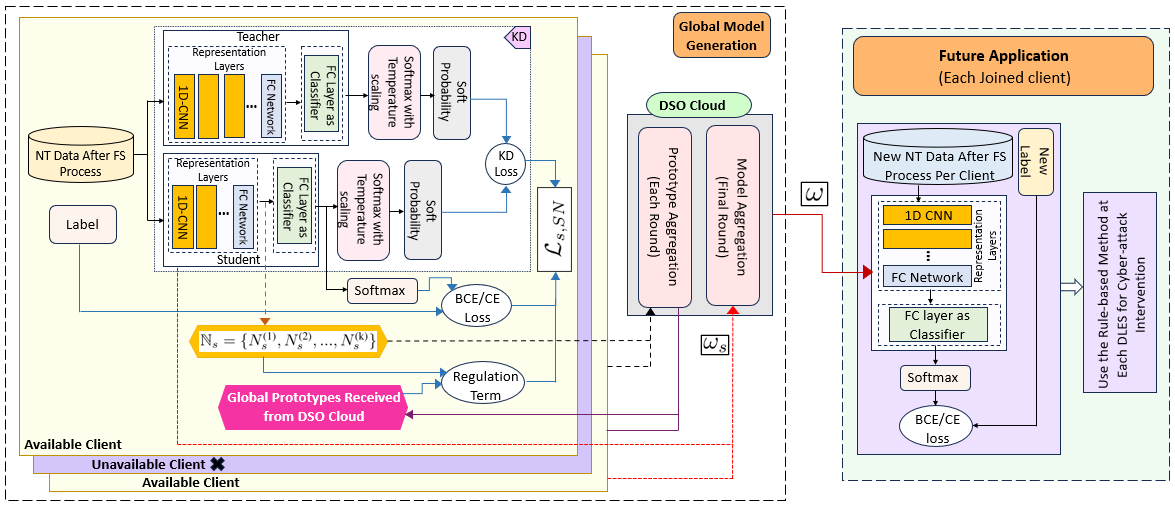}  
	\captionsetup{font=small} 
	\caption{Overall architecture of the proposed E-FPKD approach towards cyber-attacks prevention in both detection and intervention.}
	\label{fig_proposed_method_E_FPKD} 
	\vspace{-0.3cm}
\end{figure*}
\par
Fig. \ref{fig_Overall_Architecture_Proposed_Solution} shows the structure of the proposed solution, which will be illustrated next. Firstly, feature selection (FS) process for network traffic (NT) data preprocessing is employed since FS is beneficial for increasing accuracy \cite{T_Wisanwanichthan_Double_Layered_Hybrid_Approach_NIDS}. Secondly, considering the privacy issue per prosumer and the heterogeneous data distribution among each prosumer, an edge-assisted federated prototype knowledge distillation (E-FPKD) approach is proposed for cyber-attacks prevention, where this method is designed based on the incorporation of federated learning, knowledge distillation (KD), prototype aggregation and rule-based method \cite{M_Habiba_Edge_Intelligence_Network_Intrusion_Prevention}. Particularly, we deploy the DL-based CDM on each DLES (dedicated local edge server) and treat each DLES as an E-FPKD client. To be more specific, the proposed E-FPKD approach contains two parts: cyber-attacks detection by solving eq. \eqref{Opt_5}, and cyber-attacks intervention. To detect cyber-attacks, we combine KD and prototype aggregation with the FL approach. Besides, in practice, it is unrealistic to suppose all clients are always able to offer the training service \cite{T_Huang_efficiency_boosting_client_Selection_fairness}, which may cause some clients to be unwilling to join the training process. Thus, each client's availability is adopted in this article. Accordingly, as shown in Fig. \ref{fig_Overall_Architecture_Proposed_Solution}, we introduce the local client by the following order: 1) the availability of each client for joining the FL process, 2) how to use KD to construct the local client (teacher network and student network), 3) the prototypes gained by each local SNet, and 4) the objective of each local SNet. As for the DSO cloud, availability-based aggregation process for cyber-attacks detection is conducted, which contains 1) prototype aggregation in each round, 2) model aggregation in the final round for future application, and 3) the global objective. After detection, the detected cyber attacks will be intervened at each edge via the rule-based method. It is noteworthy that the proposed solution is advantageous in handling NT data with non-IID nature and can protect privacy to a certain extent. The reasons are as follows:
\begin{itemize}
\item The proposed E-FPKD approach is constructed based on the FL. For FL, it can ensure data privacy by saving data and training the model on each client locally \cite{X_Shang_FEDIC_FL_NonIID_Long_Tailed_Data}. Thus, the proposed E-FPKD approach can also assure the data privacy issue caused by central data collection.
	\item However, FL faces the challenge of handling non-IID data \cite{X_Shang_FEDIC_FL_NonIID_Long_Tailed_Data}. Since applying KD in FL can help deal with the data heterogeneity issue \cite{J_Tang_FedRAD_Heterogeneous}, we introduce KD into FL to build the proposed E-FPKD approach. Motivated by \cite{C_Wu_Communication_efficient_FL_KD} that considers each client to consist of a mentee model and a mentor model, we employ a teacher model and a student model to form each client. Hence, each client will have its own teacher model, which can better adapt to the local datasets' characteristics. 		
	\item Apart from KD, prototype aggregation proposed by \cite{Y_Tan_FedProto_Federated_Prototype_Learning} is also employed to create the proposed E-FPKD approach. Prototype aggregation has the following goodness given by \cite{Y_Tan_FedProto_Federated_Prototype_Learning}: 1) through prototype aggregation, the identical label space can enable the embedding space of the participating client to be shared, 2) the information is capable of exchanging efficiently across all the participating heterogeneous clients, and 3) the heterogeneous class spaces can be supported. Hence, integrating prototype aggregation can further help process non-IID data.	
	\end{itemize} 
\par
The sub-sections \ref{sub_section_A_Feature_Selection} $\sim$ \ref{Rule_Based_Cyber_Attack_Intervention_Application_Stage} are organized by $\encircle{A} \sim \encircle{G}$ shown in Fig. \ref{fig_Overall_Architecture_Proposed_Solution}, while sub-section \ref{sub_section_H_Algorithm_E_FPKD} summarizes the proposed E-FPKD algorithm towards cyber-attack prevention. Fig. \ref{fig_proposed_method_E_FPKD} shows the overall architecture of the proposed E-FPKD approach. The detailed contents will be described next.   

\subsection{Feature Selection for NT Data Preprocessing}
\label{sub_section_A_Feature_Selection}
Feature selection is a critical step for optimizing learning complexity through prioritizing the features, especially for the network traffic data \cite{M_Nakashima_Automated_FS_NTD}. Through the feature selection process, the irrelevant features can be excluded \cite{T_Wisanwanichthan_Double_Layered_Hybrid_Approach_NIDS}.  

\par
In this article, Pearson Correlation Coefficient (PCC) is employed for feature selection. Particularly, PCC was introduced in 1895 by Pearson \cite{G_Li_PCC_Performance_Enhancement_Broad}, which adopts bivariate analysis \cite{T_Wisanwanichthan_Double_Layered_Hybrid_Approach_NIDS} that is capable of evaluating the strength of the relationship between two vectors \cite{Y_Mu_PCC_decision_tree}. Let $\mathcal{T} = \{1, 2, ..., T\}$ denote all time slots of network traffic (NT) data. As described in Section \ref{Cyber_Attack_Detection_Model_Subsection}, we consider the NT data of each time slot $t$ belongs to $\mathbb{R}^d$ (d-dimensional Euclidean space), without loss of generality, we denote the NT data at time slot $t$ as $x^\textrm{t}=\{x_{1}^\textrm{t}, x_{2}^\textrm{t}, ..., x_{d}^\textrm{t}\}$, $t \in \mathcal{T}$. Then the NT data of all time slots can be denoted as $\mathcal{X} = \{x^\textrm{1}, x^\textrm{2}, ..., x^\textrm{t}\}$. For two features of $\mathcal{X}$, $i$ and $j$, the vector of feature $i$ can be represented as $x_{i}=\{x_{i}^\textrm{1}, x_{i}^\textrm{2}, ..., x_{i}^\textrm{t}\}$, while the vector of feature $j$ can be denoted as $x_{j}=\{x_{j}^\textrm{1}, x_{j}^\textrm{2}, ..., x_{j}^\textrm{t}\}, t \in \mathcal{T}$. Then we can calculate the PCC between feature $i$ and feature $j$, $\lambda_{ij}$, as follows \cite{T_Wisanwanichthan_Double_Layered_Hybrid_Approach_NIDS}:
\begin{subequations}\label{Opt_9}
	\vspace{-0.3cm}
\setlength{\abovedisplayskip}{3.2pt}
\setlength{\belowdisplayskip}{3.2pt}
\begin{align}
\lambda_{ij} = \frac{\sum_{t=1}^{|\mathcal{T}|}(x_{i}^\textrm{t}-\overline{x_{i}})(x_{j}^\textrm{t}-\overline{x_{j}})}{\sqrt{\sum_{t=1}^{|\mathcal{T}|}(x_{i}^\textrm{t}-\overline{x_{i}})^\textsuperscript{2}}\sqrt{\sum_{t=1}^{|\mathcal{T}|}(x_{j}^\textrm{t}-\overline{x_{j}})^\textsuperscript{2}}}.
\tag{9}
\end{align}	
\end{subequations}
Here, $\overline{x_{i}} = \frac{\sum_{t=1}^{|\mathcal{T}|}x_{i}^\textrm{t}}{|\mathcal{T}|}$ and $\overline{x_{j}} = \frac{\sum_{t=1}^{|\mathcal{T}|}x_{j}^\textrm{t}}{|\mathcal{T}|}$ \cite{H_Zhou_sampling_method_filter_PCC}. As per \cite{G_Li_PCC_Performance_Enhancement_Broad}, the range of PCC value is $[-1, +1]$. In particular, PCC in the range $\pm0.1\!-\!\pm0.3$ means weak relation, in the range $\pm0.3\!-\!\pm0.5$ represents moderate relation, in the range $\pm0.5\!-\!\pm1.0$ indicates strong relation, and if PCC is equal to $\pm1$ means perfect relation \cite{P_Rungskunroch_Benchmarking_Socio}. As a result, we consider $0.1\leq $ PCC $\leq 1$ or $-1 \leq$ PCC $\leq-0.1$ as the rule for choosing features. 
\begin{figure}[t!]
	\vspace{-0.3cm}	
	\begin{algorithm}[H]	
		\renewcommand{\algorithmicrequire}{\textbf{Input:}}
		\renewcommand{\algorithmicensure}{\textbf{Output:}}
		\caption{Feature Selection with Pearson Correlation Coefficient}
		\label{alg_1}
			\begin{algorithmic}[1]
				\REQUIRE Network traffic (NT) data: \\ \hspace{0.55cm} $\mathcal{X}=\{x^\textrm{1}, x^\textrm{2}, ..., x^\textrm{t}\}$	
				\ENSURE $\widehat{\mathcal{X}}$			
				\STATE Get NT data vector for each feature based on $\mathcal{X}$
				\STATE Calculate PCC for each two features via \eqref{Opt_9}
				\STATE Check whether $0.1\leq $ PCC $\leq 1$ or $-1 \leq$ PCC $\leq-0.1$
				\STATE Retain features according to the introduced heuristic rule
				\STATE Get NT data after feature selection process: $\widehat{\mathcal{X}}$   
				\STATE \textbf{return} NT data after feature selection: $\widehat{\mathcal{X}}$
			\end{algorithmic}  
		\end{algorithm} 
		\vspace{-0.7cm}
	\end{figure}
\par
Algorithm \ref{alg_1} lists the process of the feature selection process via PCC. Specifically, in line $1$, we generate each feature vector of the network traffic data. Then using eq. \eqref{Opt_9}, the PCC of each two features will be computed (line $2$). In line $3$, we check whether the calculated PCCs satisfy the condition. It is noteworthy that instead of finding the minimum feature set for CDM model generation in this article, to reduce the complexity, we only focus on the effectiveness of the feature selection with the PCC approach for the proposed E-FPKD approach. Thus, the following rule is introduced heuristically (line $4$): 1) obtain and sort the No. of PCC in $[-1, -0.1] \cup [0.1, 1]$ per feature, 2) retain the features corresponding to the top few numbers empirically. Next, the NT data of a client after the feature selection process will be obtained (line $5$). Finally, we can get NT data after feature selection in line $6$. 
\par
Notably, NT data handled by PCC will be used as the input of the proposed E-FPKD approach that will be illustrated next.

\subsection{Availability of Each Client for Joining the FL process}
\label{Availability_of_Clients}
In realistic, as per \cite{T_Huang_efficiency_boosting_client_Selection_fairness}, it is unrealistic to presume all clients can always offer the services for training. This may affect the willingness of some clients to join in the FL process. Motivated by \cite{T_Huang_efficiency_boosting_client_Selection_fairness} which takes into account the availability of clients for joining the training process, each client's availability is considered in this work. Concretely, each client can choose whether to participate in the FL process freely. To this end, we assume each client is able to report its availability before joining the FL process to the DSO cloud. The availability per client can be represented as below:
\begin{subequations}\label{Opt_10}
	\begin{align}
		\!\!\!\!\!\alpha_s \!=\! \left\{ 
		\begin{array}{rcl}
			\!\!	1, & \!\!\!\!\textrm{if client s is available to join}, \\
			\!\!	0,& \!\!\!\!\textrm{otherwise.}  \\
		\end{array} \right. \tag{10}
	\end{align}
\end{subequations} 

\subsection{KD for Each Local Client Construction Toward Cyber-attack Detection}
In this article, since KD in FL is capable of assisting the data heterogeneity issue \cite{J_Tang_FedRAD_Heterogeneous}, knowledge distillation (KD) is adopted to build each local client towards cyber-attack detection. Specifically, as mentioned before, identical to \cite{C_Wu_Communication_efficient_FL_KD} which considers using a mentor model and a mentee model to form each client, we consider each client to be constructed with a teacher model (TM) and a student model (SM). Particularly, TM is pre-trained and the knowledge of TM will be distilled to guide SM placed on the same DLES.
\begin{figure} [htpb]
	\centering
	\includegraphics[scale = .42]{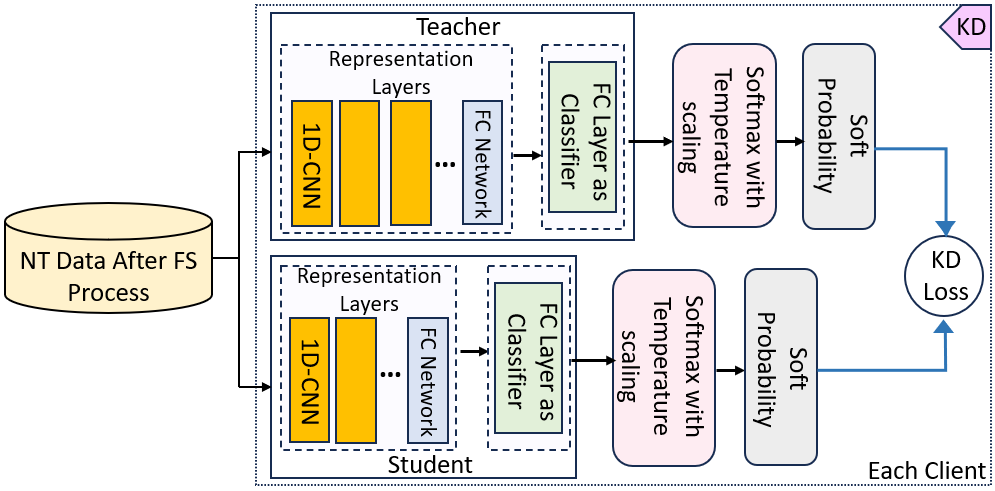}
	\captionsetup{font=small}
	\caption{KD applied to per client in the proposed E-FPKD approach.}
	\label{fig_KD} 
\end{figure}
\par
In Fig. \ref{fig_KD}, we demonstrate the KD applied to each client (i.e., DLES $s$) in the proposed E-FPKD approach, which is a part of Fig. \ref{fig_proposed_method_E_FPKD}. To be specific, each client has two network: teacher network (TNet) and student network (SNet). TNet is used to generate the teacher model, while SNet is utilized to produce the student model. In addition, the network traffic data after feature selection process will be used as the input of both TNet and SNet. For TNet and SNet, we adopt deep learning to create them. Thus, they both contain the representation layers and classifier \footnote{As mentioned in Section \ref{Cyber_Attack_Detection_Model_Subsection}, the deep learning (DL) model generally consists of the representation layers and the classifier. In this article, we adopt deep learning to build both TNet and SNet. Thus, both TNet and SNet are composed of the representation layers and the classifier.}. For the sake of simplicity, we use the combination of one-dimensional (1D) convolutional neural network (CNN) and fully connected (FC) network to build the representation layers of TNet and SNet, as shown in Fig. \ref{fig_KD}. It is noteworthy that the reason for utilizing 1D-CNN is that we represent a network traffic record by a vector. For the classifier, we simply adopt an FC layer. The idea of KD can be the idea of leveraging the TNet's soft probabilities to supervise the SNet \cite{J_H_Cho_Efficacy_KD}, where the soft probabilities can be calculated via \textit{Softmax with Temperature scaling} \cite{G_Hinton_Distilling_Knowledge_NN} (softmax\_T). Accordingly, we consider using softmax\_T to convert the output of the classifier into soft probability in this work. Then the soft probability of TNet and SNet will be used to determine the KD loss.

\par	
Given the network traffic data at time slot $t$, $x_s^\textrm{t}$, the output of TNet per client $s$ can be denoted as $b_s(x_s^\textrm{t}) = [b_s^\textrm{1}(x_s^\textrm{t}), b_s^\textrm{2}(x_s^\textrm{t}), ..., b_s^\textrm{k}(x_s^\textrm{t})]$. Here, $k$ represents class $k$. Then we can calculate the soft probability of TNet, $\varrho_{s}^\textrm{k}(x_s^\textrm{t})$, as \cite{G_Hinton_Distilling_Knowledge_NN}:
\begin{subequations}\label{Opt_11}
	\vspace{-0.22cm}
	\setlength{\abovedisplayskip}{3pt}
	\setlength{\belowdisplayskip}{3pt}
	\begin{align}
		&\label{Opt_11:const1}
		\varrho_{s}^\textrm{k}(x_s^\textrm{t}) = \frac{e^{b_s^\textrm{k}(x_s^\textrm{t})/\zeta}}{\sum_{u} e^{b_s^\textrm{u}(x_s^\textrm{t})/\zeta}}, \tag {11}
	\end{align}	
\end{subequations}
where $b_s^\textrm{u}(x_s^\textrm{t})$ denotes $u$-th elements of vector $b_s(x_s^\textrm{t})$. Notably, $\zeta$ is the temperature and the $\zeta$ used in \eqref{Opt_7} and \eqref{Opt_11} of this work is the same (identical to \cite{B_Peng_Correlation_Congruence_ICCV}). Similarly, the soft probability of SNet, $\widehat{\varrho_{s}^\textrm{k}}(x_s^\textrm{t})$, can also be obtained by softmax\_T.
\par	
In KD, an important point is the KD loss which will be used as a part of the loss function when training the student network of local client. To gain the KD loss (denoted as $\mathcal{L}_{s, SN}^\textrm{KD}$), Kullback-Leibler (KL) divergence can be employed. Together with considering the availability of that client (i.e., $\alpha_s$), according to \cite{B_Peng_Correlation_Congruence_ICCV} and eq. \eqref{Opt_7}, it can be calculated as:
\begin{subequations}\label{Opt_12}
	\vspace{-0.1cm}
	\setlength{\abovedisplayskip}{3pt}
	\setlength{\belowdisplayskip}{3pt}
	\begin{align}
		&\label{Opt_12:const1}
		\mathcal{L}_{s, SN}^\textrm{KD} = \alpha_s \frac{1}{|\mathcal{X}_s|}\sum_{t=1}^{|\mathcal{X}_s|}\zeta^2KL(\varrho_{s}^\textrm{k}(x_s^\textrm{t}), \widehat{\varrho_{s}^\textrm{k}}(x_s^\textrm{t})), \tag {12}
	\end{align}	
\end{subequations}
where $\mathcal{X}_s$ is the set of total network traffic data of client $s$ ($\mathcal{X}_s = \widehat{\mathcal{X}}_s \cup \widetilde{\mathcal{X}}_s$ as defined in Section \ref{Malicious_Benign_Boundary_Model_Section_3_B}).
\par
In this work, we consider using SNet of each local client to join in the FL process, where the prototypes produced by each student network will be used in the FL process. Thus, in the following, we will illustrate the contents of the prototypes.
	
\subsection{Prototypes Generation in Each Local Student Network}
\label{section_6_prototypes_each_local_client}
\begin{figure} [htpb]
	\centering
	\includegraphics[scale = .55]{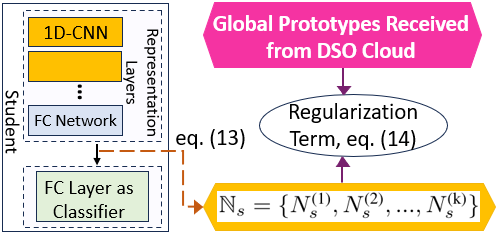} 
	\captionsetup{font=small}
	\caption{Prototypes Generation in the local student network.}
	\label{fig_Prototype} 
\end{figure}

In this part, we will go through the process regarding the prototypes generation in each local student network, as the prototypes will be utilized in the FL process. The prototypes generation process is shown in Fig. \ref{fig_Prototype}. Next, we will illustrate corresponding contents in detail.
\par	
As mentioned in Section \ref{Section_5_Backgound_Prototype_Aggregation}, \textit{Prototype} is the mean value of the embedding vectors for the instances that belong to class $k$ \cite{Y_Tan_FedProto_Federated_Prototype_Learning}. Here, an embedding vector is the output of the representation layers for an input instance. Let $\mathbb{N}_s=\{N_s^\textrm{(1)}, N_s^\textrm{(2)}, ..., N_s^\textrm{(k)}\}$ denote the prototype set generated by the student network of each participating client $s$ (i.e., $\alpha_s=1$), as shown in Fig. \ref{fig_Prototype} (a part of Fig. \ref{fig_proposed_method_E_FPKD}). Taking the availability of client $s$ (i.e., $\alpha_s=1$) into account, as per \cite{Y_Tan_FedProto_Federated_Prototype_Learning}, the prototype $N_s^\textrm{(k)}$ of a participating client $s$ can be represented by,
\begin{subequations}\label{Opt_13}
	\setlength{\abovedisplayskip}{3pt}
	\setlength{\belowdisplayskip}{3pt}
	\begin{align}
		&\label{Opt_13:const1}
		N_s^\textrm{(k)} = \alpha_s \frac{1}{|\mathcal{D}_{s,k}|}\sum_{(x_s^\textrm{t}, y_s^\textrm{t})\in\mathcal{D}_{s,k}} \Omega_s(\theta_s; x_s^\textrm{t}), \tag {13}
	\end{align}	
\end{subequations}
where $\mathcal{D}_{s,k}$ is composed of the training network traffic data and labels that belong to the $k^{th}$ category, which is the subset of the local network traffic dataset ($\mathcal{D}_\textrm{s} = \{(x_s^\textrm{1}, y_s^\textrm{1}), (x_s^\textrm{2}, y_s^\textrm{2}), ..., (x_s^\textrm{t}, y_s^\textrm{t})\}, x_s^\textrm{t} \in \mathcal{X}_s, y_s^\textrm{t} \in \mathcal{Y}_s $). $\Omega_s(\theta_s; x_s^\textrm{t})$ is the embedding function for the representation layer with the parameter $\theta_s$. After obtaining the prototypes in each local client, those prototypes will be transmitted to the DSO cloud for aggregation to produce the global prototypes. The aggregation process will be explained in Section \ref{Section_6_Availability_Aggregation_Process}.
\par
During student network training, according to \cite{D_Cheng_ProtoHAR_Prototype_FL}, in order to achieve a better representation net, the regularization term for the student network, $\mathcal{L}_{s, SN}^\textrm{R}$, can be considered and this regularization term will be added to the loss function of the local student network. Specifically, the regularization term can be obtained through punishing the distance between two types of prototypes (i.e., local prototype and global prototype) \cite{D_Cheng_ProtoHAR_Prototype_FL}. Particularly, it is noteworthy that for each local SNet, we use the prototypes generated in the current round and the global prototypes obtained in the previous round to calculate the regularization term which can be given as below \cite{Y_Tan_FedProto_Federated_Prototype_Learning}:
\begin{subequations}\label{Opt_14}
	\begin{align}
		&\label{Opt_14:const1}
		\mathcal{L}_{s, SN}^\textrm{R} = \sum_{k}distance(N_s^\textrm{(k)}, \overline{N}^\textrm{(k)}), \tag {14}
	\end{align}	
\end{subequations}
where $\overline{N}^\textrm{(k)}$ is the global prototype for class $k$ which will be given in Section \ref{Section_6_Availability_Aggregation_Process}. For the distance metric, $distance(\cdot, \cdot)$, it can be gained by using \textit{L1}, \textit{L2 distance} and the forth \cite{D_Cheng_ProtoHAR_Prototype_FL}.

\subsection{Local Objective of Each Client}
\label{Section_6_Local_objective}
Naturally, many loss functions have been designed by researchers to raise classification accuracy 	\cite{Y_Tian_Recent_Advances_on_loss_functions}. Loss function optimization is usually expressed as a minimization problem \cite{Torre_Weighted_Kappa_Loss_Function_Classification}. Motivated by these, to maximize the overall detection correctness of all prosumers (ODC) (i.e., \eqref{Opt_5}), minimizing the loss function for each local SNet is considered. Before defining the local loss function of this work, it is valuable to mention that multiple types of cyberattacks may exist in the real world. Thus, without loss of generality, we consider both binary and multi-class classification in this work. Particularly, we only consider two classes: malicious (i.e., anomaly) class and benign (i.e., normal) class for binary classification, while we consider a normal class and various cyber-attacks classes for multi-class classification. In the following, we will illustrate the considered local loss function in detail.
\par
Concretely, the local loss function of student network per client $s$ (denoted as $\mathcal{L}_{s, SN}$) is defined as the linear combination of the loss for the supervised learning of SNet (denoted as $\mathcal{L}_{s, SN}^\textrm{SL}$) and the regularization term. Regarding $\mathcal{L}_{s, SN}^\textrm{SL}$, we define it as the linear combination of the binary cross entropy (BCE) \cite{H_X_Gao_Hybrid_ConvLSTM_Anomaly} loss and KD loss for the binary classification, and the linear combination of the cross entropy (CE) loss \cite{Y_Wang_Symmetric_Cross_Entropy_Noisy} and KD loss for the multi-class classification \footnote{Binary cross entropy is the loss function used in binary classification \cite{H_X_Gao_Hybrid_ConvLSTM_Anomaly}, where the label is defined as $0$ and $1$. Cross entropy loss is one of the most widely utilized loss function in the process of deep neural network model training, especially for multi-class classification problems \cite{E_Gordon_Rodriguez_Uses_Abuses_Cross_Entropy_Loss}.}. For the readability of this paper, we use $\mathcal{L}_{s, SN}^\textrm{entropy}$ to represent BCE loss and CE loss. Namely, $\mathcal{L}_{s, SN}^\textrm{entropy}$ is BCE loss for binary classification, and is CE loss for multi-class classification. As a result, we give the loss function of the student network of client $s$ as below:
\begin{subequations}\label{Opt_15}
	\begin{align}
		&\label{Opt_15:const1}
		\hspace{-0.3cm}\mathcal{L}_{s, SN} = \mathcal{L}_{s, SN}^\textrm{SL} + \gamma\mathcal{L}_{s, SN}^\textrm{R}  \notag \\ & \qquad		
		= \psi\mathcal{L}_{s, SN}^\textrm{entropy} + (1-\psi)\mathcal{L}_{s, SN}^\textrm{KD} + \gamma \mathcal{L}_{s, SN}^\textrm{R}, \tag {15}
	\end{align}	
\end{subequations}
where $\psi$ and $\gamma$ are hyperparameters. Consequently, the local objective of each local student network can be given as below: 
\begin{subequations}\label{Opt_16}
	\vspace{-0.4cm}
	\begin{align}
		\underset{\theta_s, \vartheta_s, N_s^\textrm{(k)}} \min 
		&\; \mathcal{L}_{s, SN}, \tag {16}
	\end{align}	
\end{subequations}
where $\vartheta_s$ denotes the parameter of the classifier. As for the teacher network, to pre-train it, we simply use BCE loss for binary classification, and CE loss for multi-class classification.

\subsection{Availability-based Aggregation Process (Global Side) for Cyber-Attacks Detection}
\label{Section_6_Availability_Aggregation_Process}
\par
To handle non-IID data issue, in addition to KD, we also employ the prototype aggregation technique \cite{Y_Tan_FedProto_Federated_Prototype_Learning} at each round to assist non-IID data processing. In order to facilitate future applications, we also take into account aggregating the local models received from the SNet of each participating client for global model generation. The details will be explained next.
\par
\textbf{\textit{Prototype aggregation at each round.}} \qquad  After DSO cloud receives the prototypes generated by the student network of each participating client at each round, the prototype aggregation process will be occurred. Considering the availability of each client $\alpha_s$, according to \cite{Y_Tan_FedProto_Federated_Prototype_Learning} and eq. \eqref{Opt_8}, the global prototype (denoted as $\overline{N}^\textrm{(k)}$) for class $k$ can be obtained by,
\begin{subequations}\label{Opt_17}
	\begin{align}
		&\label{Opt_17:const1}
		\overline{N}^\textrm{(k)} = \frac{1}{|\mathcal{M}_k|}\sum_{s\in\mathcal{M}_k} \frac{\alpha_s|D_{s,k}|}{\sum_{s\in\mathcal{M}_k}|D_{s,k}|} N_s^\textrm{(k)}. \tag {17}
	\end{align}	
\end{subequations}
As denoted before, $\mathcal{M}_k$ is a set of the participating clients that own class $k$. $|D_{s,k}|$ is the size of $D_{s,k}$.

\par
\textbf{\textit{Model aggregation at the final round for future application.}} \qquad Through the training process of the E-FPKD approach, although the heterogeneous data issue can be handled, there is another problem that needs to be solved if the aggregation is only based on the prototypes. The problem is that with prototype aggregation, each client will have its own personal model. Thus, if a new client joins the process, even though the global prototypes can be downloaded from the DSO cloud, it requires the newly joined client to do the fine-tuning process by adjusting its local prototypes to match the downloaded global prototypes as much as possible. To tackle this problem, to be simplified, firstly, we assume each participating client agrees with using the same structure. Secondly, we consider using the prototype aggregation for training in all rounds, and then, different from \cite{Y_Tan_FedProto_Federated_Prototype_Learning}, we introduce model aggregation in the final round to generate the global model. Then we consider using this aggregated global model for application (i.e., cyber-attack detection). In such a way, since the obtained global model can be directly utilized by each client, if there is a newly joined client, the fine-tuning process can be eliminated for that client. Let $\omega_s$ indicate the model of the student network of client $s$ that consists of the representation layer parameter $\theta_s$ and the classifier parameter $\vartheta_s$. Namely, $\omega_s=(\theta_s, \vartheta_s)$. Then considering the availability per client simultaneously, as per \cite{L_Zou_When_HFL_Urban_Prosumers}, the global model $\omega$ in the final round (denoted as $Q$) can be calculated as, 
\begin{subequations}\label{Opt_18}
	\vspace{-0.1cm}
	\begin{align}
		&\label{Opt_18:const1}
		\omega = \sum_{s\in\mathcal{S}}\alpha_s\frac{|\mathcal{D}_s|}{\hat{H}}\omega_s, \tag {18}
	\end{align}	
\end{subequations}
where $\hat{H} = \sum_{s \in \mathcal{S}}\alpha_s|\mathcal{D}_s|$ is the total size of dataset owned by the participating clients. To obtain $\omega_s$, as per \cite{H_B_McMahan_Communication_Efficient_Learning_Deep_Networks}, one step of the gradient descent can be performed on the current model by using the local data. Concretely, $\omega_s$ can be calculated by $\omega_s \leftarrow \omega_s - \widehat{\eta}\bigtriangledown\mathcal{L}_{s, SN}$ \cite{H_B_McMahan_Communication_Efficient_Learning_Deep_Networks}.  $\widehat{\eta}$ denotes the learning rate \cite{H_B_McMahan_Communication_Efficient_Learning_Deep_Networks}.

\par
\textbf{\textit{Global objective. }} \quad For global objective, identical to \texttt{FedProto} \cite{Y_Tan_FedProto_Federated_Prototype_Learning} which considered minimizing the summation of loss towards the local learning tasks of all clients, we also adopt the same manner. Taking the per client's availability $\alpha_s$ into consideration, according to \cite{Y_Tan_FedProto_Federated_Prototype_Learning}, the global objective of this work can be defined as,
\begin{subequations}\label{Opt_19}
	\begin{align}
		\underset{\{\overline{N}^{(k)}\}_{k=1}^{K}} \argmin 
		&\; \sum_{s=1}^{|\mathcal{S}|}\frac{\alpha_s|\mathcal{D}_{s}|}{U}\mathcal{L}_{s, SN}^\textrm{SL} + \notag \\ &
		\gamma \cdot \sum_{k=1}^{K}\sum_{s=1}^{|\mathcal{S}|} \frac{\alpha_s|\mathcal{D}_{s,k}|}{U_k}distance(\overline{N}^{(k)}, N_s^{(k)}).\tag {19}
	\end{align}	
\end{subequations} 
Here, $K$ is the total number of classes. Since one prototype corresponds to one category, $K$ can also be recognized as the total number of global prototypes. $U$ is leveraged to represent the overall number of all participating clients' instances, where $U=\sum_{k}^{K}U_k$ and $U_k = \sum_{s\in\mathcal{M}_k}|D_{s,k}|$ denotes the number of instances belonging to class $k$ over all participating clients. As we define the supervised learning loss ($\mathcal{L}_{s, SN}^\textrm{SL}$) as $\mathcal{L}_{s, SN}^\textrm{SL} = \psi\mathcal{L}_{s, SM}^{BCE} + (1-\psi)\mathcal{L}_{s, SM}^{KD}$, we can rewrite \eqref{Opt_19} as below:
\begin{subequations}\label{Opt_20}
	\vspace{-0.1cm}
	\begin{align}
		\underset{\{\overline{N}^{(k)}\}_{k=1}^{K}} \argmin 
		&\; \sum_{s=1}^{|\mathcal{S}|}\frac{\alpha_s|\mathcal{D}_{s}|}{U}\Big[\psi\mathcal{L}_{s, SM}^{BCE} + (1-\psi)\mathcal{L}_{s, SM}^{KD}\Big] + \notag \\ &
		\gamma \cdot \sum_{k=1}^{K}\sum_{s=1}^{|\mathcal{S}|} \frac{\alpha_s|\mathcal{D}_{s,k}|}{U_k}distance(\overline{N}^{(k)}, N_s^{(k)}).\tag {20}
	\end{align}	
\end{subequations} 

\subsection{Rule-based Method for Cyber-Attacks Intervention}
\label{Rule_Based_Cyber_Attack_Intervention_Application_Stage}
\par
The global model generated by the model aggregation at the final round will be utilized for cyber-attack detection in the future application stage. In this stage, after detecting the cyber attacks, a cyber-attacks intervention process on each dedicated local edge server (DLES) will be performed, aiming to protect the considered prosumer-based EVCSs from being attacked by those detected cyber attacks. 
\par 
To do this, \textit{rule-based method} \cite{S_Wang_Integrated_IDS_Cluster_Wireless} is introduced. Concretely, to express the rule, a basic method called \textit{``if...then"} \cite{S_Wang_Integrated_IDS_Cluster_Wireless} is adopted, which means if ``condition" is true, then ``conclusion" will be performed. Inspired by \cite{M_Habiba_Edge_Intelligence_Network_Intrusion_Prevention} which considers triggering a preventive event in the edge service when DDoS attack is identified, we consider the following rules:
\begin{itemize}
\item If the detected result belongs to the malicious category, then the preventive event \cite{M_Habiba_Edge_Intelligence_Network_Intrusion_Prevention} will be triggered at each DLES automatically. For simplicity, we consider blocking the malicious NT directly as the preventive event.
\item If the detected result is normal, then we do not kick-start any preventive event at DLES.
\end{itemize}

\subsection{Algorithm for the Proposed E-FPKD Approach Towards Cyber-Attacks Prevention}
\label{sub_section_H_Algorithm_E_FPKD}
\par 
Algorithm \ref{alg_2} shows global model generation process in E-FPKD approach, while Algorithm \ref{alg_3} shows the proposed E-FPKD for cyber-attack prevention in future application stage. 
\begin{figure}[t!]	
	\begin{algorithm}[H]	
		\renewcommand{\algorithmicrequire}{\textbf{Input:}}
		\renewcommand{\algorithmicensure}{\textbf{Output:}}
		\caption{Global Model Generation in E-FPKD for Cyber-attack Detection}
		\label{alg_2}
			\begin{algorithmic}[1]
				\REQUIRE Each client's dataset: $\mathcal{D}_s$, client availability: $\alpha_s, s \in \mathcal{S}$, training round: $Q$
				\ENSURE Global model $\omega$
				\STATE Feature selection via \textbf{Algorithm \ref{alg_1}}
				\STATE Pre-train TM per client  \\
				\hspace{-0.55cm} \textbf{Process in the DSO cloud:}
				\STATE Initialize global model $\omega$
				\STATE Initialize each client's student network model: $\omega_s \leftarrow \omega$
				\STATE Initialize the set of global prototype $\{\overline{N}^\textrm{(k)}\}$ for all classes
				\FOR{each round $q = 1, 2, ..., Q$}
				\STATE Get participating client by checking $\alpha_s$
				\FOR{each participating client $s$ \textbf{in parallel}}
				\STATE Submit the global prototype set $\{\overline{N}^\textrm{(k)}\}$ to the participating client $s$
				\STATE $\{N_s^\textrm{(k)}\} \leftarrow \textrm{ClientUpdate(s, } \{\overline{N}^\textrm{(k)}\} \textrm{, q)},$ if $q \neq Q,$  \\  $\omega_s$, $\{N_s^\textrm{(k)}\} \leftarrow \textrm{ClientUpdate(s, } \{\overline{N}^\textrm{(k)}\} \textrm{, q)},$ if $q = Q$ 
				\ENDFOR
				\STATE Update global prototype via eq. \eqref{Opt_17}: $\{\overline{N}^\textrm{(k)}\}$
				\IF{$q == Q$}
				\STATE Generate global model $\omega$ by aggregating each local student network model $\omega_s$
				\ENDIF
				\ENDFOR
				\STATE \textbf{return} the obtained global model $\omega$
				\hfill \break
				\\
				\hspace{-0.55cm} \textbf{ClientUpdate}(s, $\{\overline{N}^\textrm{(k)}\}$, q):
				\FOR{each epoch}
				\FOR{batch $(x_s, y_s) \in \mathcal{D}_s $}
				\STATE Get embedding vectors from representation layer
				\STATE  Compute KD loss: $\mathcal{L}_{s, SN}^\textrm{KD}$
				\STATE Calculate BCE/CE loss: $\mathcal{L}_{s, SN}^\textrm{entropy}$
				\STATE Calculate regularization term: $\mathcal{L}_{s, SN}^\textrm{R}$
				\STATE Calculate local loss: $\mathcal{L}_{s, SN}$
				\STATE $\omega_s \leftarrow \omega_s - \widehat{\eta}\bigtriangledown\mathcal{L}_{s, SN}$
				\ENDFOR		    
				\ENDFOR
				\STATE Get the prototype set $\{N_s^\textrm{(k)}\}$ via eq. \eqref{Opt_13}
				\IF {$q \neq Q$} 	    
				\STATE \textbf{return} $\{N_s^\textrm{(k)}\}$
				\ELSE 
				\STATE \textbf{return} $\omega_s$, $\{N_s^\textrm{(k)}\}$
				\ENDIF				
			\end{algorithmic}  
		\end{algorithm} 
		\vspace{-0.9cm}
	\end{figure}
\par
In Algorithm \ref{alg_2}, in line $1$, we use Algorithm \ref{alg_1} to execute feature selection process. In line $2$, we pre-train the TM of each client. Lines $3-17$ show the process in the DSO cloud. Concretely, in lines $3-5$, we initialize the global model, the student network model per client and the global prototype set for all classes. Next, for each round (line $6$), we check the availability of each client and get the participating clients in line $7$. Then for each participating client $s$ (line $8$), the global prototype set will be submitted to it (line $9$). In line $10$, if it is not the final round (the final round is denoted as $Q$), only the prototypes of SNet of the participating client $s$ will be received by DSO cloud. Otherwise, both the prototypes and SNet model of client $s$ will be received. Those prototypes and the local student network model are generated in lines $18-33$. After receiving the prototypes from all participating clients, the global prototype will be updated with eq. \eqref{Opt_17} as shown in line $12$. Finally, in lines $13-15$, we check whether the current round is the ultimate round, if it is (line $13$), we will aggregate the student network models received from all participating clients to generate the global model (line $14$) which will be used to detect cyber-attacks.
\begin{figure}[t!]
	\begin{algorithm}[H]	
		\caption{E-FPKD for Cyber-attack Prevention (Future Application Stage)}
		\label{alg_3}
		\begin{algorithmic}[1]	
			\STATE Feature selection via \textbf{Algorithm \ref{alg_1}}	
			\FOR{each joined client}
			\STATE Load the global model $\omega$ obtained by \textbf{Algorithm \ref{alg_2}}
			\STATE Get detection results by executing the global model
			\STATE Block the malicious NT data via the rule-based method
			\ENDFOR
		\end{algorithmic}  
	\end{algorithm} 
	\vspace{-0.7cm}
\end{figure}
\par
In Algorithm \ref{alg_3}, we list the process of applying the obtained global model for cyber-attack prevention in the future application stage. In line $1$, we do feature selection for new NT data via Algorithm \ref{alg_1}. Then for each joined client (line $2$), cyber-attack prevention will be performed in lines $3-5$. Concretely, we load the global model obtained by Algorithm \ref{alg_2} to each joined client in line $2$. In line $3$, the detection results will be obtained. Based on the detection results, the rule-based method will be performed to block the malicious NT data in line $4$.
\begin{remark}
\textbf{\textit{Trade-off between Security and Computational Efficiency. }} \quad The proposed E-FPKD approach is designed based on FL. It is possible to support EVCSs with large-scale prosumers. When the No. of prosumers increases, we merely need to raise the No. of the clients in the proposed method. Like FL, accessing the local data owned by each client is not required by the server. Thus, compared with centralized manner, the data privacy issue caused by centrally gathering data can be protected to a certain extent. Hence, for an environment with high data privacy requirements, the proposed method can be preferred over centralized methods. As for the independent method (training model individually), the average performance across all the clients may be degraded, compared with the proposed method. Because of the existence of the data non-IID issue, some clients may not have abundant data. This can affect the overall performance, because insufficient data can hinder the performance as per \cite{W_Yoon_CollaboNet_collaboration_DNN}. Thus, there is a trade-off between the proposed method and the independent methods. Regarding the proposed method, the additional computational overhead may happen at each client during teacher model (TM) pre-training phase. However, this only occurs in the beginning. Because we do not need to train TM every round, thereby, the computation overhead of the proposed method can be still smaller than the one with training TM every round.
\par
In addition, we introduce prototype aggregation (PA) in each round, while only in the final round, both PA and model aggregation (MA) are used. Owing to this manner, the risk of inferring original input data may be reduced, compared with using MA in each round. Because the input data is able to be inferred through the given model \cite{Z_Xiong_Privacy_Thread_FL_NonIID_AIoT}. As for each prototype, it is an average value of the embedding vectors and a prototype corresponds to a class, which can increase the difficulty of reasoning the data structure (e.g., data vector length). Accordingly, since we only adopt MA in the final round, we can infer that the data leakage risk of the proposed method can be smaller than using MA per round. Moreover, since KD and prototype aggregation are beneficial for non-IID data processing, the proposed method is advantageous in processing heterogeneous data. As a result, considering the security issue (data leakage) and the potential of handling non-IID data, it is reasonable to employ the proposed method for cyber-attack prevention.
\end{remark}
\par
\textbf{\textit{Time Complexity (TC). }} \quad  
The proposed E-FPKD approach is an FL-based approach composed of a server (i.e., DSO cloud) and multiple clients, where each client consists of TNet and SNet. In particular, TNet is pre-trained and SNet is used for training. We adopt the combination of CNN and fully connected (FC) network to form each TNet and SNet. For TNet, same as \cite{P_J_Freire_Performance_Versus_Complexity}, we calculate its time complexity by summing up the time complexity of CNN and FC networks. Specifically, the TC of a convolutional layer of TNet can be given as: $O\Big(G^\textrm{ne} G^\textrm{nf} G^\textrm{ks} (G^\textrm{ts} - G^\textrm{ks} + 1)\Big)$ \cite{P_J_Freire_Performance_Versus_Complexity}, where $G^\textrm{ne}$, $G^\textrm{nf}$, $G^\textrm{ks}$ and $G^\textrm{ts}$ denote the number of elements of the input vector, the number of CNN filters, kernel size and the number of time steps (time slots in this work) \cite{P_J_Freire_Performance_Versus_Complexity}. As shown in \cite{K_He_CNN_Constrained_Time_Cost}, the overall time complexity of all the convolutional layers is given by summarizing the TC of each convolutional layer. Thus, the time complexity of all the convolutional layers for each TNet is $O\Big(\sum_{e_{t}} G^\textrm{ne} G^\textrm{nf} G^\textrm{ks} (G^\textrm{ts} - G^\textrm{ks} + 1)\Big)$. Here, $e_{t}$ denotes $(e_{t})^\textrm{th}$ convolutional layer of TNet. As for FC network, the time complexity can be given as $O\Big(\sum_{\widehat{e_{t}}} G_{\widehat{e_{t}}-1}^{\textrm{nu}} G_{\widehat{e_{t}}}^{\textrm{nu}}\Big)$ \cite{L_Zou_When_HFL_Urban_Prosumers}, where $G_{\widehat{e_{t}}-1}^{\textrm{nu}}$ is the number of the neural units (NUs) in the FC layer $\widehat{e_{t}}-1$ and $G_{\widehat{e_{t}}}^{\textrm{nu}}$ indicates the number of the NUs in the FC layer $\widehat{e_{t}}$. Accordingly, the time complexity of each client's TNet is $O\Big(\sum_{e_{t}} G^\textrm{ne} G^\textrm{nf} G^\textrm{ks} (G^\textrm{ts} - G^\textrm{ks} + 1) + \sum_{\widehat{e_{t}}} G_{\widehat{e_{t}}-1}^{\textrm{nu}} G_{\widehat{e_{t}}}^{\textrm{nu}}\Big)$. In contrast with TNet, during training, SNet per client $s$ needs to compute the local prototypes. Thus, for SNet, the time complexity contains the time complexity of both the network and local prototypes generation. Considering each SNet generates $K_s$ prototypes, the time complexity of each SNet at most is $O\Big(\widehat{E}(\widehat{Z} + K_s|\mathcal{D}_s|)\Big)$ \cite{C_Pan_Fair_Graph_FL_Incentive_Mechanisms}. Here, $\widehat{E}$ denotes the local iteration, and $\widehat{Z}$ is the time complexity of the backbone network (i.e., CNN + FC network). Similar as TNet, $\widehat{Z}$ can be given as $O\Big(\sum_{e_{s}} G^\textrm{ne} G^\textrm{nf} G^\textrm{ks} (G^\textrm{ts} - G^\textrm{ks} + 1) + \sum_{\widehat{e_{s}}} G_{\widehat{e_{s}}-1}^{\textrm{nu}} G_{\widehat{e_{s}}}^{\textrm{nu}}\Big)$, where $e_{s}$ and $\widehat{e_{s}}$ represent the $(e_{s})^\textrm{th}$ convolutional layer and $(\widehat{e_{s}})^\textrm{th}$ FC layer of SNet, respectively. $G_{\widehat{e_{s}}-1}^{\textrm{nu}}$ and $G_{\widehat{e_{s}}}^{\textrm{nu}}$ separately denote the NUs in the FC layer $\widehat{e_{s}}-1$ and in the layer $\widehat{e_{s}}$ of FC network. In regard to the server, we perform prototype aggregation in each round and both prototype and model aggregation in the final round. For model aggregation, FedAvg \cite{H_B_McMahan_Communication_Efficient_Learning_Deep_Networks} is adopted in this work. Let $M$ denote the number of participating clients in each round, the time complexity for server to perform FedAvg is bounded as $O(M)$ \cite{X_Li_FedGTA_Topology_Aware_FGL}. As for prototype aggregation, in each round, the time complexity of updating the global prototype through prototype aggregation is no more than $O(KM)$ \cite{C_Pan_Fair_Graph_FL_Incentive_Mechanisms}. Accordingly, the TC of server is $O(M + KM)$ in the final round, while TC of server is $O(KM)$ in other round. Thus, the overall TC of the proposed method for SNet per client is $O\Big(\widehat{E}(\widehat{Z} + K_s|\mathcal{D}_s|) + (KM + M)\Big)$ \cite{C_Pan_Fair_Graph_FL_Incentive_Mechanisms} in the final round. As for other round, the overall time complexity for SNet of each client is $O\Big(\widehat{E}(\widehat{Z} + K_s|\mathcal{D}_s|) + KM\Big)$.

\section{Experimental Results and Analysis}
\label{Section_7_experimental_result_analysis}
In this section, we conduct experiments for the proposed method, where both quantitative and qualitative analyses are considered. The detailed contents will be shown from Section \ref{Experimental_Setup_VII} to \ref{Evaluation_via_Qualitative_Analysis_VII}. Notably, for cyber attack intervention, we simply consider directly blocking the malicious NT by a basic rule-based method when an attack is detected. Accordingly, for simplicity, the evaluation for this process is ignored.
\subsection{Experimental Setup}
\label{Experimental_Setup_VII}
\begin{table}[t!]
	\caption{Parameter Settings}
	\begin{center}
		\begin{tabular}{|m{3.8 cm}|m{4 cm}|}
			\hline
			\hfil \textbf{Parameter Description} & \hfil \textbf{Value} \\ \hline \hline
			\hfil No. of clients & \hfil $10$ \cite{O_Jogunola_Prosumers_Matching}, 20, 50 \\ \hline		
			\hfil Dirichlet parameter & \hfil 0.5, 0.9 \cite{H_Li_FedTP_Federated_Learning_Transformer}, 5 \\ \hline
			\multicolumn{2}{|c|}{\textbf{Training Round}} \\ \hline
			\hfil NSL-KDD dataset ($N$)  & \hfil $\{10, 20, 50, 100\}$ \\ \hline
			\hfil UNSW-NB15 dataset ($U$)  & \hfil $\{5, 10, 20, 50\}$ \\ \hline
			\hfil IoTID20 dataset ($I$)  & \hfil $\{5, 10, 15, 20\}$ \\ \hline
			\multicolumn{2}{|c|}{\textbf{Teacher Network}} \\ \hline
			\hfil Learning rate & \hfil $0.001$ \\ \hline
			\hfil Batch size & \hfil $32$ ($N$) / $512$ ($U$) / $128$ ($I$) \\ \hline
			\hfil Local epoch & \hfil $5$ \\ \hline
			No. of channels for the input of each convolutional layer & \hfil $\{1, 512, 1024\}$ \\ \hline
			No. of channels generated by each convolutional layer & \hfil $\{512, 1024, 2048\}$ \\ \hline
			\hfil Input unit for the $1^{st}$ FC layer & No. of features for a record $\times$ $2048$ \\ \hline
			\hfil Output unit for the $1^{st}$ FC layer & \hfil $512$ \\ \hline
			Input unit for the $2^{nd}$ FC layer & \hfil $512$ \\ \hline
			Output unit for the $2^{nd}$ FC layer & \hfil No. of classes \\ \hline	
			\hfil Optimizer & \hfil Adam \\ \hline
			\multicolumn{2}{|c|}{\textbf{Student Network}} \\ \hline
			\hfil Initial learning rate & \hfil $0.0001$ \\ \hline
			\hfil Learning rate decay & \hfil $0.97$ \cite{G_Yan_Seizing_Critical_Learning} \\ \hline
			\hfil Batch size ($\{N, U, I\}$ dataset) & \hfil $32$ ($N$) / $512$ ($U$) / $128$ ($I$) \\ \hline
			\hfil Local epoch & \hfil $5$ \\ \hline
			Temperature ($\{N, U, I\}$ dataset) & \hfil $0.5$ ($I$) \cite{K_Kwon_Adaptive_KD_Entropy}  / $0.1$ ($U, I$) \cite{T_Liang_Compressing_Multiobject_Tracking_Model} \\ \hline
			\hfil Hyperparameters $\psi$ and $\gamma$ & \hfil $0.1$ ($\psi$), $1$ ($\gamma$)\\ \hline
			No. of input channels for each 1D-convolutional layer & \hfil  $\{1, 64\}$ \\  \hline
			No. of output channels for each 1D-convolutional layer & \hfil  $\{64, 128\}$ \\ \hline
			\hfil Input unit for the $1^{st}$ FC layer & No. of features for a record $\times$ $128$  \\ \hline
			\hfil Output unit for the $1^{st}$ FC layer & \hfil $64$ \\ \hline
			Input unit for the $2^{nd}$ FC layer & \hfil $64$ \\ \hline
			Output unit for the $2^{nd}$ FC layer & \hfil No. of classes \\ \hline
			\hfil Optimizer & \hfil SGD \\ \hline
		\end{tabular}
		\label{sec_7_tab1}
	\end{center}
	\vspace{-0.6cm}
\end{table}
\subsubsection{Model Setup}
In this article, we consider the teacher network consists of three 1D-convolutional layers with two fully connected (FC) layers. Particularly, the number of channels for the input of the convolutional layers is separately $1$, $512$ and $1024$, while the number of channels generated by the convolutional layers are $512$, $1024$, and $2048$, respectively. Each convolutional layer is considered to connect with a Batch Normalization and a ReLU activation function. As for the first FC layer, the input unit is \textit{the features' number for a record $\times$ $2048$}, and the output unit is $512$. For the second FC layer, the input unit is $512$, while the output unit is the number of classes. In the case of the student model, we consider it owns two 1D-convolutional layers and two FC layers. Specifically, regarding the convolutional layers, the number of input channels for each layer are separately $1$ and $64$, while the output units are $64$ and $128$, respectively. Besides, the input for the first FC layer is \textit{the number of features for a record $\times$ $128$}, and the corresponding output unit is $64$. The second FC layer's input unit is $64$ and the output unit is the number of classes. In addition, in regard to the optimizers, we adopt Adam for the teacher network and SGD for the student network. Additionally, identical to \cite{B_Salehi_FLASH_FL_Automated_Selection} which considers randomly initializing the weight of the deep neural network (DNN)-based network at the first iteration, we consider initializing the weights of TNet and SNet randomly in the beginning round. Moreover, regarding the availability of the clients, inspired by \cite{Y_Shang_Information_Security_Vehicle_to_Grid} which considers randomly selecting EV charging stations to join the FL process in each round, we also adopt the stochastic selection for determining the availability of the clients (i.e., DLESs of prosumers) in each round. The settings of the parameters are summarized in Table \ref{sec_7_tab1}. Notably, for datasets, we adopt NSL-KDD \cite{Dataset_NSL_KDD}, UNSW-NB15 \cite{Dataset_UNSW_NB15_1, Dataset_UNSW_NB15_2, Dataset_UNSW_NB15_3, Dataset_UNSW_NB15_4, Dataset_UNSW_NB15_5}, and IoTID20 dataset \cite{Dataset_IoTID20} (describe in Section \ref{datasets_description_section_7_3}). \textit{The features' number for a record} depends on the dataset, which will be illustrated in Section \ref{Preprocessing_FC_experimental}.

\subsubsection{Baselines}
To evaluate the effectiveness of the proposed method, we compare it with the following benchmarks: 1) \textit{\texttt{the E-FPKD approach without feature selection (FS)}}, where we remove FS but keep other parts same as the proposed E-FPKD method. 2) \textit{\texttt{The E-FPKD approach without learning rate (LR) decay}}, where we discard the LR decay ($0.97$ \cite{G_Yan_Seizing_Critical_Learning}) per round but embrace other parts of the proposed method. Particularly, we adopt $LR = LR \times 0.97^{\textrm{q}-1}$ to calculate LR in $q^{th}$ round. 3) \textit{\texttt{The E-FPKD approach with DNN-based representation layers}}, where DNN is used to create the representation layers. In particular, for handling the NSL-KDD dataset, the network per client contains two FC layers as representation layers, and one FC layer as classifier. The input units for each FC layer are separately \textit{the number of features for a record}, $128$ and $64$, while the output units for each FC layer are $128$, $64$ and \textit{No. of classes}, respectively. To address the UNSW-NB15 and IoTID20 datasets, we consider using two FC layers, where one layer is regarded as representation layer, and the other is treated as classifer. The input units for each layer are \textit{the number of features for a record} and $32$, respectively. The output units per layer are separately $32$ and \textit{No. of classes}. 4) \textit{\texttt{Independent CNN}}, where we apply CNN for each client independently (the CNN is the same as the SNet adopted by the proposed method); 5) \textit{\texttt{Independent KD}}, where KD technique is leveraged for each client independently (the KD is identical to the proposed method; 6) \textit{\texttt{FedKD}}: where we retain the KD part of the proposed method and use mode aggregation in each round. 7) some other FL-based framework: \textit{\texttt{FedAvg}} \cite{H_B_McMahan_Communication_Efficient_Learning_Deep_Networks}, \textit{\texttt{FedProx}} \cite{T_Li_Federated_Optimization_Heterogeneous}, \textit{\texttt{FedProto}} \cite{Y_Tan_FedProto_Federated_Prototype_Learning}, \textit{\texttt{pFedSD}} \cite{H_Jin_Personalized_self_KD}, \textit{\texttt{FedAU}} \cite{S_Wang_Lightweight_Method_Tackling}, \textit{\texttt{FedExP}} \cite{D_Jhunjhunwala_FedExP_Speeding_Up}, \textit{\texttt{FedGKD}} \cite{D_Yao_W_Pan_FedGKD_Heterogeneous_Federated_Learning}. For these methods, we embrace the student network of the proposed method to form each client. Next, we will discuss  the \textbf{limitations of baseline methods and how to deal with them with the proposed E-FPKD approach}.
\par
In regard to E-FPKD without feature selection (FS) method, lack of FS may result in irrelevant features in the input data. Large irrelevant features will cause serious issues for the existing machine-learning approaches \cite{Y_Sun_Local_Learning_Based_FS}. Thus, since FS is capable of excluding irrelevant features \cite{T_Wisanwanichthan_Double_Layered_Hybrid_Approach_NIDS}, we adopt FS in this work. With respect to the method of E-FPKD without learning rate (LR) decay, the LR of the network will continue to be fixed. According to \cite{Y_Zhang_Fault_Diagnosis_Rotating_Machinery}, if LR is too small, the model training process will converge slowly; if LR is too large, the model performance will degrade. Hence, to solve this issue, same as \cite{Y_Zhang_Fault_Diagnosis_Rotating_Machinery}, we adopt LR decay in this work, which will make the LR vary each round. When it comes to the method of E-FPKD with DNN-based representation layers (RLs), DNN is used as the RLs. For this manner, huge amount of data may be required for training. This is because to achieve the target accuracy, DNN needs to use large input data to train the model \cite{W_Xu_Parallelizing_DNN}. However, some prosumers may lack data, thus, in such a situation, it will become inappropriate to use DNN. As per \cite{Rahimilarki_CNN_wind_turbine_machine}, adding the convolutional layer can increase the network nonlinearity, which can help increase the accuracy. Therefore, instead of DNN, we employ the convolutional layers in this work. In the case of FedKD, FedAvg \cite{H_B_McMahan_Communication_Efficient_Learning_Deep_Networks}, FedProx \cite{T_Li_Federated_Optimization_Heterogeneous}, pFedSD \cite{H_Jin_Personalized_self_KD}, FedAU \cite{S_Wang_Lightweight_Method_Tackling}, FedExP \cite{D_Jhunjhunwala_FedExP_Speeding_Up} and FedGKD \cite{D_Yao_W_Pan_FedGKD_Heterogeneous_Federated_Learning}, all of these methods utilize model aggregation method. With model aggregation, as the input data is capable of being inferred through a given model \cite{Z_Xiong_Privacy_Thread_FL_NonIID_AIoT}, the input data inference problem needs to be focused. For FedProto \cite{Y_Tan_FedProto_Federated_Prototype_Learning}, despite FedProto leverages prototype aggregation rather than model aggregation, when a new client joins, fine-tuning process is required \cite{Y_Tan_FedProto_Federated_Prototype_Learning}. To balance the model inference and fine-tuning issues, in the proposed method, we conduct prototype aggregation in each round, and model aggregation in the final round to generate the global model for future use.
\subsubsection{Datasets Description}
\label{datasets_description_section_7_3}
To verify the goodness of the proposed approach, we use three datasets including \textbf{NSL-KDD dataset} \cite{Dataset_NSL_KDD} (i.e., KDDTrain+.csv and KDDTest+.csv), \textbf{UNSW-NB15 dataset} \cite{Dataset_UNSW_NB15_1, Dataset_UNSW_NB15_2, Dataset_UNSW_NB15_3, Dataset_UNSW_NB15_4, Dataset_UNSW_NB15_5} (UNSW\_NB15\_training-set.csv and UNSW\_NB15\_testing-set.csv is adopted), and \textbf{IoTID20 dataset} \cite{Dataset_IoTID20}.
\par 
\textbf{\textit{NSL-KDD dataset \cite{Dataset_NSL_KDD}}} \quad  KDDTrain+.csv is employed for global model generation and KDDTest+.csv is used in application stage. This dataset is a new version of the KDD dataset that is composed of the chosen records from the full KDD dataset \cite{M_Tavallaee_KDD99_NSL_KDD}. As described in \cite{A_Bhardwaj_Hyperband_Tuned}, for the NSL-KDD dataset, it contains $43$ features, where $41$ features are the input traffic features and the other two features are the label of the category (i.e., normal or attack) and the score used to indicate the severity of the attack. For simplicity, the score feature is abandoned in this article. In addition, this dataset is labeled under \texttt{normal} and four categories, viz., \texttt{DoS, probe, U2R} and \texttt{R2L}. It is worth noting that the NSL-KDD dataset has several advantages. As per \cite{M_Tavallaee_KDD99_NSL_KDD}, those advantages are given below: 1) in the training dataset, there are no redundant records, 2) for the test dataset, it does not contain duplicate records, 3) in both the training and test datasets, the number of records is reasonable (i.e., KDDTest+.csv has $22544$ records, while KDDTrain+.csv has $125973$ records). As aforementioned, we consider both binary and multi-class classification. Thus, in \textit{binary classification}, we regard DoS, probe, U2R, and R2L as \texttt{anomaly} class, and normal as \texttt{normal} class. In \textit{multi-class classification}, we consider five classes, i.e., \texttt{DoS, probe, U2R, R2L} and \texttt{normal}.
\par
\textbf{\textit{UNSW-NB15 dataset \cite{Dataset_UNSW_NB15_1, Dataset_UNSW_NB15_2, Dataset_UNSW_NB15_3, Dataset_UNSW_NB15_4, Dataset_UNSW_NB15_5}}} \quad UNSW-NB15 data is generated by \cite{Dataset_UNSW_NB15_1} from the real-world network flows \cite{Y_Yang_Research_Dung_Beetle_Optimization_Stacked}. As highlighted before, UNSW\_NB15\_training-set.csv and UNSW\_NB15\_testing-set.csv are adopted for evaluation process. In particular, UNSW\_NB15\_training-set.csv is utilized for producing the global model, while UNSW\_NB15\_testing-set.csv is adopted for application stage. For these two files, there are $175,341$ and $82,332$ records for training part and testing part, respectively. In the case of the labels for two files, $10$ labels are included, namely, \texttt{Analysis, Backdoor, DoS, Exploits, Fuzzers, Generic, Reconnaissance, Shellcode, Worms}, and \texttt{Normal}. Thus, in \textit{binary classification}, the record with the labels of Analysis, Backdoor, DoS, Exploits, Fuzzers, Generic, Reconnaissance, Shellcode, and Worms are treated as \texttt{anomaly} category, while the records with Normal label are regarded as \texttt{normal} category. In \textit{multi-class classification}, we consider these $10$ labels as $10$ categories for detection.
\begin{figure*} [t!]
	\centering
	\includegraphics[scale = .49]{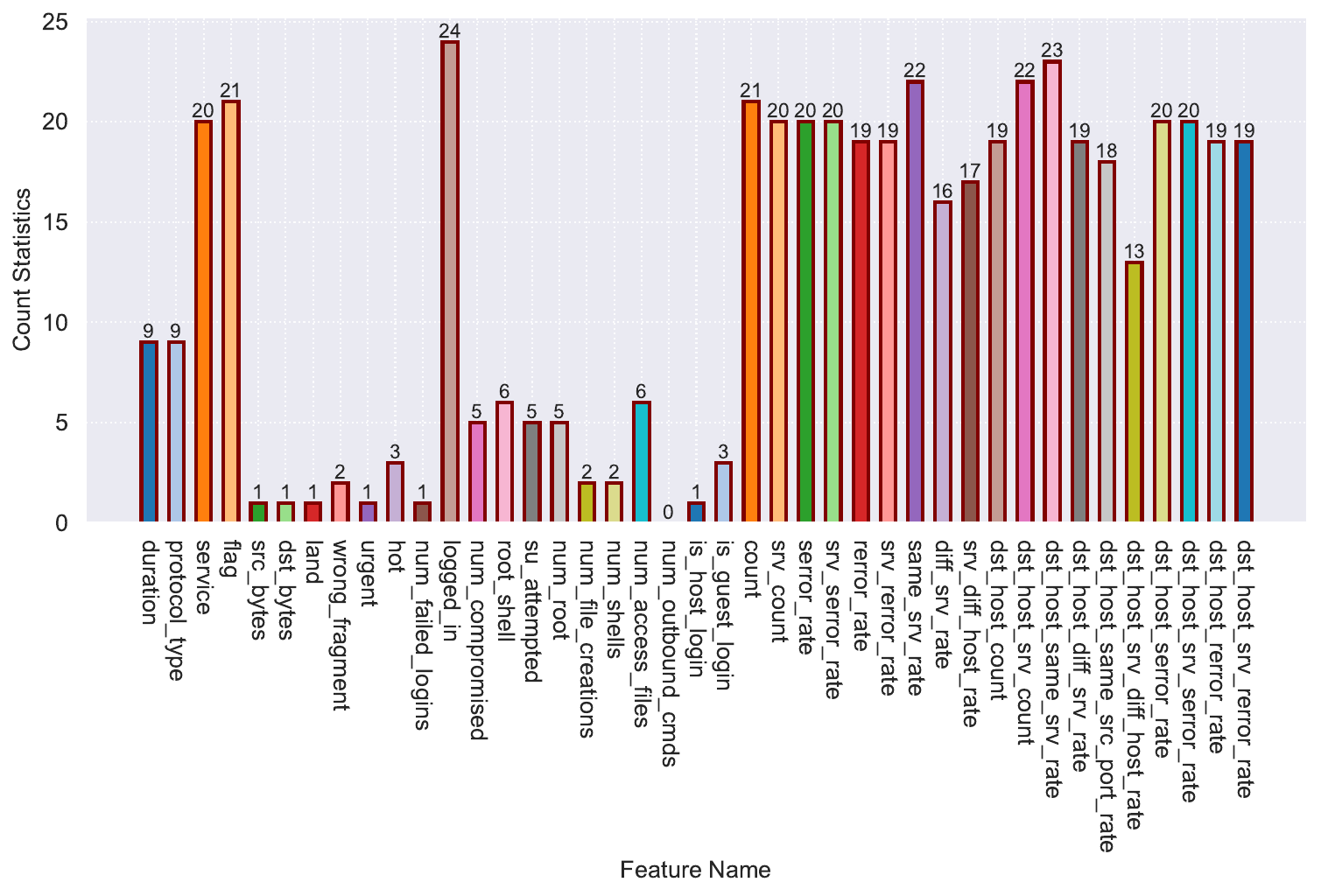}
	\vspace{-0.15cm}   
	\captionsetup{font=small}      
	\caption{The number of PCC values for each feature versus other features that satisfy $PCC \in \{-1, 0.1\} \cup \{0.1, 1\}$ (NSL-KDD Dataset).}
	\label{fig_feature_selection_1} 
	\vspace{-0.4cm}
\end{figure*}
\begin{figure*} [t!]
	\centering
	\includegraphics[scale = .52]{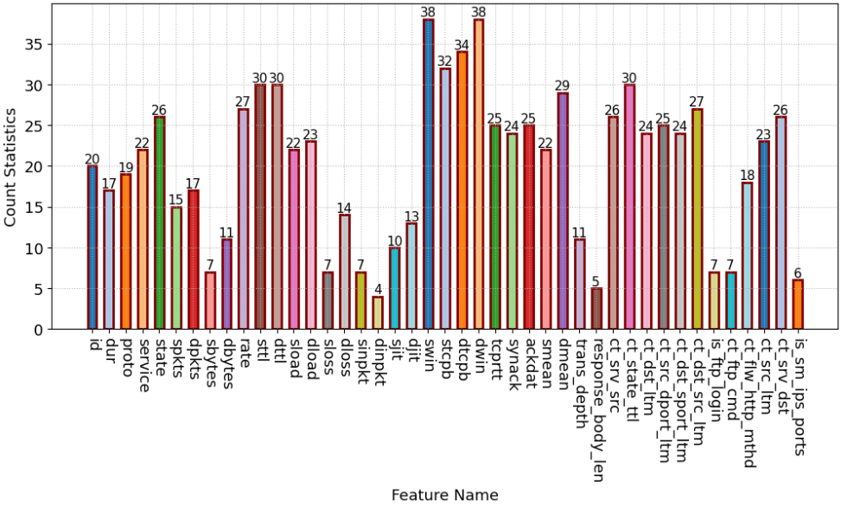}
	\vspace{-0.15cm}   
	\captionsetup{font=small}      
	\caption{The number of PCC values for each feature versus other features that satisfy $PCC \in \{-1, 0.1\} \cup \{0.1, 1\}$ (UNSW-NB15 Dataset).}
	\label{fig_feature_selection_2} 
	\vspace{-0.4cm}
\end{figure*}
\par
\textbf{\textit{IoTID20 dataset \cite{Dataset_IoTID20}}} \quad IoTID20 dataset was proposed by \cite{Dataset_IoTID20}, which contains a huge number of the general network features as well as flow-based features \cite{I_Ullah_Technique_Botnet}. For this dataset, it is produced from a real-time scenario by IoT device \cite{A_A_Alsulami_ID_Classification_System} and can be used for detecting the anomalous activity in the IoT networks \cite{I_Ullah_Technique_Botnet}. Besides, it contains $86$ columns and $625,783$ records (i.e., rows) \cite{A_A_Alsulami_ID_Classification_System}. Considering this dataset contains duplicate records, same as \cite{A_A_Alsulami_ID_Classification_System}, we consider removing those duplicate records. Consequently, only $461,696$ \cite{A_A_Alsulami_ID_Classification_System} records are left. In addition, for IoTID20, there exists several \textit{infinite values} \cite{V_Surya_Effective_Machine_Learning_IoT_SMOTE}. It can be known from \cite{V_Surya_Effective_Machine_Learning_IoT_SMOTE} that we can remove those infinite values. Hence, for simplicity, the records with infinite value are discarded in this paper. Additionally, some features in IoTID20 dataset belong to nominal type, which are highly regarding the testbed's network configuration \cite{M_Omar_Toward_lightweight_ML_Cyber_Intrusions_IoT}. Those features include $\{Flow\_ID, Src\_IP, Dst\_IP, Timestamp\}$ \cite{M_Omar_Toward_lightweight_ML_Cyber_Intrusions_IoT}. As mentioned in \cite{M_Omar_Toward_lightweight_ML_Cyber_Intrusions_IoT}, those features can be discarded. Thus, we get rid of those features for convenience. For global model generation and testing in the application phase, we consider using around $70\%$ and $30\%$ data, respectively. In IoTID20 dataset, $3$ features that are related to label are provided, where those features are named \textit{Label, Cat, and Sub\_Cat}, respectively. To be specific, the \textit{Label feature} is composed of \{\texttt{Anomaly, Normal}\} \cite{V_Surya_Effective_Machine_Learning_IoT_SMOTE} these two classes, while the \textit{Cat feature} consists of \{\texttt{DoS, Mirai, MITM ARP Spoofing (MAS), Scan, Normal}\} \cite{V_Surya_Effective_Machine_Learning_IoT_SMOTE} these five classes. As for the \textit{Sub\_Cat feature}, it consists of more classes. Thus, to be convenient, the \textit{Label feature} is used for binary classification in this work, while the \textit{Cat feature} is leveraged for multi-class classification. 

\begin{figure*} [t!]
	\centering
	\includegraphics[scale = .45]{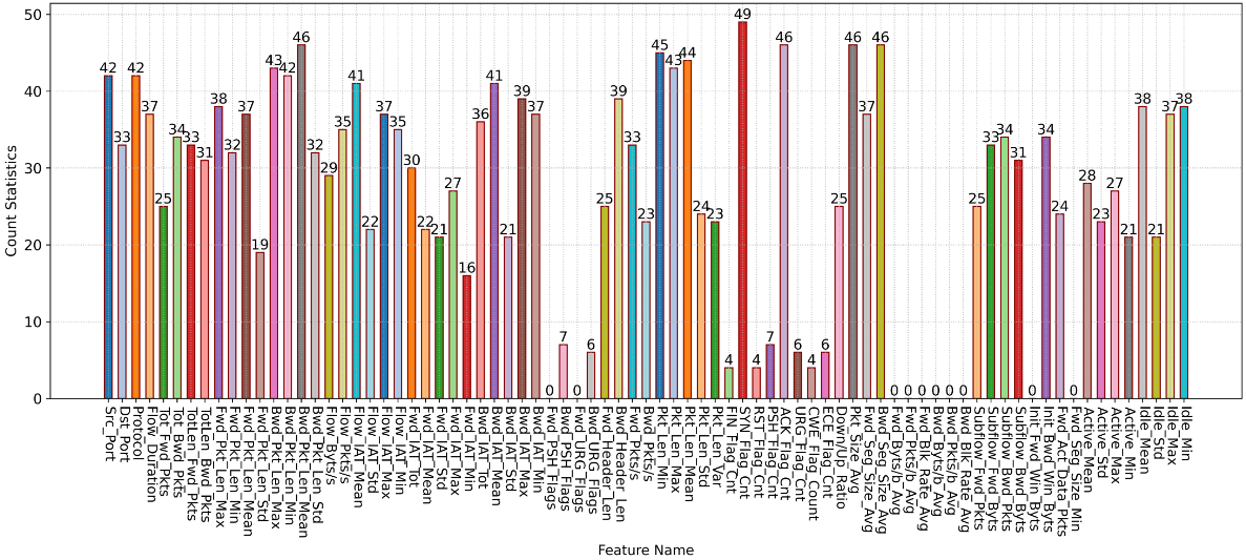}
	\vspace{-0.15cm}   
	\captionsetup{font=small}      
	\caption{The number of PCC values for each feature versus other features that satisfy $PCC \in \{-1, 0.1\} \cup \{0.1, 1\}$ (IoTID20 Dataset).}
	\label{fig_feature_selection_3} 
	\vspace{-0.5cm}
\end{figure*}
\subsubsection{Data Preprocessing via Feature Selection (FS)}
\label{Preprocessing_FC_experimental}
\par
The original NSL-KDD and  UNSW-NB15 datasets have training and test parts. For simplicity, we perform FS process with Pearson Correlation Coefficient (PCC) for the training part of NSL-KDD and UNSW-NB15. For IoTID20, FS process is executed for the whole dataset. As mentioned in Section \ref{sub_section_A_Feature_Selection}, we consider PCC in the range of $[-1, -0.1]\cup[0.1, 1]$ for feature selection. From Fig. \ref{fig_feature_selection_1} to Fig. \ref{fig_feature_selection_3}, we list the number of PCC values for each feature versus other features with $0.1\leq $ PCC $\leq 1$ or $-1 \leq$ PCC $\leq-0.1$. Notably, the \textit{Label feature} is excluded. Since we only focus on the effectiveness of the feature selection with the PCC approach for the proposed E-FPKD method rather than finding the minimum feature set, for simplicity, we retain the features with the larger number of PCC values in the range of $[-1, -0.1]\cup[0.1, 1]$. Empirically, we select the top $22$ features with the largest number of PCC values that meet the conditions for NSL-KDD dataset. Similarly, the top $27$ features of UNSW-NB15 dataset are selected, while the top $40$ features of IoTID20 dataset are chosen. The name of those selected features can be found in Fig. \ref{fig_feature_selection_1}, Fig. \ref{fig_feature_selection_2} and Fig. \ref{fig_feature_selection_3}, respectively.

\subsubsection{Scenario and Condition} 
As the NT data per prosumer may exist non-IID phenomenon, and various levels of data heterogeneity may pose challenges for constructing the cyber attack prevention mechanism in real-life deployments, it is imperative to investigate a mechanism that can shed light on the effectiveness across different heterogeneity degrees towards real-world scenarios. To this end, for the sake of exploring the robustness of the proposed method, to tail the dataset for each client with heterogeneous characteristics, the \textit{Dirichlet Distribution} with parameter $\delta=0.5$ \cite{H_Li_FedTP_Federated_Learning_Transformer}, $\delta=0.9$ \cite{H_Li_FedTP_Federated_Learning_Transformer} and $\delta=5$ are employed. In addition, to further exhibit the goodness, various levels of clients are taken into consideration. To be specific, we consider $10$, $20$, and $50$ clients, respectively. As mentioned in Section \ref{Section_6_Local_objective}, there may be sorts of cyber attacks in the real-world. Consequently, for the considered prosumer-based EV charging stations, to achieve the purpose of cyber attacks prevention, we consider both binary and multi-class classification. As a result, in this section, those two types of classification are taken into account for evaluating the potential advantage of the proposed method. In terms of the \textit{binary classification}, we only consider two classes: $\{anomaly, normal\}$, while the classes in \textit{multi-class classification} will be varied according to the dataset being used in this work. It is noteworthy that in Section \ref{Section_4_System_Model_PD}, we consider the ODC is calculated based on two categories: the malicious traffic and benign traffic. Hence, although multi-class classification is considered, when computing the ODC, various detected anomaly classes will be counted as the category of malicious traffic, while normal traffic belongs to the category of benign traffic. As for the \textbf{condition} towards the considered prosumer-based EVCSs, in order to avoid the fine-tuning process of each newly joined client, as mentioned before, we assume those clients are not against using the same model structure. 

\subsection{Evaluation via Quantitative Performance Metrics Analysis}
In this subsection, firstly, we will show the quantitative performance metrics (QPMs) that will be used in this work. Secondly, we will do the efficiency comparisons among the proposed method and the baselines via those metrics.
\subsubsection{Quantitative Performance Metrics (QPMs) for efficiency analysis} \label{QPM_Efficiency}
QPMs can be \textbf{accuracy, precision, recall and F1-score} \cite{O_Hoekstra_Healthcare_prediction_textual_data}. Particularly, accuracy, precision, recall and F-measure are the typical metrics for evaluating the efficiency of the FL-based intrusion detection system (IDS) \cite{Fedorchenko_Comparative_Review_IDS}. As F1-score is a type of F-measure, we adopt those four terms to identify the efficiency of the proposed method for cyber-attack (i.e., cyber-intrusion) detection. In addition, \textbf{false alarm rate (FAR)} can also be considered as the quantitative performance metric as per \cite{T_Alldieck_Context_Aware_Fusion_RGB}, where FAR is another metric to evaluate the efficiency of IDS \cite{Milan_Reducing_False_Alarm_IDS}. Therefore, we adopt accuracy, precision, recall, F1 score, and FAR as QPMs to analyze the efficiency of the proposed method towards cyber-attack detection. Moreover, based on the first four metrics, we additionally define \textbf{average accuracy (AA), average precision (AP), average recall (AR), and average F1 Score (AFS)} for quantitative analysis. Detailed information will be given next.
\par
\textbf{\textcircled{1} Binary Classification.} \qquad \label{sec7_B1_binary_classification}
In binary classification, we adopt the metrics of: 1) accuracy, 2)precision, 3) recall, 4) F1 score, 5) FAR, 6) AA, AP, AR and AFS. To obtain those metrics, there are four terms that need to be known. Identical to \cite{D_Jing_SVM_NID_UNSW_NB15, J_Kim_Method_intrusion_detection}, we present those four terms as below: 1) True Positive (TP) means correctly detected as attacks (actual is attack traffic), 2) True Negative (TN) indicates correctly detected as normal traffic (actual is normal traffic), 3) False Positive (FP) represents the normal traffic is incorrectly detected as attacks, and 4) False Negative (FN) denotes the attack traffic is incorrectly detected as the normal traffic. For easier understanding, we list the meaning of TP, TN, FP and FN in Table \ref{sec_7_tab2_confusion_matrix} which is based on \cite{D_Jing_SVM_NID_UNSW_NB15, J_Kim_Method_intrusion_detection}.
\begin{table}[htbp]
\renewcommand{\thetable}{\Roman{table}}
\setlength{\extrarowheight}{1pt}
\caption{Confusion Matrix for Binary Classification}
\vspace{-0.35cm}
\begin{center}
\begin{tabular}{|c|c|c|c|}\hline
	\multicolumn{2}{|c|}{ \multirowcell{2.2}{Class}}
	& \multicolumn{2}{c|}{\bfseries Detected Class} \\
	\cline{3-4}
	\multicolumn{2}{|c|}{} & \thead{Anomaly (Positive)} & \thead{Normal (Negative)}\\
	\hline
	\multirowcell{2.7}{\bfseries Actual Class}
	& \thead{Anomaly} & TP & FN\\
	\cline{2-4}
	& \thead{Normal} & FP & TN\\
	\hline
\end{tabular}
\end{center}
\label{sec_7_tab2_confusion_matrix}
\vspace{-0.45cm}
\end{table}
\par
The concrete mathematical calculation for the considered metrics will be clarified next. \textbf{Accuracy}, $\Upsilon_{Acc}$, is the ratio of the total number of correct detections to the total number of detections. It can be calculated as follows \cite{T_Anwar_deep_learning_diagnosis_COVID_19}:
\begin{subequations}\label{Opt_21}
	\vspace{-0.1cm}
	\begin{align}
		&\label{Opt_21:const1}
		\Upsilon_{Acc} = \frac{TP+TN}{FP+TP+FN+TN}. \tag {21}
	\end{align}	
\end{subequations}
For \textbf{average accuracy} $\overline{\Upsilon}_{Acc}$, it is defined as the mean accuracy of all participating clients. Thus, we can calculate it by,
\begin{subequations}\label{Opt_22}
	\vspace{-0.1cm}
	\setlength{\abovedisplayskip}{3pt}
	\setlength{\belowdisplayskip}{3pt}
	\begin{align}
		&\label{Opt_22:const1}
		\overline{\Upsilon}_{Acc} = \frac{\sum_{s\in\mathcal{S}}\alpha_s \cdot \Upsilon_{Acc}^s}{\sum_{s\in\mathcal{S}}\alpha_s}, \tag {22}
	\end{align}	
\end{subequations}
where $\Upsilon_{Acc}^s$ represents the accuracy of client $s$ that can be computed via eq. \eqref{Opt_21}.
\par
Regarding \textbf{precision} (denoted as $\Upsilon_{Prec}$), it can be calculated as below \cite{J_Zhang_MFCP_Net}:
\begin{subequations}\label{Opt_23}
	\begin{align}
		&\label{Opt_23:const1}
		\Upsilon_{Prec} = \frac{TP}{TP+FP}. \tag {23}
	\end{align}	
\end{subequations}
For \textbf{average precision} $\overline{\Upsilon}_{Prec}$, we define it as the mean precision of all participating clients which can be given by,
\begin{subequations}\label{Opt_24}
	\begin{align}
		&\label{Opt_24:const1}
		\overline{\Upsilon}_{Prec} = \frac{\sum_{s\in\mathcal{S}}\alpha_s \cdot \Upsilon_{Prec}^s}{\sum_{s\in\mathcal{S}}\alpha_s}, \tag {24}
	\end{align}	
\end{subequations}
where $\Upsilon_{Prec}^s$ denotes the precision of client $s$ that can be computed via eq. \eqref{Opt_23}. 
\par
With respect to \textbf{recall}, $\Upsilon_{Rcl}$, it can be obtained by \cite{J_Zhang_MFCP_Net},
\begin{subequations}\label{Opt_25}
	\vspace{-0.1cm}
	\setlength{\abovedisplayskip}{3pt}
	\setlength{\belowdisplayskip}{3pt}
	\begin{align}
		&\label{Opt_25:const1}
		\Upsilon_{Rcl} = \frac{TP}{TP+FN}. \tag {25}
	\end{align}	
\end{subequations}
Accordingly, the \textbf{average recall} $\overline{\Upsilon}_{Rcl}$, defined as the mean recall across all participating clients, can be computed by,
\begin{subequations}\label{Opt_26}
	\vspace{-0.1cm}
	\setlength{\abovedisplayskip}{3pt}
	\setlength{\belowdisplayskip}{3pt}
	\begin{align}
		&\label{Opt_26:const1}
		\overline{\Upsilon}_{Rcl} = \frac{\sum_{s\in\mathcal{S}}\alpha_s \cdot \Upsilon_{Rcl}^s}{\sum_{s\in\mathcal{S}}\alpha_s}, \tag {26}
	\end{align}	
\end{subequations}
where $\Upsilon_{Rcl}^s$ indicates the recall of each client $s$, which can be obtained by eq. \eqref{Opt_25}.
\par
\textbf{F1 score} is used to represent the harmonic average value of the precision and the recall \cite{T_Anwar_deep_learning_diagnosis_COVID_19}, which can be given as:

\begin{subequations}\label{Opt_27}
	\vspace{-0.1cm}
	\setlength{\abovedisplayskip}{3pt}
	\setlength{\belowdisplayskip}{3pt}
	\begin{align}
		&\label{Opt_27:const1}
		\Upsilon_{F1\_score} = \frac{2\Upsilon_{Prec}\Upsilon_{RCL}}{\Upsilon_{Prec}+\Upsilon_{RCL}}=\frac{2TP}{2TP+FP+FN}. \tag {27}
	\end{align}	
\end{subequations}
In terms of the \textbf{average F1 score} $\overline{\Upsilon}_{F1\_score}$, it is defined as the mean F1 score of all participating clients. Hence, we can give the following calculation manner:
\begin{subequations}\label{Opt_28}
	\vspace{-0.1cm}
	\setlength{\abovedisplayskip}{3pt}
	\setlength{\belowdisplayskip}{3pt}
	\begin{align}
		&\label{Opt_28:const1}
		\overline{\Upsilon}_{F1\_score} = \frac{\sum_{s\in\mathcal{S}}\alpha_s \cdot \Upsilon_{F1\_score}^s}{\sum_{s\in\mathcal{S}}\alpha_s}, \tag {28}
	\end{align}	
\end{subequations}
where the $\Upsilon_{F1\_score}^s$ denotes the F1 score of client $s$ that is capable of being obtained through eq. \eqref{Opt_27}.
\par
\textbf{False alarm rate (FAR)} is used to measure the proportion of normal instances that are detected as attacks. This can be known through \cite{J_Kim_Method_intrusion_detection}. The FAR is computed by \cite{J_Kim_Method_intrusion_detection},
\begin{subequations}\label{Opt_29}
	\vspace{-0.1cm}
	\setlength{\abovedisplayskip}{3pt}
	\setlength{\belowdisplayskip}{3pt}
	\begin{align}
		&\label{Opt_29:const1}
		\Upsilon_{FAR} = \frac{FP}{TN+FP}. \tag {29}
	\end{align}	
\end{subequations}

\textbf{\textcircled{2} Multi-class Classification.} \qquad For simplicity, we only use accuracy to evaluate the proposed solution in multi-class classification. The accuracy ($\Upsilon_{Acc}$) is defined as the quotient between the number of the correct detected results (denoted as $\Xi_{cdr}$) and the overall number of network traffic data in a dataset (represented as $\Xi_{overall}$).

\begin{subequations}\label{Opt_30}
\vspace{-0.1cm}
\setlength{\abovedisplayskip}{3pt}
\setlength{\belowdisplayskip}{3pt}
\begin{align}
	&\label{Opt_30:const1}
	\Upsilon_{Acc} = \frac{\Xi_{cdr}}{\Xi_{overall}}. \tag {30}
\end{align}	
\end{subequations}

\subsubsection{Efficiency Comparisons via QPMs} \label{Quantitative_Performance_Metrics_QPM}
According to \cite{O_Hoekstra_Healthcare_prediction_textual_data, Fedorchenko_Comparative_Review_IDS, T_Alldieck_Context_Aware_Fusion_RGB, Milan_Reducing_False_Alarm_IDS}, we summarize the adopted QPMs in Section \ref{QPM_Efficiency} for evaluating the efficiency of the proposed method. Thus, in this part, we will conduct efficiency comparisons between the proposed method and baselines via considered QPMs.
\par
\textbf{\textit{Various Scenarios.} \quad} In Table \ref{sec_7_tab3_variety_Scenarios_conditions}, we perform the proposed E-FPKD approach on three datasets for binary and multi-class classification, where we choose different parameter for \textit{Dirichlet Distribution} (i.e., $\delta = \{0.5, 0.9, 5\}$). It is noted that the smaller $\delta$ means higher data heterogeneity \cite{H_Li_FedTP_Federated_Learning_Transformer}. Through Table \ref{sec_7_tab3_variety_Scenarios_conditions}, it can be observed that for binary classification over the considered three datasets, the higher the data heterogeneity (i.e., smaller $\delta$), the higher the accuracy achieved by the proposed method. However, when it comes to multi-class classification, it can be seem that for both UNSW-NB15 \cite{Dataset_UNSW_NB15_1, Dataset_UNSW_NB15_2, Dataset_UNSW_NB15_3, Dataset_UNSW_NB15_4, Dataset_UNSW_NB15_5} and IoTID20 \cite{Dataset_IoTID20} datasets, the accuracy increases as $\delta$ decreases. Whereas, concerning the NSL-KDD dataset \cite{Dataset_NSL_KDD}, when $\delta$ changes from $5$ to $0.9$, the accuracy rises, but when it drifts from $0.9$ to $0.5$, the accuracy declines. Thus, we can conjecture that the proposed method can be adapted to Non-IID data environment within a certain range. For the convenience of evaluating the proposed method, we select $\delta=0.9$ in the following.

\begin{table}[htpb]
\renewcommand{\thetable}{\Roman{table}}
\setlength{\extrarowheight}{1 pt}
\caption{Accuracy achieved by the proposed E-FPKD when considering different $\delta$ for Dirichlet Distribution.}
\begin{center}
\begin{tabular}{p{0.07\textwidth}>{\centering}p{0.07\textwidth}>{\centering}p{0.11\textwidth}>{\centering\arraybackslash}p{0.11\textwidth}}
	\hline	
	\multirow{3}{*}{Dataset}&\multirow{3}{*}{$\delta$} &\multicolumn{2}{c}{Accuracy}\\\cline{3-4}
	& & Binary Classification & Multi-class Classification\\
	\hline
	\multirow{3}{*}{NSL-KDD} & $0.5$ & $\textbf{0.7972}$ & $0.6877$\\
	& $0.9$ & $0.7629$ & $\textbf{0.7170}$ \\
	& $5$ & $0.7295$ &  $0.6802$ \\ \hline
	\multirow{3}{*}{UNSW-NB15} & $0.5$ & $\textbf{0.7131}$ & $\textbf{0.6286}$ \\
	& $0.9$ & $0.6834$ & $0.6119$ \\
	& $5$ & $0.5832$ & $0.5459$ \\ \hline
	\multirow{3}{*}{IoTID20} & $0.5$ & $\textbf{0.9688}$ & $\textbf{0.7725}$ \\
	& $0.9$ & $0.9581$ & $0.7710$ \\
	& $5$ & $0.9543$ & $0.7648$ \\
	\hline
\end{tabular}
\end{center}
\label{sec_7_tab3_variety_Scenarios_conditions}
\end{table}

\par
\textbf{\textit{Scalability. }} \quad Similar to \cite{Basset_Efficient_Lightweight_Convolutional_Networks} which evaluates the scalability of their proposed solution by considering \textit{various number of clients on two datasets}, in this work, we assess the scalability of the proposed approach by considering the \textit{various scale of clients on three datasets}. Particularly, we consider both binary and multi-class classification scenarios. For each scenario, we consider three cases which consists of  $10$, $20$ and $50$ clients, respectively, where $\delta=0.9$ is adopted for Dirichlet Distribution. In Table \ref{sec_7_tab4_Scalability}, we exhibit the accuracy realized by the proposed method. For instance, for multi-class classification, considering $10$ clients, the accuracy for NSL-KDD \cite{Dataset_NSL_KDD}, UNSW-NB15 \cite{Dataset_UNSW_NB15_1, Dataset_UNSW_NB15_2, Dataset_UNSW_NB15_3, Dataset_UNSW_NB15_4, Dataset_UNSW_NB15_5} and IoTID20 \cite{Dataset_IoTID20} datasets (test part) are $0.7170$, $0.6119$ and $0.7710$, respectively. Through Table \ref{sec_7_tab4_Scalability}, it can be known that when the number of clients is on an upward trend, the accuracy is on a downward trend. The reasons for this phenomenon are speculated as follows: 1) the total data quantity of those three datasets is fixed regardless of the number of clients, 2) additionally, for each round,  it may be not all clients are willing to participate in model generation process. As observed in Table \ref{sec_7_tab4_Scalability}, the proposed approach can support varying scale of clients in both binary and multi-class classification. Thus, we can infer that the proposed method has the characteristic of scalability. For simplicity, next we will consider $10$ clients for future evaluation.  

\begin{table}[htpb]
\renewcommand{\thetable}{\Roman{table}}
\setlength{\extrarowheight}{1 pt}
\caption{Accuracy achieved by the proposed E-FPKD approach when considering various numbers of clients.}
\begin{center}
\begin{tabular}{|c|c|c|c|c|}\hline
	\multirowcell{2.5}{\bfseries Classification} & \multirowcell{2.5}{\bfseries \# Clients}
	& \multicolumn{3}{c|}{\bfseries Accuracy} \\
	\cline{3-5}
	& & \thead{\bfseries NSL-KDD} & \thead{\bfseries UNSW-NB15} & \thead{\bfseries IoTID20}\\
	\hline
	\multirowcell{4}{Binary}
	& \thead{10} & \hfil $0.7629$ & \hfil $0.6834$ & $0.9581$\\%
	\cline{2-5}
	& \thead{20} & \hfil $0.7372$ & \hfil $0.6243$ & $0.9228$ \\%
	\cline{2-5}
	& \thead{50} & \hfil $0.7299$ & \hfil $0.5843$ & $0.8849$ \\%
	\hline
	\multirowcell{4}{Multi-class}
	& \thead{10} & \hfil $0.7170$ & \hfil $0.6119$ & $0.7710$ \\%
	\cline{2-5}
	& \thead{20} & \hfil $0.6627$ & \hfil $0.5470$ & $0.7464$ \\%
	\cline{2-5}
	& \thead{50} & \hfil $0.6219$ & \hfil $0.5371$ & $0.7107$ \\%
	\hline
\end{tabular}
\end{center}
\label{sec_7_tab4_Scalability}
\end{table}

\begin{table*}[t!] 
\renewcommand{\thetable}{\Roman{table}}
\setlength{\extrarowheight}{1 pt}
\caption{Accuracy for three datasets (test part) achieved by various method. Bold numbers indicate the best results.}
\vspace{-0.15cm}
\begin{center}
\begin{tabular}{|c|c|m{2.4cm}|m{2.4cm}|m{2.4cm}|m{2.4cm}|}\hline
	\multirowcell{2.5}{\bfseries Classification} & \multirowcell{2.5}{\bfseries Dataset}
	& \multicolumn{4}{c|}{\bfseries Accuracy} \\%
	\cline{3-6}
	& & \thead{\bfseries Without FS} & \thead{\bfseries Without LR Decay}& \thead{\bfseries DNN-based RL} &\thead{\bfseries Ours}\\%
	\hline
	\multirowcell{4}{Binary}
	& \thead{NSL-KDD} & \hfil $0.7438$ & \hfil $0.7370$ & \hfil $0.7384$ &\hfil $\textbf{0.7629}$\\%
	\cline{2-6}
	& \thead{UNSW-NB15} & \hfil $0.6766$ & \hfil $0.6469$ & \hfil $0.5506$ &\hfil $\textbf{0.6834}$\\%
	\cline{2-6}
	& \thead{IoTID20} & \hfil $0.8553$ & \hfil $0.9145$ & \hfil $0.9014$ &\hfil $\textbf{0.9581}$\\%
	\hline
	\multirowcell{4}{Multi-class}
	& \thead{NSL-KDD} & \hfil $0.4481$ & \hfil $0.6971$ & \hfil $0.3308$ &\hfil $\textbf{0.7170}$ \\%
	\cline{2-6}
	& \thead{UNSW-NB15} & \hfil $0.2668$ & \hfil $0.3830$ & \hfil - &\hfil $\textbf{0.6119}$ \\%
	\cline{2-6}
	& \thead{IoTID20} & \hfil $0.5580$ & \hfil $0.3303$ & \hfil $0.5581$ &\hfil $\textbf{0.7710}$ \\%
	\hline
\end{tabular}
\end{center}
\label{sec_7_tab5_without_with_FS_LR_Decay_DNN}
\vspace{-0.35cm}
\end{table*}

\par
In Table \ref{sec_7_tab5_without_with_FS_LR_Decay_DNN}, we compare the performance of the proposed E-FPKD approach with other three methods, i.e., 1) the E-FPKD without feature selection (FS) method, 2) the E-FPKD without learning rate (LR) decay method, 3) the E-FPKD framework but adopting DNN as the representation layers (RL), in terms of binary and multi-class classification. It is noteworthy that we comprehensively employ the FS, LR decay, and CNN-based representation layers in our proposed method. Through Table \ref{sec_7_tab5_without_with_FS_LR_Decay_DNN}, it can be observed that the proposed method can achieve the highest accuracy among the four methods. Specifically, compared with the one without FS, taking multi-classification as an example, the accuracy obtained by the one without FS is $0.4481$, $0.2668$, and $0.5580$ for NSL-KDD, UNSW-NB15 and IoTID20 test dataset, respectively, while the accuracy gained by the proposed method is separately $0.7170$, $0.6119$ and $0.7710$. The reason for this phenomenon is that FS is helpful in increasing the accuracy \cite{T_Wisanwanichthan_Double_Layered_Hybrid_Approach_NIDS}. Compared with the one without LR decay, for instance, for IoTID20 dataset, the accuracy achieved by the proposed method is $\frac{0.9581}{0.9145}=1.0477\times$ greater than that of the one without LR decay in binary classification, while that in multi-class classification is $\frac{0.7710}{0.3303} = 2.3342\times$. As mentioned in \cite{Y_Zhang_Fault_Diagnosis_Rotating_Machinery}, large LR will cause the degradation of the model performance, while small LR will lead to slow convergence for model training. Since LR decay is adopted by the proposed method ($LR = LR \times 0.97^{\textrm{q}-1}$), the LR will be variable in each round, which can make the LR in the proposed method more suitable for the considered datasets. When comparing with the one adopts DNN-based representation layers (RLs), for binary classification,  it can be seen that the accuracy achieved by the proposed method on NSL-KDD, UNSW-NB15 and IoTID20 datasets are separately $\frac{0.7629-0.7384}{0.7384} \times 100\%= 3.3180\%$, $\frac{0.6834-0.5506}{0.5506} \times 100\% = 24.1191\%$ and $\frac{0.9581-0.9014}{0.9014} \times 100\% = 6.2902\%$ higher than those obtained by the DNN-based E-FPKD approach. In terms of multi-classification, for both NSL-KDD and IoTID20 datasets, it can be seen that the proposed method can acquire higher accuracy. As for the UNSW-NB15 dataset, when leveraging DNN as the representation layer, the achieved accuracy is very low, thereby we omit to mark the corresponding value in the Table. This phenomenon occurs because the convolutional layer is adopted by the proposed method, which can increase the network nonlinearity such that the accuracy can be increased \cite{Rahimilarki_CNN_wind_turbine_machine}.

\begin{figure*}[h]
\centering
\begin{subfigure}{0.26\textwidth} 
\includegraphics[width=1.\linewidth]{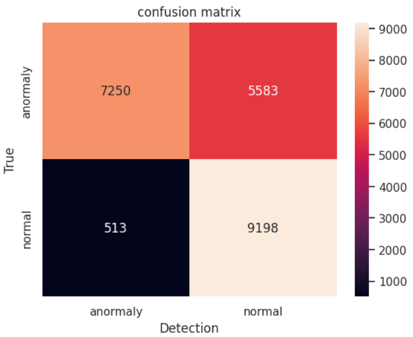}
\captionsetup{font=small} 
\caption{FedProx}
\label{FedProx_NSL_KDD}
\end{subfigure} \hfil
\begin{subfigure}{0.26\textwidth} 
\includegraphics[width=1.\linewidth]{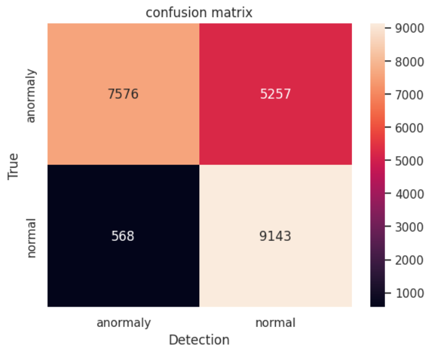}
\captionsetup{font=small} 
\caption{FedKD}
\label{FedKD_NSL_KDD}
\end{subfigure}
\hfil
\begin{subfigure}{0.26\textwidth} 
\includegraphics[width=1.\linewidth]{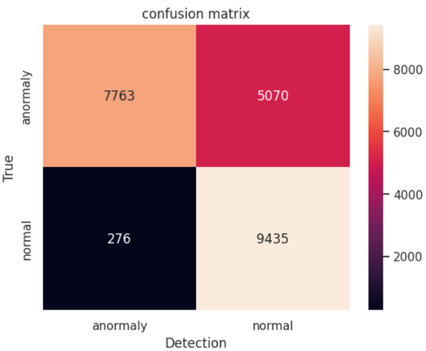}
\captionsetup{font=small} 
\caption{Ours}
\label{Ours_NSL_KDD}
\end{subfigure} \hfil
\begin{subfigure}{0.26\textwidth} 
\includegraphics[width=1.\linewidth]{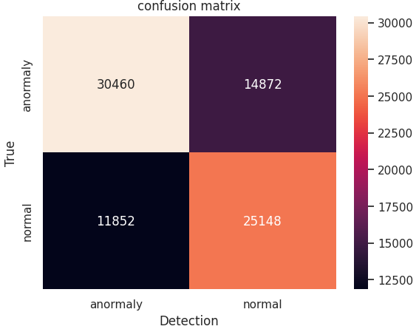}
\captionsetup{font=small} 
\caption{FedProx}
\label{FedProx_UNSW_NB15}
\end{subfigure} \hfil
\begin{subfigure}{0.26\textwidth} 
\includegraphics[width=1.\linewidth]{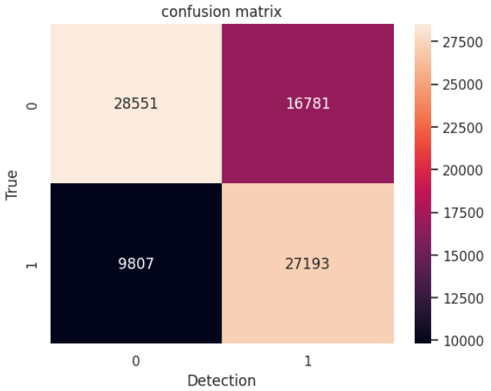}
\captionsetup{font=small} 
\caption{FedKD}
\label{FedKD_UNSW_NB15}
\end{subfigure}
\hfil
\begin{subfigure}{0.26\textwidth} 
\includegraphics[width=1.\linewidth]{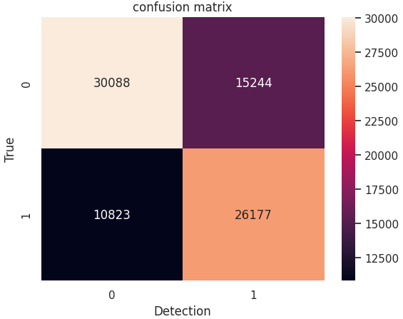}
\captionsetup{font=small} 
\caption{Ours}
\label{Ours_UNSW_NB15}
\end{subfigure}
\vspace{-0.1cm}
\captionsetup{font=small} 
\caption {Confusion matrix for binary classification using the test dataset of NSL-KDD and UNSW-NB15 dataset, respectively.} 
\label{Confusion_Matrix_Binary_Classification}
\vspace{-0.4cm}
\end{figure*}	

\begin{figure*}[h]
\centering
\begin{subfigure}{0.25\textwidth} 
\includegraphics[width=1.\linewidth]{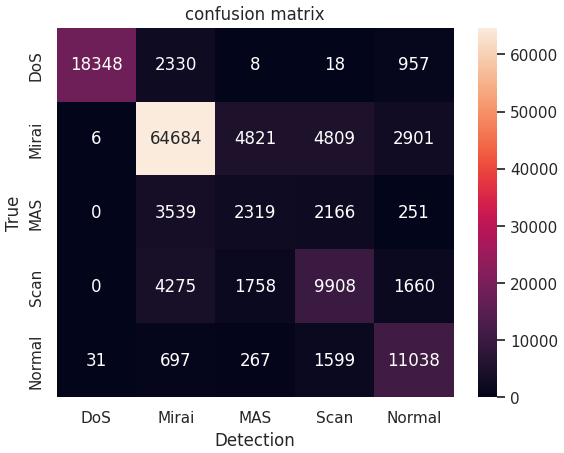}
\captionsetup{font=small} 
\caption{FedProto}
\label{FedProto_IoTID20}
\end{subfigure} \hfil
\begin{subfigure}{0.25\textwidth} 
\includegraphics[width=1.\linewidth]{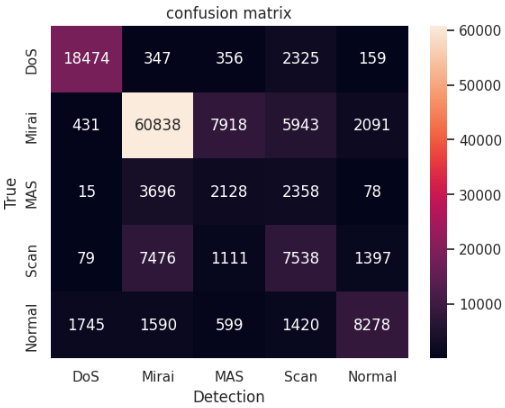}
\captionsetup{font=small} 
\caption{pFedSD}
\label{pFedSD_IoTID20}
\end{subfigure}
\hfil
\begin{subfigure}{0.25\textwidth} 
\includegraphics[width=1.\linewidth]{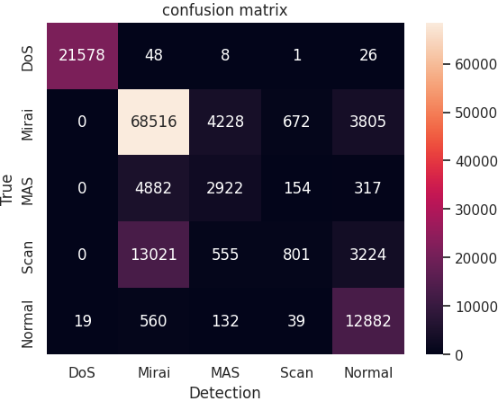}
\captionsetup{font=small} 
\caption{Ours}
\label{Ours_IoTID20}
\end{subfigure} \hfil
\vspace{-0.1cm}
\captionsetup{font=small} 
\caption {Confusion matrix for multi-class classification using the test dataset of IoTID20 dataset.}
\label{Confusion_Matrix_Multi_Class_Classification}
\vspace{-0.5cm}
\end{figure*}

\par
To reduce the occupied space in this work, and for convenience, we show the confusion matrix (CM) for binary classification by using the NSL-KDD and UNSW-NB15 datasets only (as shown in Fig. \ref{Confusion_Matrix_Binary_Classification}), while we show CM for multi-class classification with IoTID20 dataset (shown as Fig. \ref{Confusion_Matrix_Multi_Class_Classification}). According to Fig. \ref{Confusion_Matrix_Binary_Classification} and Section \ref{sec7_B1_binary_classification}, we derive Table \ref{sec_7_tab8_FedProx_FedKD_Ours_Binary}. From this Table, compared with FedProx and FedKD, for NSL-KDD dataset, it can be seen that the accuracy, precision, recall, F1 score and AA achieved by the proposed method are the highest, while the achieved FAR is the lowest. As for UNSW-NB15 dataset, although the FAR achieved by the proposed method is not the smallest and the achieved precision and recall are not the largest, the achieved accuracy, F1 Score and AA are the largest. For multi-class classification, according to Fig. \ref{Confusion_Matrix_Multi_Class_Classification}, the accuracy achieved by FedProto, pFedSD and the proposed method can be calculated as $0.7681$, $0.7028$, and $0.7710$, respectively. The reasons for such phenomenon are as follows: 1) compared with FedProx, FedKD and pFedSD, only the proposed E-FPKD approach adds the regularization term (eq. \eqref{Opt_14}) to each local loss function. As stated in \cite{D_Cheng_ProtoHAR_Prototype_FL}, the regularization term can help to gain a better representation net. 2) Regarding Fedproto, although it also uses the regularization term, KD is not adopted. Integrating KD into FL is an effective manner to handle data heterogeneity \cite{J_Tang_FedRAD_Heterogeneous}. Thus, owing to KD, the performance of the proposed method in handling non-IID data is improved. Accordingly, we can infer that the proposed method is an effective way to detect cyber-attacks.

\begin{table}[htbp]
\caption{Various metrics comparison among FedProx, FedKD, and the proposed method in binary classification.}
\renewcommand\arraystretch{1}
\begin{center}
\begin{tabular}{|c|m{1.1cm}||m{1.1cm}|m{1.1cm}|m{1.1cm}|}
	\hline
	\hfil \textbf{Dataset} & \hfil \textbf{Metric} & \hfil \textbf{FedProx} & \hfil \textbf{FedKD} &  \hfil  \textbf{Ours}\\ \hline
	\hfil \multirowcell{6}{NSL-KDD}& \hfil Accuracy & \hfil $0.7296$ & \hfil $0.7416$ & \hfil $\textbf{0.7629}$\\ \cline{2-5}
	& \hfil FAR & \hfil $0.0528$ & \hfil $0.0585$ & \hfil $\textbf{0.0284}$ \\ \cline{2-5}	
	& \hfil Precision & \hfil $0.9339$ & \hfil $0.9303$ & \hfil $\textbf{0.9657}$ \\ \cline{2-5}		
	& \hfil Recall & \hfil $0.5649$ & \hfil $0.5904$ & \hfil $\textbf{0.6049}$ \\ \cline{2-5}	
	& \hfil F1 Score & \hfil $0.7040$ & \hfil $0.7223$ & \hfil $\textbf{0.7439}$ \\ \cline{2-5}
	& \hfil AA & \hfil $0.7374$ & \hfil $0.7484$ & \hfil $\textbf{0.7709}$ \\ \hline \hline	
	\multirowcell{6}{UNSW-NB15}& \hfil Accuracy & \hfil $0.6754$ & \hfil $0.6771$ & \hfil $\textbf{0.6834}$\\ \cline{2-5}	
	& \hfil FAR & \hfil $0.3203$ & \hfil $\textbf{0.2651}$ & \hfil $0.2925$ \\ \cline{2-5}	
	& \hfil Precision & \hfil $0.7199$ & \hfil $\textbf{0.7443}$ & \hfil $0.7355$ \\ \cline{2-5}		
	& \hfil Recall & \hfil $\textbf{0.6719}$ & \hfil $0.6298$ & \hfil $0.6637$ \\ \cline{2-5}	
	& \hfil F1 Score & \hfil $0.6951$ & \hfil $0.6823$ & \hfil $\textbf{0.6977}$ \\ \cline{2-5}
	& \hfil AA & \hfil $0.6490$ & \hfil $0.6564$ & \hfil $\textbf{0.6594}$ \\ \hline							
\end{tabular}
\label{sec_7_tab8_FedProx_FedKD_Ours_Binary}
\end{center}
\vspace{-0.3cm}
\end{table}

\begin{figure}[htpb]
\centering
\begin{subfigure}{0.31\textwidth} 
\includegraphics[width=1.\linewidth]{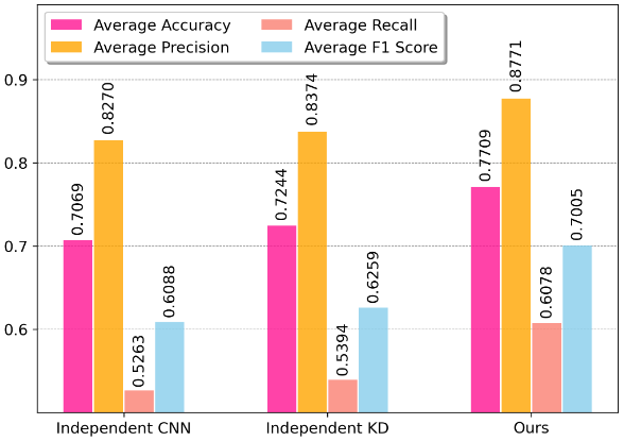}
\captionsetup{font=small}
\caption{NSL-KDD}
\label{Comparison_Among_CNN_KD_Ours_NSL_KDD}
\end{subfigure} \\
\begin{subfigure}{0.31\textwidth} 
\includegraphics[width=1.\linewidth]{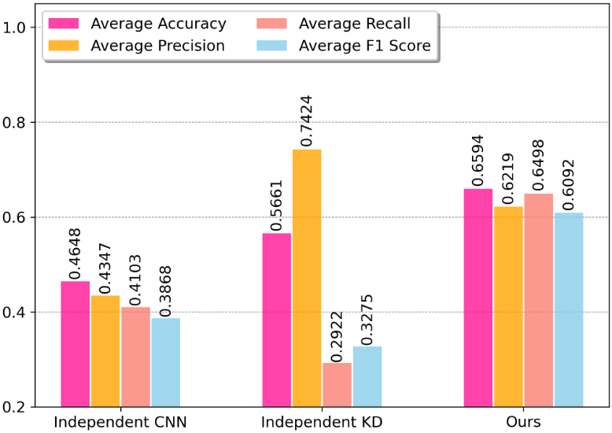}
\captionsetup{font=small}
\caption{UNSW-NB15}
\label{Comparison_Among_CNN_KD_Ours_UNSW_NB15}
\end{subfigure}
\vspace{-0.1cm}
\captionsetup{font=small}
\caption {Comparison among CNN, KD and the proposed method using the test dataset of NSL-KDD and UNSW-NB15 dataset.}
\label{Comparison_Among_CNN_KD_Ours}
\vspace{-0.45cm}
\end{figure}

\par
Next, we compare our proposed method with Independent CNN (ICNN) and Independent KD (IKD) approach regarding average accuracy, average precision, average recall, and average F1 score by using NSL-KDD and UNSW-NB15 datasets, as shown in Fig. \ref{Comparison_Among_CNN_KD_Ours}. Notably, for ICNN and IKD approaches, CNN and KD are independently applied to each client. In addition, compared with CNN, the difference of KD is that KD has one more teacher network than CNN. Through Fig. \ref{Comparison_Among_CNN_KD_Ours}, the following contents can be observed: 1) for NSL-KDD dataset, four metrics achieved by the proposed method are separately $0.7709$, $0.8771$, $0.6078$, and $0.7005$, which are the highest among those three methods, followed by IKD approach and ICNN approach. 2) For the UNSW-NB15 dataset, we can observe that the proposed method can realize the highest AA, AR, and AFS, however, the achieved AP is the second-largest. The reason for this phenomenon is: some participating clients may only have a small amount of data, as per \cite{W_Yoon_CollaboNet_collaboration_DNN}, lack of data is capable of affecting the performance. Regarding the proposed method, benefiting from the prototype aggregation process, the performance of handling heterogeneous data can be improved such that comparable results can be gained.

\begin{table}[htpb]
\renewcommand{\thetable}{\Roman{table}}
\renewcommand\arraystretch{0.75}
\setlength{\extrarowheight}{1 pt}
\caption{Accuracy comparison among various methods in both binary and multi-class classification.}
\vspace{-0.3cm}
\begin{center}
\begin{tabular}{c | c |>{\centering} *2{m{1.8cm}}}
	\Xhline{0.8pt} \XGap{1pt}	
	\multirow{3.5}{*}{Dataset}&\multirow{3.5}{*}{Method} &\multicolumn{2}{c}{Accuracy}\\\cline{3-4}
	& &  \multicolumn{2}{c}{Classification}\\ \cline{3-4}
	& &  Binary &  Multi-class \\
	\Gap \hline \Gap
	\multirow{9}{*}{NSL-KDD} & FedAvg & \hfil $0.6899$ & \hfil $0.6363$ \\ 
	& FedProto & \hfil $0.7537$ & \hfil  $0.6856$ \\
	& FedAU & \hfil $0.7337$ & \hfil $0.6433$ \\
	& FedExP & \hfil $0.7076$ & \hfil $0.6112$ \\ \Gap \cline{2-4} \Gap
	& FedGKD (2) & \hfil $0.6462$ & \hfil $0.5946$ \\
	& FedGKD (5) & \hfil $0.6769$ & \hfil $0.6629$ \\
	& pFedSD & \hfil $0.7451$ & \hfil $0.7078$\\ \Gap \cline{2-4} \Gap
	& Ours & \hfil $\textbf{0.7629}$ & \hfil $\textbf{0.7170}$  \\ \Gap \hline \Gap
	\multirow{9}{*}{UNSW-NB15} & FedAvg & \hfil $0.5645$ & \hfil $0.4904$ \\
	& FedProto & \hfil $0.6629$ & \hfil $0.5358$  \\
	& FedAU & \hfil $0.6537$ & \hfil $0.6065$ \\
	& FedExP & \hfil $0.6238$ & \hfil $0.3943$ \\ \Gap \cline{2-4} \Gap
	& FedGKD (2) & \hfil $0.5732$  & \hfil $0.4268$ \\
	& FedGKD (5) & \hfil $0.6539$ & \hfil $0.4487$ \\
	& pFedSD & \hfil $0.6757$ & \hfil $0.5565$ \\ \Gap \cline{2-4} \Gap
	& Ours & \hfil $\textbf{0.6834}$ & \hfil $\textbf{0.6119}$  \\ \Gap \hline \Gap
	\multirow{9}{*}{IoTID20} & FedAvg & \hfil $0.9015$ & \hfil $0.6706$ \\
	& FedProto & \hfil $0.9151$ & \hfil $0.7681$ \\
	& FedAU & \hfil $0.9445$  & \hfil $0.7382$ \\
	& FedExP & \hfil $0.8253$ & \hfil $0.4715$ \\ \Gap \cline{2-4} \Gap
	& FedGKD (2) & \hfil $0.8715$ & \hfil $0.4916$	 \\
	& FedGKD (5) & \hfil $0.8953$ & \hfil $0.7080$  \\
	& pFedSD & \hfil $0.9479$ & \hfil $0.7028$ \\ \Gap \cline{2-4} \Gap
	& Ours & \hfil $\textbf{0.9581}$ & \hfil $\textbf{0.7710}$ \\ \Gap 
	\hline
\end{tabular}
\end{center}
\label{sec_7_tab9_variety_Scenarios_conditions}
\vspace{-0.7cm}
\end{table}

\par
To further reveal the goodness of the proposed E-FPKD approach, we conduct the comparison with other methods over three datasets, as shown in Table \ref{sec_7_tab9_variety_Scenarios_conditions}. In particular,  FedGKD (2) \cite{D_Yao_W_Pan_FedGKD_Heterogeneous_Federated_Learning}, FedGKD (5) \cite{D_Yao_W_Pan_FedGKD_Heterogeneous_Federated_Learning}, and pFedSD \cite{H_Jin_Personalized_self_KD} are KD-based FL approaches. Notably, (2) and (5) separately represent two and five latest generated global models (including the global model generated in the current round) that will be averaged to produce the ensemble model. Through this Table, we can see the accuracy achieved by the proposed method in both binary and multi-class classification are the highest for the considered three datasets. Particularly, considering the IoTID20 dataset and binary classification as the example, the accuracy achieved by the proposed method is $\frac{0.9581}{0.9015} = 1.0628\times$, $1.0470\times$, $1.0144\times$, $1.1609\times$, $1.0994\times$, $1.0701\times$, and $1.0108\times$ greater than FedAvg \cite{H_B_McMahan_Communication_Efficient_Learning_Deep_Networks}, FedProto \cite{Y_Tan_FedProto_Federated_Prototype_Learning}, FedAU \cite{S_Wang_Lightweight_Method_Tackling}, FedExP \cite{D_Jhunjhunwala_FedExP_Speeding_Up}, FedGKD (2) \cite{D_Yao_W_Pan_FedGKD_Heterogeneous_Federated_Learning}, FedGKD (5) \cite{D_Yao_W_Pan_FedGKD_Heterogeneous_Federated_Learning} and pFedSD \cite{H_Jin_Personalized_self_KD}, respectively. Accordingly, we can infer the proposed method can realize the most significant effect in terms of accuracy among all the methods shown in this Table. The reasons for obtaining those results are that: 1) compared with FedProto, we additionally integrate KD, where as mentioned previously, adding KD to FL is helpful for handling the non-IID data. 2) In contrast with other methods, a regularization term is added to the proposed approach, which can help to learn a better representation net. Consequently, the proposed method can gain the highest accuracy.
\begin{table*}[htbp]
\renewcommand{\thetable}{\Roman{table}}
\setlength{\extrarowheight}{1 pt}
\caption{Evaluation of three methods over the test part of NSL-KDD, UNSW-NB15, and IoTID20 at different rounds.}
\begin{center}
\begin{tabular}{|c|c|m{1.3cm}|m{1.45cm}|m{1.3cm}|m{1.45cm}|m{1.3cm}|m{1.45cm}|m{1.3cm}|m{1.45cm}|}\hline
	\multirowcell{2.5}{\bfseries Dataset} & \multirowcell{2.5}{\bfseries Method}
	& \multicolumn{2}{c|}{\bfseries \# Round = 10} & \multicolumn{2}{c|}{\bfseries \# Round = 20} & \multicolumn{2}{c|}{\bfseries \# Round = 50} & \multicolumn{2}{c|}{\bfseries \# Round = 100} \\
	\cline{3-10}
	& & \thead{\bfseries Binary} & \thead{\bfseries Multi-class} & \thead{\bfseries Binary} & \thead{\bfseries Multi-class} & \thead{\bfseries Binary} & \thead{\bfseries Multi-class} & \thead{\bfseries Binary} & \thead{\bfseries Multi-class}\\
	\hline
	\multirowcell{4.5}{NSL-KDD}
	& \thead{FedAvg} & \hfil $0.6597$ & \hfil $0.6098$ & \hfil $0.6710$ & \hfil $0.6183$ & \hfil $0.6826$ & \hfil $0.6173$ & \hfil $0.6899$ & \hfil $0.6363$ \\
	\cline{2-10}
	& \thead{pFedSD} & \hfil $0.7061$ & \hfil $0.6685$ & \hfil $0.7105$ & \hfil $0.6761$ & \hfil $0.7157$ & \hfil $0.6969$ & \hfil $0.7451$ & \hfil $0.7078$ \\
	\cline{2-10}
	& \thead{Ours} & \hfil $\textbf{0.7105}$ & \hfil $\textbf{0.6833}$  & \hfil $\textbf{0.7121}$ & \hfil $\textbf{0.6970}$ & \hfil $\textbf{0.7168}$ & \hfil $\textbf{0.7037}$ & \hfil $\textbf{0.7629}$ & \hfil $\textbf{0.7170}$ \\%
	\hline \hline
	\multirowcell{2.5}{\bfseries Dataset} & \multirowcell{2.5}{\bfseries Method}
	& \multicolumn{2}{c|}{\bfseries \# Round = 5} & \multicolumn{2}{c|}{\bfseries \# Round = 10} & \multicolumn{2}{c|}{\bfseries \# Round = 20} & \multicolumn{2}{c|}{\bfseries \# Round = 50} \\%
	\cline{3-10}
	& & \thead{\bfseries Binary} & \thead{\bfseries Multi-class} & \thead{\bfseries Binary} & \thead{\bfseries Multi-class} & \thead{\bfseries Binary} & \thead{\bfseries Multi-class} & \thead{\bfseries Binary} & \thead{\bfseries Multi-class} \\%
	\hline
	\multirowcell{4.5}{UNSW-NB15}
	& \thead{FedAvg} & \hfil $0.5193$ & \hfil $0.3402$ & \hfil $0.5489$ & \hfil $0.4110$ & \hfil $0.5505$ & \hfil $0.4159$ & \hfil $0.5645$ & \hfil $0.4904$ \\
	\cline{2-10}
	& \thead{pFedSD} & \hfil $0.4350$ & \hfil $\textbf{0.5072}$ & \hfil $0.4579$ & \hfil $0.5343$ & \hfil $0.6307$ & \hfil $0.5437$ & \hfil $0.6757$ & \hfil $0.5565$ \\%
	\cline{2-10}
	& \thead{Ours} & \hfil $\textbf{0.5506}$ & \hfil $0.4488$ & \hfil $\textbf{0.5973}$ & \hfil $\textbf{0.5390}$ & \hfil $\textbf{0.6821}$ & \hfil $\textbf{0.5701}$ & \hfil $\textbf{0.6834}$ & \hfil $\textbf{0.6119}$ \\%
	\hline \hline
	\multirowcell{2.5}{\bfseries Dataset} & \multirowcell{2.5}{\bfseries Method}
	& \multicolumn{2}{c|}{\bfseries \# Round = 5} & \multicolumn{2}{c|}{\bfseries \# Round = 10} & \multicolumn{2}{c|}{\bfseries \# Round = 15} & \multicolumn{2}{c|}{\bfseries \# Round = 20} \\%
	\cline{3-10}
	& & \thead{\bfseries Binary} & \thead{\bfseries Multi-class} & \thead{\bfseries Binary} & \thead{\bfseries Multi-class} & \thead{\bfseries Binary} & \thead{\bfseries Multi-class} & \thead{\bfseries Binary} & \thead{\bfseries Multi-class} \\%
	\hline
	\multirowcell{4.5}{IoTID20}
	& \thead{FedAvg} & \hfil $0.9004$ & \hfil $0.5580$ & \hfil $0.9009$ & \hfil $0.6053$ & \hfil $0.9015$ & \hfil $0.6314$ & \hfil $0.9172$ & \hfil $0.6706$ \\
	\cline{2-10}
	& \thead{pFedSD} & \hfil $0.9259$ & \hfil $0.3955$  & \hfil $0.9317$ & \hfil $0.6112$ & \hfil $0.9364$ & \hfil $0.6676$ & \hfil $0.9479$ & \hfil $0.7028$ \\
	\cline{2-10}
	& \thead{Ours} & \hfil $\textbf{0.9459}$ & \hfil $\textbf{0.7154}$  & \hfil $\textbf{0.9512}$ & \hfil $\textbf{0.7617}$ & \hfil $\textbf{0.9529}$ & \hfil $\textbf{0.7484}$ & \hfil $\textbf{0.9581}$ & \hfil $\textbf{0.7710}$ \\%
	\hline
\end{tabular}
\end{center}
\label{sec_7_tab10_Robustness}
\end{table*}

\begin{table*}[htbp]
\renewcommand{\thetable}{\Roman{table}}
\setlength{\extrarowheight}{1 pt}
\caption{ODC achieved by eq. \eqref{Opt_5} realized by various methods on three test datasets.}
\begin{center}
\begin{tabular}{|c|c|m{2.1cm}|m{1.7cm}|m{2.1cm}|m{1.7cm}|m{2.1cm}|m{1.7cm}|}\hline
	\multirowcell{2.7}{\bfseries Classification} & \multirowcell{2.7}{\bfseries Method}
	& \multicolumn{2}{c|}{\bfseries NSL-KDD} & \multicolumn{2}{c|}{\bfseries  UNSW-NB15} & \multicolumn{2}{c|}{\bfseries  IoTID20} \\%
	\cline{3-8}
	& & \thead{\bfseries Ground Truth} & \thead{\bfseries ODC} & \thead{\bfseries Ground Truth} & \thead{\bfseries ODC} & \thead{\bfseries Ground Truth} & \thead{\bfseries ODC} \\%
	\hline
	\multirowcell{12}{Binary}
	& \thead{FedAvg} & \hfil \multirowcell{24}{$22544$} & \hfil $15553$ & \hfil \multirowcell{24}{$82332$} & \hfil $46478$  & \hfil \multirowcell{24}{$138390$} & \hfil $126929$ \\%
	\cline{2-2} \cline{4-4} \cline{6-6} \cline{8-8}
	& \thead{FedProto} & \hfil  & \hfil $16849$ & \hfil & \hfil $54578$ & \hfil  & \hfil $126634$ \\%
	\cline{2-2} \cline{4-4} \cline{6-6} \cline{8-8}
	& \thead{FedGKD(2)} & \hfil  & \hfil $14568$ & \hfil & \hfil $47190$ & \hfil  & \hfil $120610$ \\%
	\cline{2-2} \cline{4-4} \cline{6-6} \cline{8-8}
	& \thead{FedGKD(5)} & \hfil  & \hfil $15260$ & \hfil & \hfil $53837$  & \hfil  & \hfil $123896$ \\%
	\cline{2-2} \cline{4-4} \cline{6-6} \cline{8-8}
	& \thead{pFedSD} & \hfil  & \hfil $16798$ & \hfil & \hfil $55632$ & \hfil  & \hfil $131178$ \\%
	\cline{2-2} \cline{4-4} \cline{6-6} \cline{8-8}
	& \thead{FedAU} & \hfil  & \hfil $16541$ & \hfil & \hfil $53824$ & \hfil  & \hfil $130706$ \\%
	\cline{2-2} \cline{4-4} \cline{6-6} \cline{8-8}
	& \thead{FedExP} & \hfil  & \hfil $15953$ & \hfil & \hfil $51362$ & \hfil  & \hfil $114209$ \\%
	\cline{2-2} \cline{4-4} \cline{6-6} \cline{8-8}
	& \thead{Ours} & \hfil  & \hfil $\textbf{17198}$  & \hfil  & \hfil $\textbf{56265}$ & \hfil  & \hfil $\textbf{132596}$  \\%
	\hhline{|==|~|=|~|=|~|=|}
	\multirowcell{12}{Multi-class}
	& \thead{FedAvg} & \hfil  & \hfil $14345$ & \hfil & \hfil $40375$ & \hfil  & \hfil $92803$  \\%
	\cline{2-2} \cline{4-4} \cline{6-6} \cline{8-8}
	& \thead{FedProto} & \hfil  & \hfil $15456$  & \hfil & \hfil $44111$ & \hfil  & \hfil $106297$ \\%
	\cline{2-2} \cline{4-4} \cline{6-6} \cline{8-8}
	& \thead{FedGKD(2)} & \hfil  & \hfil $13404$  & \hfil & \hfil $35138$ & \hfil  & \hfil $68031$ \\%
	\cline{2-2} \cline{4-4} \cline{6-6} \cline{8-8}
	& \thead{FedGKD(5)} & \hfil  & \hfil $14944$ & \hfil & \hfil $36942$ & \hfil  & \hfil $97983$ \\%
	\cline{2-2} \cline{4-4} \cline{6-6} \cline{8-8}
	& \thead{pFedSD} & \hfil  & \hfil $15956$ & \hfil & \hfil $45814$  & \hfil  & \hfil $97256$  \\%
	\cline{2-2} \cline{4-4} \cline{6-6} \cline{8-8}
	& \thead{FedAU} & \hfil  & \hfil $14502$  & \hfil & \hfil $49935$ & \hfil  & \hfil $102156$  \\%
	\cline{2-2} \cline{4-4} \cline{6-6} \cline{8-8}
	& \thead{FedExP} & \hfil  & \hfil $13778$ & \hfil & \hfil $32464$ & \hfil  & \hfil $65256$ \\%
	\cline{2-2} \cline{4-4} \cline{6-6} \cline{8-8}
	& \thead{Ours} & \hfil  & \hfil $\textbf{16343}$  & \hfil  & \hfil $\textbf{50380}$ & \hfil  & \hfil $\textbf{106699}$  \\%
	\hline
\end{tabular}
\end{center}
\label{sec_7_tab11_ODC}
\vspace{-0.6cm}
\end{table*}

\par
\textbf{\textit{Robustness of the Proposed Method.}} \quad To evaluate the robustness of the proposed method, same as \cite{D_Yao_W_Pan_FedGKD_Heterogeneous_Federated_Learning}, we also consider demonstrating the achieved accuracy across different rounds (as shown in Table \ref{sec_7_tab10_Robustness}). In this Table, compared with FedAvg and pFedSD, for NSL-KDD and IoTID20 datasets, the accuracy obtained by the proposed method is the best in both binary and multi-class classification at various rounds. As for UNSW-NB15 dataset, the proposed method can realize the maximum accuracy in binary classification. In multi-class classification, in comparison with FedAvg, the achieved accuracy via the proposed method is higher. However, compared with pFedSD, the accuracy obtained via the proposed approach is lower when considering $5$ rounds. This is because the model training is still in the exploration stage in the initial rounds, it is possible to get lower accuracy via the proposed method than that obtained by pFedSD for multi-class classification of the UNSW-NB15 dataset. This trend is changed when the round is increased to $10$, $20$ and $50$. Hence, overall, it can be conjectured that the robustness of the proposed method is significant to some extent, compared to FedAvg and pFedSD. 
\par
In Table \ref{sec_7_tab11_ODC}, we clarify the ODC (eq. \eqref{Opt_5}) obtained by various FL-based approach. From this Table, it can be known that the proposed method can get maximum ODC for the adopted three datasets among all the listed FL-based approaches. To be specific, the ODC obtained via the proposed method is separately $17198$, $56265$ and $132596$ for NSL-KDD, UNSW-NB15, and IoTID20 datasets (test part) in binary classification, while the obtained ODC is $16343$, $50380$ and $106699$, respectively, for multi-class classification. The reason for getting these results is that the proposed method can achieve the highest accuracy compared with other baselines (the achieved accuracy can be seen in Table \ref{sec_7_tab9_variety_Scenarios_conditions}). The higher the  accuracy, the higher the  ODC. Thus, we can infer that the proposed method outperforms other listed FL-based methods.

\subsection{Evaluation via Qualitative Analysis}
\label{Evaluation_via_Qualitative_Analysis_VII}
\subsubsection{Qualitative analysis by analyzing several advantages} Identical to \cite{Rehman_Blockchain_Based_Reputation_Aware_FL} and \cite{Abbas_UAVs_Blockchain_Synergy_Secure_Reputation} which made the qualitative comparison by listing several advantages, in this work, the same manner is adopted. As can be seen in Table \ref{qualitative_analysis_by_advantages}, we compare the proposed method with FedAvg, pFedSD, FedProto and FedGKD. Specifically, since FL can handle the data heterogeneity issue and all the considered methods are FL-based approaches, we can know all the methods in Table \ref{qualitative_analysis_by_advantages} can address NT data heterogeneity issues. Particularly, because the proposed method can achieve the maximum ODC (i.e., \eqref{Opt_5}) and the highest accuracy (as shown in Section \ref{Quantitative_Performance_Metrics_QPM}), it can be inferred that the performance of the proposed method in addressing the non-IID NT data of this work is the best among those methods. In addition, all of them can solve the data privacy issue caused by centralized gathering, since centralized data collection is not required. Considering the input data can be deduced through the given model \cite{Z_Xiong_Privacy_Thread_FL_NonIID_AIoT}, data leakage is taken into account. Among the considered methods, only FedProto and the proposed method can reduce data leakage risk caused by model inverse inference. This is because prototype aggregation is employed in each round for FedProto, while model aggregation (MA) is only adopted in the final round for the proposed method. Through such a manner, the model inverse inference can be mitigated. Apart from this, for future application, when a new client joins, only FedProto requires new client to conduct a fine-tuning process. The reason is that MA is adopted by FedAvg, pFedSD, and FedGKD in each round, and employed by the proposed method in the final round. Thereby, the final generated global model can be directly used by the newly joined client without the need of fine-tuning. As for FedProto, it generates global prototypes which can not be directly used as a model. Further, only FedGKD requires the server buffer to save historical global models. Because FedGKD needs to save several historical global models for generating the ensemble model.  

\begin{table*}[htbp]
	\caption{Qualitative Comparison Among Various Methods Regarding Advantages.}
	\setlength{\tabcolsep}{5.1mm}
	\begin{center}
		\begin{tabular}{|m{7.3cm}|m{0.6cm}|m{0.7cm}|m{0.8cm}|m{0.8cm}|m{1.3cm}|}			
			\hline
			\hfil \textbf{Advantages} & \textbf{FedAvg} & \textbf{pFedSD} & \textbf{FedProto} &  \textbf{FedGKD} & \hfil \textbf{Ours} \\
			\hline	
			\hfil Support NT data heterogeneity issue (i.e., non-IID NT data) &\hfil Yes &\hfil Yes &\hfil Yes &\hfil Yes & Yes \& Best\\			
			\hline		
			\hfil Solve data privacy issue caused by centralized gathering  &\hfil Yes &\hfil Yes &\hfil Yes &\hfil Yes  &\hfil Yes \\			
			\hline
		    \hfil Reduce data leakage caused by model inverse inference &\hfil No &\hfil No &\hfil Yes &\hfil No &\hfil Yes \\			
			\hline	
		    \hfil Eliminate the fine-tuning process for newly joined client &\hfil Yes  &\hfil Yes &\hfil No &\hfil Yes &\hfil Yes \\			
			\hline
			\hfil Need server buffer to save historical global models &\hfil No &\hfil No &\hfil No  &\hfil Yes &\hfil No \\			
			\hline		
		\end{tabular}
		\label{qualitative_analysis_by_advantages}
	\end{center}
	\vspace{-0.5cm}
\end{table*}  	 
\subsubsection{Qualitative analysis by analyzing Computing Resource Usage}
In this part, we will compare the computing resource usage of the proposed method with FedAvg, pFedSD, FedProto, and FedGKD. In particular, we compare the server and each client of the proposed method with those baselines in terms of computing resource usage. Notably, after detecting the cyber-attacks with each method, we simply adopt a basic rule-based method to directly block the malicious network traffic. Thus, for simplicity, we only consider the computing resource utilization for cyber-attack detection. Specifically, compared with FedAvg and FedProto, for each client, pre-training TNet is required by the proposed method, which causes the proposed method to consume additional computing resources. During the client's student network training, FedProto needs to compute the regularization term and the local prototypes, but the client of FedAvg does not need to. In contrast to FedProto, the proposed method also needs to calculate the KD loss. Accordingly, the client of proposed method requires the most computational resources compared with FedAvg and FedProto. Regarding the server, model aggregation (MA) is adopted by FedAvg in each round, while prototype aggregation (PA) is employed by FedProto. When it comes to the proposed method, the PA is performed in each round, and the MA is only executed in the final round. Hence, we can infer the computing resource usage of the server of the proposed method is in the middle of FedAvg and FedProto. As for FedGKD and pFedSD, compared with FedAvg, those methods need to additionally compute the KD loss. In addition to KD loss, as the proposed method also needs to calculate the prototype in each client, the client computation resource utilization of the proposed method is higher than FedGKD's client and pFedSD's client. Moreover, both FedGKD and pFedSD adopt model aggregation in each round. For FedGKD, it is necessary to calculate the ensemble model by using several models. As such, among FedGKD, pFedSD and the proposed method, the server computing resource usage of the proposed method is the smallest, and that of FedGKD is the largest. Accordingly, the order of server's computing resource usage is: FedGKD $>$ pFedSD $=$ FedAvg $>$ E-FPKD (Ours) $>$ FedProto. Considering the purpose of this work is to maximize the ODC \eqref{Opt_5}, for the proposed method, although the computing resources separately used by each client and the server are not the smallest, its efficiency (as shown in Section \ref{Quantitative_Performance_Metrics_QPM}) is the best in achieving \eqref{Opt_5}. Therefore, it is feasible to adopt the proposed E-FPKD for cyber-attack detection.

\section{Discussion}
\label{Section_8_Discussion}
\par
\textbf{\textit{Limitations. }} \quad In this work, although the ODC achieved by the proposed solution is competitive compared to other baselines (e.g., FedProto), the proposed solution still has two limitations. Concretely, supervised learning is adopted to construct the proposed E-FPKD approach. Owing to this, to generate the model, labeled data are required. In such a way, unknown patterns will fail to be recognized \cite{M_Verkerken_Unsupervised_Machine_Learning_Network_ID}. In addition, in the final round, model aggregation is adopted to generate the global model. According to \cite{Z_Xiong_Privacy_Thread_FL_NonIID_AIoT}, with the given model, input data can be inferred. Thus, although the proposed solution can solve the data privacy issue caused by centralized gathering data, the risk of data leakage may still occur due to the potential of model inverse inference in the final round. 	
\par
\textbf{\textit{Future Potential Research Areas. }} \quad Future potential research areas can be conducted based on above two limitations. Specifically, rather than supervised learning, unsupervised learning \cite{M_Verkerken_Unsupervised_Machine_Learning_Network_ID} (UL) or semi-supervised learning \cite{Y_Gao_Novel_Semi_Supervised_Learning_NID_Cloud} (SSL) can be leveraged. Thus, combining FL with advanced UL or SSL can be a potential research direction. To mitigate the data leakage issue, model protecting technique can be considered. For instance, masks and homomorphic encryption are possible to be employed since they can prevent data from being inferred by a variety of attacks (e.g., model inversion attack) \cite{L_Zhang_Vijayakumar_Homomorphic_Encryption_Privacy_Preserving_FL}.

\section{Conclusion}
\label{Section_9_conclusion}
In this paper, cyber-attack prevention for prosumer-based EV charging stations is focused, where the network traffic data of all prosumers has heterogeneous characteristics (i.e., non-IID). In particular, the considered cyber-attack prevention process covers two aspects: cyber-attack detection and intervention. The purpose is to maximize the ODC. To tackle this aspect, an edge-assisted federated prototype knowledge distillation (E-FPKD) approach has been proposed. To improve accuracy, we have conducted feature selection with Pearson Correlation Coefficient before using the proposed E-FPKD approach. In addition, we have considered deploying each client model on a dedicated local edge server, where each client can launch its availability for participating in the FL process. Apart from this, we have integrated the KD technique for each client, which allows the student model can learn from the teacher model to gain useful information. Besides, we have combined the KD with the prototype aggregation for performance enhancement. Then, the global model obtained by averaging all the participating clients' personal models in the final round is adopted for the application stage such that the newly joined clients can directly use it without the necessity of fine-tuning. Finally, to intervene in the detected cyber attacks, a rule-based method called \textit{``if...then"} is employed. The conducted experiments clarify the practicality of the proposed method. For instance, in binary classification, the proposed method can achieve the highest ODC for the considered three datasets separately with a value of $17198$, $56265$, and $132596$, compared with other considered FL-based methods. 
\appendix[Applications in other domains]
\begin{table}[htbp]
\caption{Application in other fields via proposed E-FPKD approach ($\delta = 0.9$, \# Clients = $10$).}
\setlength{\tabcolsep}{5.1mm}
\begin{center}
\begin{tabular}{|m{4cm}|m{2cm}|}			
	\hline
	\hfil \textbf{Dataset} & \hfil \textbf{Accuracy}  \\
	\hline			
	\hfil Smart Grid Stability (Binary) & \hfil $0.8571$  \\			
	\hline
	\hfil TD\_SGE (Binary) & \hfil  $0.6626$ \\ \hline
	\hfil TD\_SGE (Multi-class) & \hfil $0.5856$    \\			
	\hline					
\end{tabular}
\label{appendix_other_domains}
\end{center}
\end{table}  	
\textbf{\textit{Applicability. }} \quad In this part, we shed light on the \textit{applicability} of the proposed E-FPKD approach. As stated before, 1D-CNN combined with FC network, and supervised learning are used to construct the proposed method. Thus, the application range of the proposed method can be: since 1D-CNN is utilized to receive the input data, if the data is labeled data and each data record can be denoted as a vector, the proposed method can be leveraged. Next, as examples, we apply the proposed approach to two cases in other domains, where the detailed scenario will be discussed based on each case.
\par
\textbf{\textit{Case 1: Smart Grid Stability Analysis.}} \quad Considering an urban area that consists of multiple regions and a distribution system operator (DSO) for buying power grid energy, particularly, each region is formed by one energy producer and multiple energy consumers. For such an urban area, we aim to analyze its smart grid stability by the proposed approach. Specifically, each region is considered as the client of the proposed approach and the aggregation process will be executed in the DSO. To perform this, Smart Grid Stability (SGS) dataset \cite{smart_grid_stability_dataset} is adopted since each record of that dataset is labeled by stable or unstable, and it contains the information regarding a energy producer and multiple energy consumers. As SGS dataset only includes two categories (i.e., \{\texttt{unstable, stable}\}), only binary classification scenario is considered in this case. Further, we use \textit{Dirichlet Distribution} with parameter $\delta = 0.9$ to prepare data for $10$ clients.
\par
\textbf{\textit{Case 2: Energy Theft Detection.}} \quad Consider a smart grid environment that is composed of multiple energy users and a DSO (same as case $1$), we aim to detect the energy theft towards such an environment with the proposed approach. Specifically, each energy user is considered as each client, and the aggregation process is performed in DSO. For evaluation, we employ Theft detection in smart grid environment dataset (TD\_SGE) \cite{theft_detection_in_smart_grid_environment_dataset} which contains energy consumption data belonging to \textit{normal} or \textit{six types of thefts}. Thus, we consider both binary (i.e., \{\texttt{Theft, Normal}\}) and multi-class classification (i.e., \{\texttt{Theft1, Theft2, Theft3, Theft4, Theft5, Theft6, Normal}\}) scenarios for TD\_SGE dataset. In addition, we also use \textit{Dirichlet Distribution} with parameter $\delta = 0.9$ to split data for $10$ clients.
\par
In Table \ref{appendix_other_domains}, we clarify the accuracy achieved by the proposed method for both Case $1$ and Case $2$. We can see that the proposed approach can also be applied to other domains.

\begin{IEEEbiography}[{\includegraphics[width=1in,height=1.25in,clip,keepaspectratio]{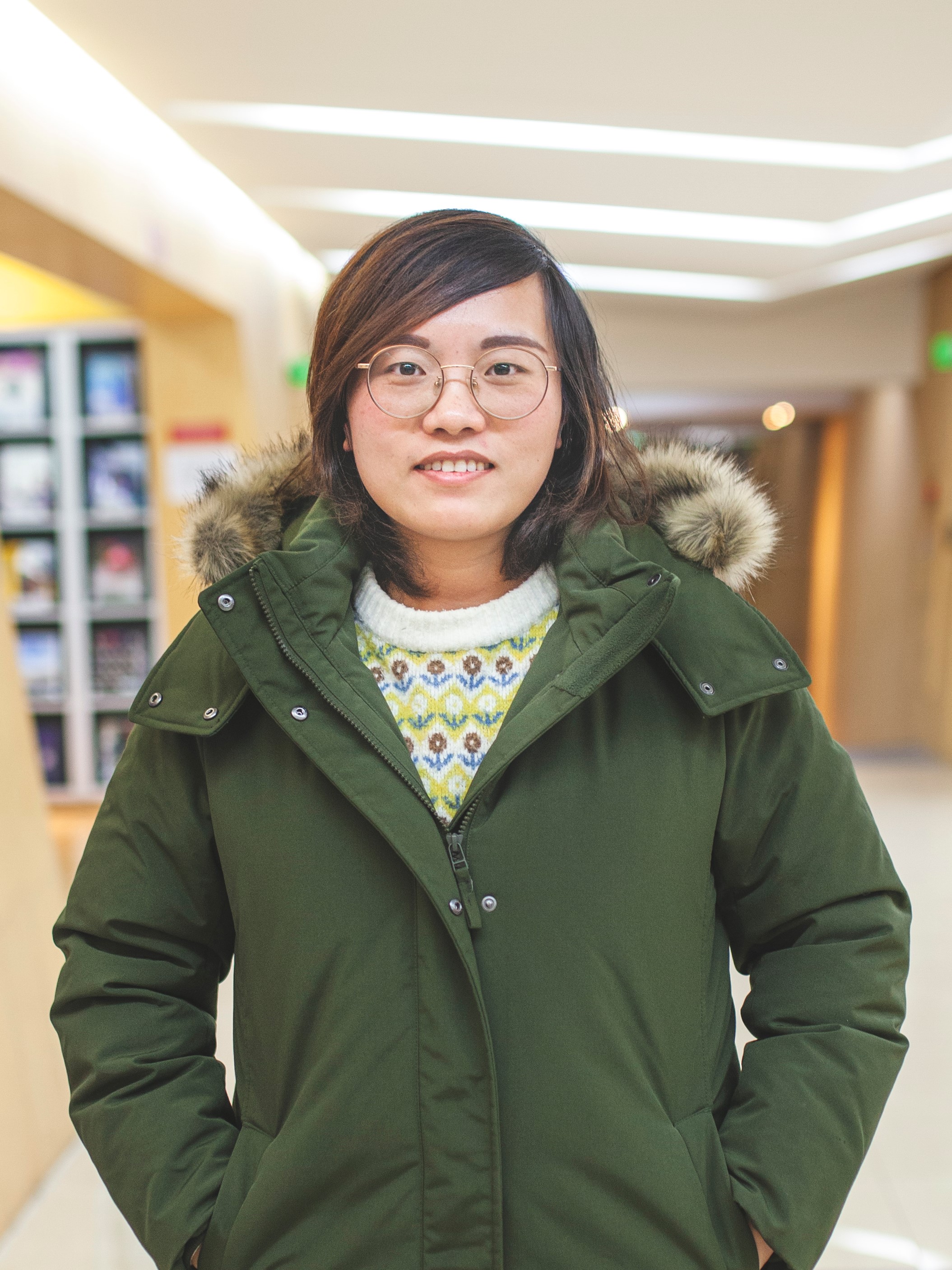}}] {Luyao Zou}

received the B.E. degree in electrical engineering and automation from NingboTech University (current name), China in 2014, the M.S. degree in the major of computer science and engineering from the Department of Computer, Information \& Communications Engineering at Konkuk University, South Korea in 2018, and the Ph.D. degree in computer engineering from the Department of Computer Science and Engineering at Kyung Hee University, South Korea in 2024, where she is currently working as a Postdoctoral Fellow. Her current research interests include edge intelligence and deep learning.

\end{IEEEbiography}
\vspace{-0.8cm}
\begin{IEEEbiography}[{\includegraphics[width=1in,height=1.25in,clip,keepaspectratio]{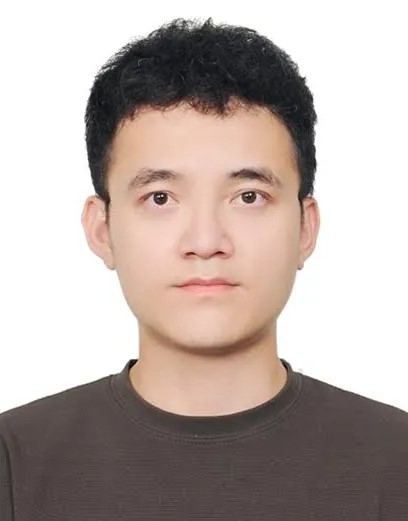}}] {Quang Hieu Vo}
received the B.S degree in Electrical Engineering from Ho Chi Minh City University of Technology, Ho Chi Minh, Vietnam, in 2017. Between 2017 and 2018, he worked at eSilicon Vietnam Inc., Ho Chi Minh, Vietnam, as an ASIC design engineer. In 2024, he received the Ph.D. degree in Computer Science and Engineering at Kyung Hee Univerity, Korea, where he was working on efficient deep learning architecture. Since 2024, he has been a research engineer at Samsung Electric, South Korea. His research interests include the area of computer vision, machine learning, VLSI system design, Deep learning hardware architecture, and automated design optimization.
\end{IEEEbiography}
\vspace{-0.5cm}
\begin{IEEEbiography}[{\includegraphics[width=1in,height=1.25in,clip,keepaspectratio]{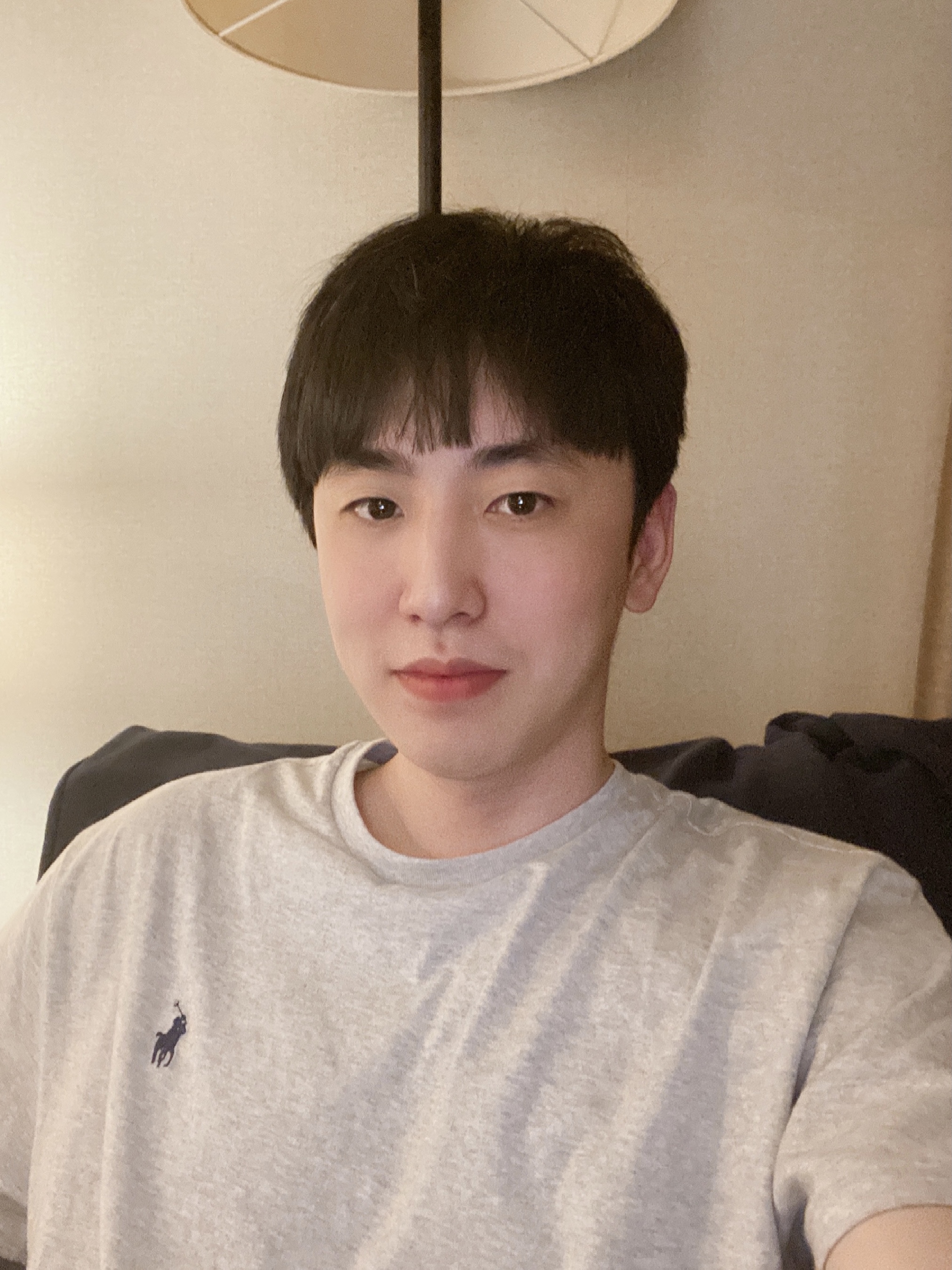}}] {Kitae Kim}
	received his B.S., M.S., and Ph.D. degrees in Computer Science and Engineering from Kyung Hee University, Seoul, South Korea, in 2017, 2019, and 2024, respectively. He is currently a Postdoctoral Researcher at the Networking Intelligence Laboratory at Kyung Hee University. His research focuses on 5G/6G wireless communication, AI/ML-based wireless networks, and channel estimation and prediction.
\end{IEEEbiography}
\vspace{-0.6cm}
\begin{IEEEbiography}[{\includegraphics[width=1in,height=1.25in,clip,keepaspectratio]{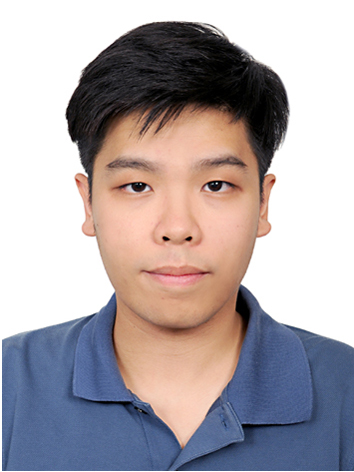}}] {Huy Q. Le}
received the B.E. degree in Control Engineering Automation from the Ho Chi Minh City University of Technology, Ho Chi Minh City, Vietnam, in 2019. He is currently pursuing a Ph.D. degree in Computer Science at Kyung Hee University, South Korea. His research interests include machine learning, federated learning, self-supervised learning, and multimodal learning.
\end{IEEEbiography}
\vspace{-0.6cm}
\begin{IEEEbiography}[{\includegraphics[width=1in,height=1.25in,clip,keepaspectratio]{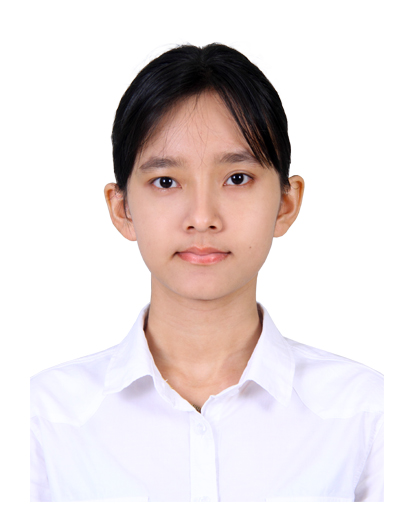}}] {Chu Myaet Thwal}
	received the B.E degree in Computer Engineering and Information Technology from Yangon Technological University, Myanmar in 2019. She is currently pursuing the Ph.D. degree in Computer Science and Engineering at Kyung Hee University, South Korea. Her research interests include Deep Learning, Artificial Intelligence, and Federated Learning.
\end{IEEEbiography}
\vspace{-1cm}
\begin{IEEEbiography}[{\includegraphics[width=1in,height=1.25in,clip,keepaspectratio]{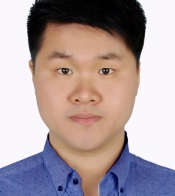}}] {Chaoning Zhang} received the B.E. and M.E. degree in Electrical Engineering from Harbin Institute of Technology, China, in 2012 and 2015, respectively, and the Ph.D. degree from KAIST in 2018. Since 2019, He has been an assistant professor at the department of Artificial Intelligence, School of Computing, Kyung Hee University. Prior to that, he worked as a post-doctoral researcher at KAIST. His research interests include but are not limited to adversarial machine learning and self-supervised learning for addressing model robustness and data efficiency issues in computer vision and beyond. 
\end{IEEEbiography}
\vspace{-1cm}
\begin{IEEEbiography}[{\includegraphics[width=1in,height=1.25in,clip,keepaspectratio]{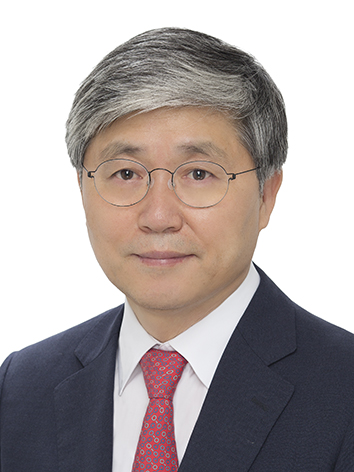}}]{Choong Seon Hong}
(S’95-M’97-SM’11-F’24) received the B.S. and M.S. degrees in electronic engineering from Kyung Hee University, Seoul, South Korea, in 1983 and 1985, respectively, and the Ph.D. degree from Keio University, Tokyo, Japan, in 1997. In 1988, he joined KT, Gyeonggi-do, South Korea, where he was involved in broadband networks as a member of the Technical Staff. Since 1993, he has been with Keio University. He was with the Telecommunications Network Laboratory, KT, as a Senior Member of Technical Staff and as the Director of the Networking Research Team until 1999. Since 1999, he has been a Professor with the Department of Computer Science and Engineering, Kyung Hee University. His research interests include future Internet, intelligent edge computing, network management, and network security. Dr. Hong is a member of the Association for Computing Machinery (ACM), the Institute of Electronics, Information and Communication Engineers (IEICE),the Information Processing Society of Japan (IPSJ), the Korean Institute of Information Scientists and Engineers (KIISE), the Korean Institute of Communications and Information Sciences (KICS), the Korean Information Processing Society (KIPS), and the Open Standards and ICT Association (OSIA). He has served as the General Chair, the TPC Chair/Member, or an Organizing Committee Member of international conferences, such as the Network Operations and Management Symposium (NOMS), International Symposium on Integrated Network Management (IM), Asia-Pacific Network Operations and Management Symposium (APNOMS), End-to-End Monitoring Techniques and Services (E2EMON), IEEE Consumer Communications and Networking Conference (CCNC), Assurance in Distributed Systems and Networks (ADSN), International Conference on Parallel Processing (ICPP), Data Integration and Mining (DIM), World Conference on Information Security Applications (WISA), Broadband Convergence Network (BcN), Telecommunication Information Networking Architecture (TINA), International Symposium on Applications and the Internet (SAINT), and International Conference on Information Networking (ICOIN). He was an Associate Editor of the IEEE TRANSACTIONS ON NETWORK AND SERVICE MANAGEMENT and the IEEE JOURNAL OF COMMUNICATIONS AND NETWORKS and an Associate Editor for the International Journal of Network Management and an Associate Technical Editor of the IEEE Communications Magazine. He currently serves as an Associate Editor for the International Journal of Network Management and Future Internet Journal.
\end{IEEEbiography}

\vfill

\end{document}